\documentclass[a4paper,11pt]{article}
\usepackage{heppub}
\pdfoutput=1
\graphicspath{{plots/}}
\usepackage{marginnote}
\usepackage{subcaption}

\usepackage{xcolor}

\newcommand{\ord}[1]{\mathcal{O}(#1)}
\newcommand{\ORd}[1]{\mathcal{O}\Bigl(#1\Bigr)}
\newcommand{\df}{\mathrm{d}}
\newcommand{\img}{\mathrm{i}}
\newcommand{\f}{\frac}

\newcommand{\MeV}{\,\mathrm{MeV}}
\newcommand{\GeV}{\,\mathrm{GeV}}
\newcommand{\TeV}{\,\mathrm{TeV}}

\newcommand{\nn}{\nonumber}

\newcommand{\eps}{\epsilon}
\newcommand{\as}{\alpha_s}

 \newcommand{\bt}{{\vec b}_T}
 \newcommand{\qt}{{\vec q}_T}

 \newcommand{\cO}{\mathcal{O}}

 \newcommand{\Ecm}{E_\mathrm{cm}}
\newcommand{\lqcd}{\Lambda_\mathrm{QCD}}
\newcommand{\qttilde}{\tilde{q}_T}

\newcommand{\sigmares}{\sigma^\res}
\newcommand{\sigmanons}{\sigma^\nons}
\newcommand{\sigmacorr}{\sigma^\mathrm{corr}}
\newcommand{\sigmasing}{\sigma^\sing}
\newcommand{\frun}{f_\mathrm{run}}
\newcommand{\grun}{g_\mathrm{run}}

\newcommand{\sigmanonstilde}{\tilde{\sigma}^\nons}
\newcommand{\sigmasingtilde}{\tilde{\sigma}^\sing}

\newcommand{\sigmaFO}{\sigma^{\FO}}
\newcommand{\muZeromin}{\mu_0^\mathrm{min}}
\newcommand{\muBmin}{\mu_B^\mathrm{min}}
\newcommand{\muSmin}{\mu_S^\mathrm{min}}
\newcommand{\nuSmin}{\nu_S^\mathrm{min}}
\newcommand{\mufmin}{\mu_f^\mathrm{min}}

\newcommand{\vary}{\mathrm{vary}}
\newcommand{\run}{\mathrm{run}}
\newcommand{\FO}{\mathrm{FO}}

\newcommand{\nons}{\mathrm{nons}}
\newcommand{\sing}{\mathrm{sing}}

\newcommand{\match}{\mathrm{match}}
\newcommand{\res}{\mathrm{res}}
\newcommand{\canon}{\mathrm{canon}}

\newcommand{\nnnllp}{N$^3$LL$'$}
\newcommand{\loone}{LO$_1$}
\newcommand{\nloone}{NLO$_1$}
\newcommand{\nnloone}{NNLO$_1$}
\newcommand{\scetlib}{\textsc{SCETlib}}
\newcommand{\geneva}{{\textsc{Geneva}}}
\newcommand{\openloops}{\textsc{OpenLoops}}

\allowdisplaybreaks[2]

\newcommand{\WidthTwoSubfigs}{0.48\textwidth}
\title{\boldmath The $q_T$ spectrum for Higgs production via heavy quark annihilation at N$^3$LL$'$+aN$^3$LO}

\author[a]{Pedro Cal,}
\author[a]{Rebecca von Kuk,}
\author[a,b]{Matthew A. Lim}
\author[a]{and Frank J. Tackmann}

\affiliation[a]{Deutsches Elektronen-Synchrotron DESY, Notkestr. 85, 22607 Hamburg, Germany}
\affiliation[b]{Department of Physics and Astronomy, University of Sussex, Sussex House, Brighton, BN1 9RH, UK\vspace{0.5ex}}

\emailAdd{pedro.cal@desy.de}
\emailAdd{rebecca.von.kuk@desy.de}
\emailAdd{m.a.lim@sussex.ac.uk}
\emailAdd{frank.tackmann@desy.de}

\abstract{%
We study the transverse momentum ($q_T$) spectrum of the Higgs boson produced
via the annihilation of heavy quarks ($s,c,b$) in proton-proton collisions.
Using soft-collinear effective theory (SCET) and working in the five-flavour
scheme, we provide predictions at three-loop order in resummed perturbation
theory (N$^3$LL$'$). We match the resummed calculation to full fixed-order results
at next-to-next-to-leading order (NNLO), and introduce a decorrelation method to
enable a consistent matching to an approximate N$^3$LO (aN$^3$LO) result. Since
the $b$-quark initiated process exhibits large nonsingular corrections, it
requires special care in the matching procedure and estimation of associated
theoretical uncertainties, which we discuss in detail. Our results constitute
the most accurate predictions to date for these processes in the small $q_T$
region and could be used to improve the determination of Higgs Yukawa couplings
from the shape of the measured Higgs $q_T$ spectrum.
}
\date{June 28, 2023}

\preprint{\vbox{%
\hbox{DESY 23-080}}
}

\begin{document}

\maketitle

%%%%%%%%%%%%%%%%%%%%%%%%%%%%%%%%%%%%%%%%%%%%%%%%%%%%%%%%%%%%%%%%%%%%%%%%%%%%%%%%
\section{Introduction}
\label{sec:intro}
%%%%%%%%%%%%%%%%%%%%%%%%%%%%%%%%%%%%%%%%%%%%%%%%%%%%%%%%%%%%%%%%%%%%%%%%%%%%%%%%

With the discovery of the Higgs boson by the ATLAS and CMS experiments at the LHC \cite{ATLAS:2012yve,CMS:2012qbp}, the precise measurement of its properties has become essential to establish the Standard Model (SM) as the true mechanism of electroweak symmetry breaking. The four main production mechanisms --  gluon fusion, vector boson fusion, Higgstrahlung, and top-quark pair associated production -- have been observed experimentally~\cite{ATLAS:2012yve,CMS:2012qbp,ATLAS:2018kot,ATLAS:2018mme,CMS:2018uxb}, while the Higgs couplings to vector bosons have been found consistent with the SM down to an accuracy of $4\%$ \cite{ATLAS:2022vkf,ATLAS:2021vrm}.

Probing the Higgs interactions with the fermionic sector is also of great importance. In the SM, the couplings of the Higgs boson to fermions, i.e.\ the Yukawa couplings $y_F$, are proportional to the fermion mass $m_F$,
$y_F^\mathrm{SM}\equiv m_F / v$, where $v$ denotes the Higgs vacuum expectation value.
This implies that the measurement of the Yukawa couplings to the heavy fermions is within the reach of the LHC. In fact, the reduced coupling strength factors $\kappa_F\equiv y_F/y_F^\mathrm{SM}$ have been measured to be $\kappa_t=0.92\pm0.06$, $\kappa_b=0.88\pm0.11$, and $\kappa_\tau=0.92 \pm 0.07$ for the top-quark, bottom-quark, and $\tau$ lepton respectively~\cite{ATLAS:2022vkf,ATLAS:2021vrm}.

The bottom-quark Yukawa coupling $y_b$ is of particular interest. For example, in SM extensions such as the two Higgs doublet model or the minimally supersymmetric SM, $y_b$ can be enhanced relative to its SM value. This coupling has been measured in $H\to b\bar{b}$ decays~\cite{ATLAS:2018kot,CMS:2018nsn}, which is challenging due to the required $b$ tagging and the huge multi-jet background. The same is true for the charm-quark Yukawa $y_c$, whose measurement from $H\to c\bar c$ decays is possible but presents an even greater challenge~\cite{ATLAS:2018mgv}. Therefore, a complementary determination of the Yukawa couplings stemming from the production process, rather than the decay, is of great interest.

In this work, we focus on Higgs production via quark annihilation $q\bar{q}\to H$, where we consider bottom, charm, and strange quarks for the incoming quarks. Of these, bottom-quark annihilation is by far the dominant process since the $b$ is the heaviest, followed by charm and then strange annihilation. Precise predictions for the $q\bar{q}\to H$ process are important, since it can in principle provide direct sensitivity to the quark Yukawa couplings from the production process. In addition, while the cross section for bottom-quark annihilation is significantly smaller than that of gluon fusion, these are often grouped together in experimental analyses, since they have very similar acceptances and are a priori hard to distinguish experimentally.

For these reasons, $q\bar{q}\to H$ production, in particular bottom-quark annihilation, has received much attention in the past, see e.g.\ \refscite{Dicus:1998hs, Balazs:1998sb,
Harlander:2003ai, Dittmaier:2003ej, Dawson:2003kb, Belyaev:2005bs, Harlander:2010cz, Harlander:2011fx,
Buhler:2012ytl, Harlander:2012pb, Harlander:2014hya, Wiesemann:2014ioa, Bonvini:2015pxa, Bonvini:2016fgf,
Forte:2015hba, Forte:2016sja, Harlander:2015xur, Lim:2016wjo, Das:2023rif}.
The $q\bar q H$ form factor and hard function have been computed up to four loops~\cite{Gehrmann:2014vha, Chakraborty:2022yan},
the total inclusive $b\bar{b}\to H$ cross section to N$^3$LO~\cite{Duhr:2019kwi, Duhr:2020kzd},
and $b\bar{b}\to H+$jet to NNLO$_1$~\cite{Mondini:2021nck}.

The direct measurement of $b\bar{b}\to H$ by tagging one or both of the accompanying $b$-jets in the final state seems to be hopeless in practice~\cite{Pagani:2020rsg}. Alternatively, one can exploit the pattern of QCD emissions from the incoming quarks and gluons to discriminate between the gluon and various quark channels in the initial state~\cite{Ebert:2016idf}. That is, the radiation pattern for different initial states yields different shapes for the transverse momentum ($q_T$) spectrum of the recoiling Higgs boson. As a result, a precise measurement and fit to the Higgs $q_T$ spectrum, especially at small $q_T$, allows one to gain sensitivity to the quark Yukawa couplings~\cite{Bishara:2016jga, Soreq:2016rae}. With sufficient statistics, this might even open a way at the LHC to obtain some constraint on the strange Yukawa coupling (or more generally the PDF-weighted sum of light-quark Yukawa couplings).

In \refscite{ATLAS:2022qef,CMS:2023gjz}, ATLAS and CMS have demonstrated that it
is already possible with existing data to obtain meaningful constraints on
$\kappa_c$ and $\kappa_b$ from the shape of the Higgs $q_T$ spectrum alone. To
fully exploit this possibility, a precise prediction of the $q_T$ spectrum for
both gluon fusion and quark annihilation is essential. At small $q_T \ll m_H$,
this requires the all-order resummation of logarithms of $q_T/m_H$ that would
otherwise spoil the convergence of perturbation theory in this regime. While a
N$^3$LL$'$+N$^3$LO resummed prediction exists for the Higgs $q_T$ spectrum in
gluon fusion~\cite{Billis:2021ecs}, which was used in \refcite{ATLAS:2022qef},
and predictions of similar accuracy also exist for
Drell-Yan~\cite{Camarda:2021ict, Re:2021con, Ju:2021lah, Neumann:2022lft, Billis:2023xxx},
no prediction of similar accuracy exists for $q\bar{q}\to H$, which so far has
only been resummed to NNLL$+$NNLO accuracy~\cite{Belyaev:2005bs,
Harlander:2014hya}.

In this paper, we fill this gap and compute the resummed $q_T$ spectrum for
$q\bar{q}\to H$ at N$^3$LL$'$ order matched to fixed NNLO and approximate
N$^3$LO. We use soft-collinear effective theory (SCET)
\cite{Bauer:2000yr,Bauer:2001ct,Bauer:2001yt,Bauer:2002nz,Beneke:2002ph} to
resum the logarithms of $q_T/m_H$. We work in the limit $m_q \ll q_T$, where we
only keep the Yukawa coupling of the annihilating quarks and otherwise take them
to be massless. For $b\bar{b}\to H$, this is commonly referred to as the
five-flavour scheme. Finite-mass effects become relevant for $m_q \sim q_T$ and
are thus necessary for a complete description of the small-$q_T$ region,
especially for $b\bar{b}\to H$~\cite{Belyaev:2005bs, Pietrulewicz:2017gxc}.
Their full treatment in the resummed $q_T$ spectrum was worked out in
\refcite{Pietrulewicz:2017gxc} and is quite involved. We therefore focus here on
the massless limit and leave the inclusion of finite-mass effects in the
resummed spectrum to future work.

This paper is organized as follows. We first provide a brief review of the
structure of $q_T$ resummation in SCET in \sec{theory}, where we also discuss
the general procedure for matching the resummed and fixed-order
calculations using profile scales and for estimating perturbative uncertainties
from profile-scale variations. In \sec{matching},
we discuss our implementation of the fixed-order results and the matching to them in some
detail. It transpires that the numerical size
of the nonsingular fixed-order corrections depends strongly on the incoming
quark flavour. In particular, they are substantially larger for $b\bar b \to H$
than what is commonly found to be the case for gluon-fusion or Drell-Yan production. This
requires additional care in the matching and some refinements to the usual
estimation of the matching uncertainties based on profile-scale variations.
Furthermore, we discuss the matching to approximate N$^3$LO, i.e., to approximate
$\ord{\as^3}$.
For this purpose, we introduce a general strategy to decorrelate the singular and
nonsingular contributions at large $q_T$, generalizing a method recently introduced in
\refcite{Dehnadi:2022prz}. This allows us to construct an approximation of the missing
$\mathcal{O}(\alpha_s^3)$ nonsingular contributions and a corresponding
approximate full NNLO$_1$ result that incorporates the exact $\ord{\as^3}$
singular contributions, which are neccessary for a consistent matching to the N$^3$LL$'$ result.
We present our numerical results for the resummed $q_T$ spectrum and its
perturbative uncertainties in \sec{results}, and offer avenues for potential
future work in \sec{conclusions}.

%%%%%%%%%%%%%%%%%%%%%%%%%%%%%%%%%%%%%%%%%%%%%%%%%%%%%%%%%%%%%%%%%%%%%%%%%%%%%%%%
\section{Theoretical framework}
\label{sec:theory}
%%%%%%%%%%%%%%%%%%%%%%%%%%%%%%%%%%%%%%%%%%%%%%%%%%%%%%%%%%%%%%%%%%%%%%%%%%%%%%%%
%\begin{itemize}
%\item Brief overview of $q_T$ resummation how we like it and how it should be done
%\item Essentially all the process-independent stuff (up to $bbH$ hard function
%\item includes citation dump for N$^3$LL$'$ ingredients
%\item introduce singular/resummed and nonsingular (power expansion)
%\item introduce need for profiles and turning off resummation, for details refer
%to next section
%\end{itemize}
%
%\PC{----------------------------------- Section goals Above -----------------------------------}

%===============================================================================
\subsection{Factorization and resummation}
%===============================================================================

We consider the cross section for an on-shell Higgs boson differential in the Higgs
rapidity $Y$ and Higgs transverse momentum $\qt$.
At small $q_T \equiv |\qt| \ll m_H$, we can expand the cross section in powers
of $q_T^2/m_H^2$ as
%%%
\begin{align} \label{eq:tmd_factorization}
\frac{\df \sigma}{\df Y \df^2 \qt}
&= \frac{\df \sigma^\sing}{ \df Y \df^2 \qt} + \frac{\df \sigma^\nons}{ \df Y \df^2 \qt}
=\frac{\df \sigma^\sing}{ \df Y \df^2 \qt}
\biggl[1+\cO\Bigl(\f{q_T^2}{m_H^2} \Bigr) \biggr]
\,.\end{align}
%%%
Here, $\df\sigma^\sing$ contains the leading-power ``singular'' contributions in the
$q_T\to 0$ limit involving $\delta(q_T)$ distributions and logarithmic plus
distributions $[\ln^n(q_T/m_H)/q_T]_+$. All remaining ``nonsingular'' contributions,
which are suppressed by $\ord{q_T^2/m_H^2}$ relative to $\df\sigma^\sing$,
are contained in $\df\sigma^\nons$.

The factorization of the leading-power $q_T$ spectrum was first established by Collins, Soper,
and Sterman~\cite{Collins:1981uk,Collins:1981va, Collins:1984kg}, and was further
elaborated upon and extended in \refscite{Catani:2000vq, deFlorian:2001zd,
Collins:2011zzd}. In this work we employ the framework of SCET,
in which $q_T$ factorization was formulated in \refscite{Becher:2010tm,
Echevarria:2011epo, Chiu:2012ir, Li:2016axz}, and which is equivalent to the
modern formulation in \refcite{Collins:2011zzd}. We employ the rapidity
renormalization group~\cite{Chiu:2012ir} together with the
exponential regulator from \refcite{Li:2016axz} for which the ingredients
required for the resummation at N$^3$LL$'$ are known.  Up to two loops it yields
the same results as the $\eta$ regulator used in \refcite{Chiu:2012ir}. In this
formulation, the singular cross section can be written in factorized form as
%%%
\begin{align} \label{eq:tmd_factorization1}
\frac{\df \sigma^\sing}{\df Y \df^2 \qt}
&= \sum_{a,b} H_{ab}(m_H^2; \mu) [B_a\otimes  B_b \otimes S_{ab}]( x_a, x_b, \qt; \mu)
\,,\end{align}
%%%
where the kinematic quantities $\omega_{a,b}$ and $x_{a,b}$ are given by
%%%
\begin{align}
\omega_{a}=m_H e^{+ Y}, \quad \omega_{b}=m_H e^{-Y} \quad \text{ and} \quad \quad x_{a,b}= \f{\omega_{a,b}}{\Ecm}
\,.\end{align}
%%%

The process dependence is encoded in the hard function $H_{ab}(m_H^2, \mu)$. It
describes the underlying hard interaction producing the Higgs boson via $a b\to
H$, with the available partonic channels being $ab= \{{q\bar{q}, \,
{\bar{q}q}}\}$. At leading order, $H^{(0)}$ corresponds to the partonic Born
squared matrix element, while at higher orders it includes the finite virtual
corrections to the Born process.

The factor $[B_a\otimes  B_b \otimes S_{ab}]$ in \eq{tmd_factorization1}
describes physics at the low scale $\mu \sim q_T$ and is defined as the
following convolution in $\qt$:
%%%
\begin{align}\label{eq:tmd_factorization2}
[B_a\otimes B_b\otimes S_{ab}](x_a, x_b, \qt; \mu)
&\equiv \int \! \df^2 \vec{k}_a \, \df^2 \vec{k}_b \, \df^2 \vec{k}_s \,
\delta^2(\qt - \vec{k}_a - \vec{k}_b - \vec{k}_s)
\\\nn & \qquad \times
B_a(x_a, \vec{k}_a; \mu, \nu/\omega_a) \, B_b(x_b, \vec{k}_b; \mu, \nu/\omega_b)\,
S_{ab}(\vec{k}_s; \mu, \nu)
\,.\end{align}
%%%
The functions in \eq{tmd_factorization2} are universal objects in $\qt$
factorization, independent of the details of the hard process. They are
renormalized, with $\mu$ and $\nu$ denoting their virtuality and rapidity
renormalization scales. The beam functions $B_{a,b}$ describe collinear
radiation with total transverse momentum $\vec{k}_{a,b}$ and longitudinal
momentum $\omega_{a,b}$, while the soft function $S_{ab}$ describes soft
radiation with total transverse momentum $\vec{k}_s$. Momentum conservation in
the transverse plane implies that the sum of $\vec{k}_{a}$, $\vec{k}_{b}$,
$\vec{k}_{s}$ must be equal to the measured Higgs transverse momentum $\qt$,
leading to the convolution structure in \eq{tmd_factorization2}.

In Fourier-conjugate $\bt$ space, the convolutions in $\qt$ in
\eq{tmd_factorization2} turn into simple products. The factorized singular cross
section in $\bt$ space then takes the form
%%%
\begin{align} \label{eq:qt_factorization_b}
\frac{\df\tilde\sigma^\sing(\vec{b}_T)}{\df Y}
&\equiv \int\!\f{\df^2\qt}{(2\pi)^2} \, e^{-\img\,\qt \cdot \bt} \,
\frac{\df \sigma^\sing}{\df Y \df^2 \qt} \nn \\
&= \sum_{a,b} H_{ab}(m_H^2; \mu)
\tilde B_a(x_a, \bt; \mu, \nu/\omega_a)\, \tilde B_b(x_b, \bt; \mu, \nu/\omega_b)\,\tilde S_{ab}(b_T; \mu, \nu)
\,,\end{align}
%%%
where $\tilde B_{a,b}$ and $\tilde S_{ab}$ are the Fourier transforms of
$B_{a,b}$ and $S_{ab}$ appearing in \eq{tmd_factorization2}.

To perform all-order resummation, each function is first evaluated at its own
natural boundary scale(s): $\mu_H$, $(\mu_B, \nu_B)$, and $(\mu_S, \nu_S)$. By
choosing appropriate values for the boundary scales close to their canonical
values (see \sec{canonical}), each function is free of large logarithms and can
therefore be evaluated in fixed-order perturbation theory. Next, all functions
are evolved from their respective boundary conditions to a common arbitary point
$(\mu, \nu)$ by solving their coupled system of renormalization group equations
(RGEs). The RGEs are themselves multiplicative in ${b}_T$ space and convolutions
in $\qt$ space. For more details we refer to \refscite{Ebert:2016gcn,
Ebert:2020dfc}.

%-------------------------------------------------------------------------------
\begin{table}
\centering
\renewcommand{\tabcolsep}{2ex}
\begin{tabular}{l | c c c | c}
\hline\hline
& Boundary & \multicolumn{2}{c|}{Anomalous dimensions} & FO matching
\\
Order & conditions & $\gamma_i$ (noncusp) & $\Gamma_\mathrm{cusp}$, $\beta$ & (nonsingular)
\\ \hline\hline
LL             & $1$  & - & 1-loop & -
\\
NLL            & $1$  & 1-loop & 2-loop & -
\\[0.25ex]
\hline
NLL$'$ $(+$NLO$)$  & $\alpha_s$ & 1-loop & 2-loop & $\alpha_s$ \\
NNLL $(+$NLO$)$ & $ \alpha_s$ & 2-loop & 3-loop & $\alpha_s$
\\[0.25ex]
\hline
NNLL$'$ $(+$NNLO$)$ & $\alpha_s^2$ & 2-loop & 3-loop & $\alpha_s^2$ \\
N$^3$LL $(+$NNLO$)$ & $\alpha_s^2$ & 3-loop & 4-loop & $\alpha_s^2$
\\[0.25ex]\hline
N$^3$LL$'$ $(+$N$^3$LO$)$ & $\alpha_s^3$ & 3-loop & 4-loop & $\alpha_s^3$ \\
N$^4$LL $(+$N$^3$LO$)$ & $\alpha_s^3$ & 4-loop & 5-loop & $\alpha_s^3$
\\\hline\hline
\end{tabular}
\caption{Definition of resummation orders.
The $(+$N$^n$LO$)$ in the order refers to whether or not the nonsingular
$\ord{\alpha_s^n}$ corrections in the last column are included.}
\label{tab:orders}
\end{table}
%-------------------------------------------------------------------------------

The resummation order is defined by the $\alpha_s$ and loop orders
to which the boundary conditions and anomalous dimensions entering the RGE are
included, as summarized in \tab{orders}.
For the resummation at N$^3$LL$'$ we require the N$^3$LO
boundary conditions for the hard function~\cite{Gehrmann:2014vha,
Ebert:2017uel}, and the beam and soft functions~\cite{Lubbert:2016rku,
Li:2016ctv, Billis:2019vxg, Luo:2019szz, Ebert:2020yqt}. We also need the 3-loop
noncusp virtuality~\cite{Lubbert:2016rku, Moch:2005id, Stewart:2010qs,
Bruser:2018rad, Billis:2019vxg} and rapidity anomalous
dimensions~\cite{Lubbert:2016rku, Li:2016ctv, Vladimirov:2016dll}, as well as
the 4-loop cusp anomalous dimension
$\Gamma_\mathrm{cusp}$~\cite{Korchemsky:1987wg, Moch:2004pa, Bruser:2019auj,
Henn:2019swt, vonManteuffel:2020vjv} and QCD $\beta$
function~\cite{Tarasov:1980au, Larin:1993tp, vanRitbergen:1997va,
Czakon:2004bu}.

%===============================================================================
\subsection{Canonical scales and resummation in \texorpdfstring{$b_T$}{bT} space}
\label{sec:canonical}
%===============================================================================

The canonical boundary scales in $b_T$ space are given by
%%%
\begin{align}\label{eq:canonical_scales}
\text{virtuality: }& \quad \mu_H = m_H, \quad \mu_B = b_0/b_T, \quad \mu_S = b_0/b_T,
\quad \mu_f = b_0/b_T,
\quad \mu_0 =  b_0/b_T, \nn \\
\text{rapidity: }& \hspace{2.65cm} \nu_B = m_H,\,  \hspace{0.66cm} \nu_S = b_0/b_T
\,,\end{align}
%%%
where $b_0\equiv 2 e^{-\gamma_E}\approx 1.12291$.
Here, $\mu_H$, $(\mu_B, \nu_B)$, and $(\mu_S, \nu_S)$ are the boundary scales for the
hard, beam, and soft functions, and $\mu_f$ is the scale at which the PDFs inside
the beam functions are evaluated. The rapidity
anomalous dimension must also be resummed and $\mu_0$ is its associated boundary
scale. When the functions in \eq{qt_factorization_b} are evolved from these
scales, the evolution resums all canonical $b_T$-space logarithms
$\ln^n[(b_0/b_T)/m_H]$.

As shown in \refcite{Ebert:2016gcn}, the exact solution for the RG evolution in
$\qt$ space in terms of distributions is equivalent to this canonical solution
in $b_T$ space modulo different conventions for the boundary conditions. Since
the latter is much easier to obtain, we also use it here, as is often done. The
resummed singular $\qt$ spectrum, $\df\sigma^\res$, is then obtained as the
inverse Fourier transform of the canonically resummed $b_T$ space result,
$\df\tilde\sigma^\res(\bt)$,
%%%
\begin{align} \label{eq:fourier_transform}
\frac{\df \sigma^\res}{\df Y \df^2 \qt}
&= \int\! \df^2 \bt \, e^{\img\,\qt \cdot \bt} \,
 \frac{\df\tilde\sigma^\res(\vec{b}_T)}{\df Y} = 2\pi \int\! \df b_T  \, b_T J_0(b_T q_T) \,
 \frac{\df\tilde\sigma^\res(\vec{b}_T)}{\df Y}
\,.\end{align}
%%%

With the canonical scales in \eq{canonical_scales}, the strong coupling and the PDFs
inside the beam functions are evaluated at $\as (b_0/b_T)$,
which means the beam and soft functions and rapidity anomalous dimension become sensitive to
nonperturbative effects for $1/b_T \lesssim \lqcd$. To perform the Fourier transform in \eq{fourier_transform}, we must therefore choose a prescription to avoid such nonperturbative scales.

The traditional approach is to perform a global replacement of $b_T$ everywhere
in the $b_T$-space cross section by a
function $b^*(b_T)$, which asymptotes to some fixed
perturbative scale $b_\mathrm{max} \lesssim 1/\lqcd$ for $1/b_T\to 0$, while away
from this limit it becomes $b_T$.
An important drawback of this global $b^*$ prescription is that it leads to much larger
than necessary distortions of the $b_T$-space cross section. This can be avoided
by applying this replacement only to the canonical scale choices~\cite{Lustermans:2019plv},
which suffices to avoid nonperturbative scales. More precisely, following \refcite{Billis:2023xxx},
we use the $\mu_*$ prescription
%%%
\begin{align}\label{eq:mu_star}
\mu_X = \mu_*\bigl(b_0/b_T, \mu_X^\mathrm{min} \bigr)
\quad\text{with}\quad
\mu_*(x,y) = \sqrt{x^2 + y^2}
\,,\end{align}
%%%
where $\mu_X$ stands for any of $\mu_S$, $\mu_B$, $\mu_0$, $\mu_f$. In principle
any function $\mu_*(x, y)$ can be used which satisfies $\mu_*(x \to 0, y)\to y$
and $\mu_*(x \gg y, y) \to x$. Under these conditions, all scales approach their
chosen minimum value $\mu_X^\mathrm{min}$ for $1/b_T\to 0$, while approaching
their canonical values away from this limit, as desired. Note that one advantage
of this prescription is that we have the option to choose different $\mu_X^\mathrm{min}$
values for different scales, which we will make use of for $\mu_f$.

%===============================================================================
\subsection{Profile scales and matching to fixed order}
\label{sec:profile}
%===============================================================================

In addition to the leading-power contributions, which are resummed with the help
of the factorization theorem in \eqs{tmd_factorization1}{tmd_factorization2}, we
also have to include the nonsingular power corrections $\df\sigma^\nons$ in
\eq{tmd_factorization}. To do so, we add them to the resummed singular
contributions to obtain the final matched result
%%%
\begin{align} \label{eq:matched}
\df \sigma &= \df \sigmares(\mu_\res) + \df \sigmanons(\mu_\FO)
\nn \\
&= \df \sigmares(\mu_\res) + \Bigl[\df
\sigmaFO(\mu_\FO) -\df \sigmasing(\mu_\FO)\Bigr]
\,.\end{align}
%%%
The first line is equivalent to \eq{tmd_factorization}, using the all-order
resummed result $\df \sigmares(\mu_\res)$ for the singular contributions. We use
the $\mu_\res$ argument here to indicate that $\df\sigmares$ is evaluated using
the resummation (boundary) scales as discussed in \sec{canonical}. The
nonsingular cross section $\df \sigmanons(\mu_\FO)$ is included at fixed order,
with its $\mu_\FO$ argument indicating that it is evaluated at the fixed-order scales
(the usual $\mu_R$ and $\mu_F$).
It is obtained as shown in the second line in \eq{matched}, i.e.\ by
using \eq{tmd_factorization} at fixed order and subtracting the fixed-order
singular terms from the full fixed-order result, where (as indicated)  both are
evaluated at common fixed-order scales $\mu_\FO$. This subtraction can be
done directly in momentum space.

For small $q_T\ll m_H$, the nonsingular terms are a small power correction
and it is sufficient to include them at fixed order despite the fact that the
singular terms are resummed there. For the nonsingular to be indeed power suppressed
it is essential that $\df\sigmaFO$ and $\df\sigmasing$ are evaluated at the same fixed order,
such that $\df\sigmasing$ exactly contains and cancels the singular terms of $\df\sigmaFO$.

On the other hand, as $q_T$ approaches $q_T \sim m_H$, the distinction between
singular and nonsingular becomes arbitrary and only the full fixed-order result
in $\df\sigmaFO$ is physically meaningful. To recover the correct $\df\sigmaFO$
in this limit, $\df\sigma^\res(\mu_\res)$ and $\df\sigma^\sing(\mu_\FO)$ must cancel each other in \eq{matched}. We require this cancellation to be exact with no leftover higher-order terms in $\alpha_s$, because for $q_T \sim m_H$ the (resummed) singular terms are unphysical and typically become
numerically much larger than the actual physical result given by $\df\sigmaFO$.
This requires the turning off of the resummation in $\df\sigma^\res(\mu_\res)$  -- in so doing, one guarantees that the result becomes equal to the fixed-order $\df\sigmasing(\mu_\FO)$.
Considering the first line of \eq{matched}, this implies that for $q_T \sim m_H$
there are typically large numerical cancellations between the singular and nonsingular contributions.

In summary, in order to have a consistent description of the cross section for all values of $q_T$,
the terms in \eq{matched} are required to satisfy two conditions: firstly, $\df\sigmasing$ and
$\df\sigmaFO$ must be evaluated at the same fixed order; secondly, $\df\sigmasing$ and
$\df\sigmares$ must become equal in the limit where the resummation in $\df\sigmares$
is turned off.

The most natural way to turn off the resummation in $\df \sigmares(\mu_\res)$ is
to set all boundary scales to the common fixed-order scales $\mu_\FO$, i.e.\ in
our notation $\mu_\res = \mu_\FO$. The second condition above thus requires
$\df\sigmares(\mu_\res = \mu_\FO) = \df\sigmasing(\mu_\FO)$.
The first condition above then requires for a given resummation order a specific
order for the nonsingular matching corrections. Namely, the $\alpha_s$ order
of the boundary conditions in the resummed result must match the
$\alpha_s$ order of the full and nonsingular results, which are the orders
shown in the last column of \tab{orders}.

In practice, we want to turn off the resummation smoothly, such that the difference $\df \sigmares(\mu_\res) - \df \sigmasing(\mu_\FO)$ vanishes equally smoothly as $q_T \to m_H$. This is conveniently achieved by using profile scales \cite{Ligeti:2008ac, Abbate:2010xh}, which provide a smooth transition for $\mu_\res$ from canonical resummation scales to the common fixed-order scales. Here we use hybrid profile scales $\mu_X(b_T, q_T)$~\cite{Lustermans:2019plv}, which depend on both $b_T$ and $q_T$ and undergo a smooth transition from their canonical $b_T$-dependence in \eq{mu_star} to the $b_T$-independent $\mu_\FO$,
with the transition happening as a function of $q_T$,
%%%
\begin{alignat}{9}
\mu_X(b_T, q_T) &= \mu_*\bigl(b_0/b_T, \mu_X^\mathrm{min} \bigr) \quad &&\text{for} \quad q_T \ll m_H
\,,\nn\\
\mu_X(b_T,q_T) &\to  \mu_\FO \quad &&\text{for} \quad q_T \to m_H
\,.\end{alignat}
%%%
We choose the central scales as
%%%
\begin{align} \label{eq:central_scales}
\mu_H = \nu_B = \mu_\FO &= m_H
\,,\nn\\
\mu_X &= m_H\, \frun\Bigl[ \frac{q_T}{m_H}, \frac{1}{m_H} \mu_*\Bigl(\frac{b_0}{b_T}, \mu_X^\mathrm{min}\Bigr)\Bigr]
\qquad \text{for}\quad \mu_X \in \{\mu_B, \mu_S, \nu_S, \mu_f\}
\,, \nn \\
\mu_0 &= \mu_*\Bigl(\frac{b_0}{b_T}, \mu_0^\mathrm{min}\Bigr)
\,,\end{align}
%%%
where $\frun$ is the hybrid profile function given by~\cite{Lustermans:2019plv}
%%%
\begin{align}
\frun(x,y)&= 1 + \grun(x)(y-1)
\,,\end{align}
%%%
where $\grun(x)$ determines the transition as a function of $x = q_T/m_H$,
%%%
\begin{align}\label{eq:def_g_run}
\grun(x) &= \begin{cases}
1 & 0 < x \leq x_1 \,, \\
1 - \frac{(x-x_1)^2}{(x_2-x_1)(x_3-x_1)} & x_1 < x \leq x_2
\,, \\
\frac{(x-x_3)^2}{(x_3-x_1)(x_3-x_2)} & x_2 < x \leq x_3
\,, \\
0 & x_3 \leq x
\,,\end{cases}
\end{align}
%%%
with the transition points $x_i$ with $i\in\{1,2,3\}$. The parameters $x_1$ and
$x_3$ determine the start and end of the transition and $x_2 =(x_1+ x_3)/2$
corresponds to the turning point. As a result the scales are canonical for $q_T
\leq x_1 m_H$ and the resummation is fully turned off for $q_T > x_3\, m_H$. The
values are usually chosen such that the transition begins somewhere in the
resummation region and is finished by the time the singular and the nonsingular
contributions are of the same size and exhibit sizeable numerical cancellations.
We will use $[x_1, x_2, x_3] = [0.1, 0.45, 0.8]$ as our central values as
explained in \sec{matchprocedure}.

For the $\mu_X^\mathrm{min}$ nonperturbative cutoffs we use
%%%
\begin{align}
\muBmin = \muSmin = \muZeromin = 1 \GeV
\,, \qquad
\nuSmin = 0
\,.\end{align}
%%%
We can set $\nuSmin = 0$ because $\nu_S$ never appears as argument of $\alpha_s$
or the PDFs. For $\mufmin$ we pick the larger of the PDF's $Q_0$ value or a
value based on the quark mass $m_q$ used by the PDF set as threshold for the
corresponding heavy-quark PDF. This choice of $\mu_f^\mathrm{min}$ avoids
running into numerical noise below the scale where the heavy-quark PDFs vanish
and where the results are in any case not particularly meaningful without the proper inclusion
of finite-mass effects, which is beyond our scope here. For the \texttt{MSHT20nnlo}
PDF set we use, this amounts to taking $\mufmin = Q_0 = 1.0 \GeV$ for
$s\bar{s}\to H$, $\mufmin = m_c = 1.4 \GeV$ for $c\bar{c}\to H$ and  $\mufmin =
5.0 \GeV$ for $b\bar{b}\to H$. The latter is chosen slightly above the actual
bottom-quark mass threshold $m_b=4.75 \GeV$ to avoid numerical instabilities.

In the fixed-order limit, we can identify $\mu_\FO \equiv \mu_R$ with the usual
renormalization scale for $\alpha_s$ and $\mu_f \equiv \mu_F$ with the usual
factorization scale at which the PDFs are evaluated. Our central choices above
correspond to $\mu_R = \mu_F = m_H$.

%===============================================================================
\subsection{Perturbative uncertainties}
\label{sec:pertunc}
%===============================================================================

To estimate the perturbative uncertainties, we vary the profile scales about
their central values given in \sec{profile}. Following
\refscite{Stewart:2013faa, Ebert:2020dfc, Billis:2023xxx}, we identify several
different sources of uncertainty, which are considered as independent and are
estimated from different suitable types of variations.
The profile scales are varied as follows:
%%%
\begin{align} \label{eq:profile_vars}
\mu_H &= \mu_\FO = 2^{w_\FO}\, m_H
\,, \nn \\
\nu_B &= \mu_\FO \, f_\vary^{v_{\nu_B}}\Bigl(\frac{q_T}{m_H}\Bigr)
\,, \nn \\
\mu_X &= \mu_\FO \, f_\vary^{v_{\mu_X}}\Bigl(\frac{q_T}{m_H}\Bigr) f_\run\biggl[
\frac{q_T}{m_H},
\frac{1}{m_H}\mu_*\Bigl(\frac{b_0}{b_T}, \frac{\mu_X^\mathrm{min}}{2^{w_\FO}f_\vary^{v_{\mu_X}} } \Bigr)
\biggr]
\quad\text{for}\quad \mu_X \in \{\mu_B, \mu_S, \nu_S\}
\,, \nn \\
\mu_f &= 2^{w_F}\, m_H\,  f_\run\biggl[
\frac{q_T}{m_H},
\frac{1}{m_H}\mu_*\Bigl(\frac{b_0}{b_T}, \frac{\mufmin}{2^{w_F}} \Bigr)
\biggr]
\,,\nn\\
\mu_0
&= \mu_*\Bigl( \frac{b_0}{b_T}, \muZeromin \Bigr)
\,.\end{align}
%%%

To estimate an uncertainty associated with the resummation $\Delta_\res$, the
beam and soft scales are varied, where the exponents $v_{\mu_B}$, $v_{\nu_B}$,
$v_{\mu_S}$, and $v_{\nu_S}$ are taken to be $v_i=\{-1,0,+1\}$ with the central
scales corresponding to $v_i = 0$. The function %%%
\begin{align}
f_\vary(x) &= \begin{cases}
2(1-x^2/x_3^2) &  0\leq x \leq x_3/2
\,, \\
1-2(1-x/x_3)^2  & x_3/2 < x \leq x_3
\,, \\
1 & x_3 \leq x
\,,\end{cases}
\end{align}
%%%
with $x \equiv q_T/m_H$
controls the size of the variations, ranging from a factor of 2 for $x = 0$ to 1
for $x \geq x_3$, where $x_3$ is the same as for $\frun(x)$. This source of
uncertainty is thus turned off for $q_T \geq x_3 m_H$ just as the resummation itself is
turned off. To estimate the resulting resummation uncertainty $\Delta_\res$ we
perform 36 variations of suitable combinations of the $v_i$ and take their
maximum envelope. For details, we refer the reader to \refcite{Ebert:2020dfc}.

For the fixed-order uncertainty $\Delta_\FO$, we vary $\mu_\FO$ by a factor of 2 by taking
$w_\FO =\{-1,0,+1\}$ everywhere. Note that $\Delta_\FO$ is not
defined to be the uncertainty in the fixed-order limit but is rather meant
to estimate a common uncertainty due to missing fixed-order contributions at any
$q_T$. It therefore contributes to both the singular and nonsingular pieces. In the
resummed singular it amounts to an overall variation of the boundary scales such
that the resummed logarithms are unchanged, which is why one can interpret it as
a fixed-order uncertainty.
Furthermore, we estimate a separate uncertainty $\Delta_{\mu_f}$ related to
the DGLAP running of the PDFs, for which we vary the PDF scale $\mu_f$ by taking
$w_F =\{-1,0,+1\}$  (where $w_F =0$ is the central value). In the nonsingular and
full fixed-order cross sections, this corresponds to taking $\mu_f \equiv \mu_F
= 2^{w_F} m_H$. The resulting $\Delta_\FO$ and $\Delta_{\mu_f}$ are then given by
the maximum envelope of the respective variations.

We obtain the total perturbative uncertainty by adding the individual uncertainties
in quadrature,
%%%
\begin{align}
\Delta_{\mathrm{total}} = \sqrt{\Delta_\FO ^2 + \Delta_{\mathrm{res}}^2 + \Delta_{\mu_f}^2 + \Delta^2_{\mathrm{match}}}.
\end{align}
%%%
The matching uncertainty $\Delta_{\mathrm{match}}$ will be discussed in \sec{matchprocedure}.

Note that in the fixed-order limit, we do not use an envelope of
$\mu_R$ and $\mu_F$ variations as is commonly done. Instead, we estimate separate
uncertainties $\Delta_\FO$ and $\Delta_{\mu_f}$ which are added in quadrature. By
separating these two uncertainties in the resummation limit, we essentially have
no choice but to do the same also at fixed order. This is not problematic, but is in fact a perfectly sensible choice for the fixed-order
prediction -- here, as in the resummation, the two variations probe two
conceptually different sources of uncertainty.

%%%%%%%%%%%%%%%%%%%%%%%%%%%%%%%%%%%%%%%%%%%%%%%%%%%%%%%%%%%%%%%%%%%%%%%%%%%%%%%%
\section{Fixed-order contributions and matching at \texorpdfstring{aN$^3$LO}{aN3LO}}
\label{sec:matching}
%%%%%%%%%%%%%%%%%%%%%%%%%%%%%%%%%%%%%%%%%%%%%%%%%%%%%%%%%%%%%%%%%%%%%%%%%%%%%%%%

In this section, we discuss several aspects specific to the $q\bar q\to H$
process we are interested in here. In \sec{fullFO}, we describe our
implementation and validation of the fixed-order calculation for the
$q\bar{q}\to H+j$ process from which we obtain the nonsingular corrections. In
\sec{matchprocedure}, we discuss how we choose the transition points for the
profile function in \eq{def_g_run}, and detail the procedure to estimate the
associated matching uncertainty, which is particularly delicate for $b\bar{b}\to
H$. In \sec{decorrelation}, we describe a general strategy to decorrelate the
singular and nonsingular contributions. Based on this, we construct
in \sec{aNNLO1} a suitable approximation for the fixed
$\ord{\alpha_s^3}$ corrections to the nonsingular and full cross sections.

%\begin{itemize}
%\item singular-nonsingular plots, how one usually chooses $x_1$
%\item $b\bar{b}\to H$ is very sensitive
%\item Why $x_1$ and $x_3$ variations are bad
%\item What we do instead: $x_2$ variations
%\end{itemize}

%``How I stopped worrying and learned to love $x_2$'' (M. Lim)

%\PC{----------------------------------- Section goals Above -----------------------------------}

%===============================================================================
\subsection{Fixed-order calculations}
\label{sec:fullFO}
%===============================================================================

As discussed in \sec{profile}, the nonsingular corrections are obtained at
fixed order by taking
%%%
\begin{align} \label{eq:nons}
\frac{\df \sigmanons}{\df q_T}
&= \frac{\df\sigmaFO}{\df q_T} - \frac{\df \sigmasing}{\df q_T}
\,,\end{align}
%%%
where $\df\sigmasing$ is obtained directly in momentum space from the
fixed-order expansion of the factorization theorem in \eq{tmd_factorization1}.
Since $\df\sigmanons$ is power suppressed, we only need it for $q_T > 0$. Hence,
to evaluate $\df\sigmaFO$ we require the fixed-order calculation for the $q_T$
spectrum in $q\bar q\to H+j$. At N$^3$LL$'$ we need $\df\sigmaFO$ at
$\ord{\as^3}$ corresponding to the $q\bar q\to H+j$ calculation at NNLO$_1$
(the subscript $1$ on the order counting indicates that it is
relative to the $H+1$-parton cross section).

%~~~~~~~~~~~~~~~~~~~~~~~~~~~~~~~~~~~~~~~~~~~~~~~~~~~~~~~~~~~~~~~~~~~~~~~~~~~~~~~
\subsubsection{\texorpdfstring{LO$_1$}{LO1} and \texorpdfstring{NLO$_1$}{NLO1}}
%~~~~~~~~~~~~~~~~~~~~~~~~~~~~~~~~~~~~~~~~~~~~~~~~~~~~~~~~~~~~~~~~~~~~~~~~~~~~~~~

For the lowest order, \loone, we have performed an analytic calculation, which
we have implemented in \scetlib~\cite{scetlib} -- the relevant details are
provided for completeness in \app{lo1}. For the \nloone~calculation, we use a
parton-level Monte Carlo calculation, which we have implemented in the \geneva\
event generator~\cite{Alioli:2015toa, Alioli:2023har} using FKS
subtractions~\cite{Frixione:1995ms}. We have used the virtual
matrix elements in analytic form, which were calculated in \refcite{DelDuca:2015zqa} and implemented in the \geneva~code in \refcite{Alioli:2020fzf}. The tree-level
double-real emission matrix elements are obtained from the \openloops\
library~\cite{Buccioni:2019sur}. Note that often only the $b\bar b\to H$ process
is considered. We therefore performed several internal cross checks to also
ensure the correct implementation of $c\bar c\to H+j$ and $s\bar s\to H+j$. At
\loone, we also checked the implementation
against our analytic implementation in \scetlib.

\begin{figure}
\includegraphics[width=\WidthTwoSubfigs]{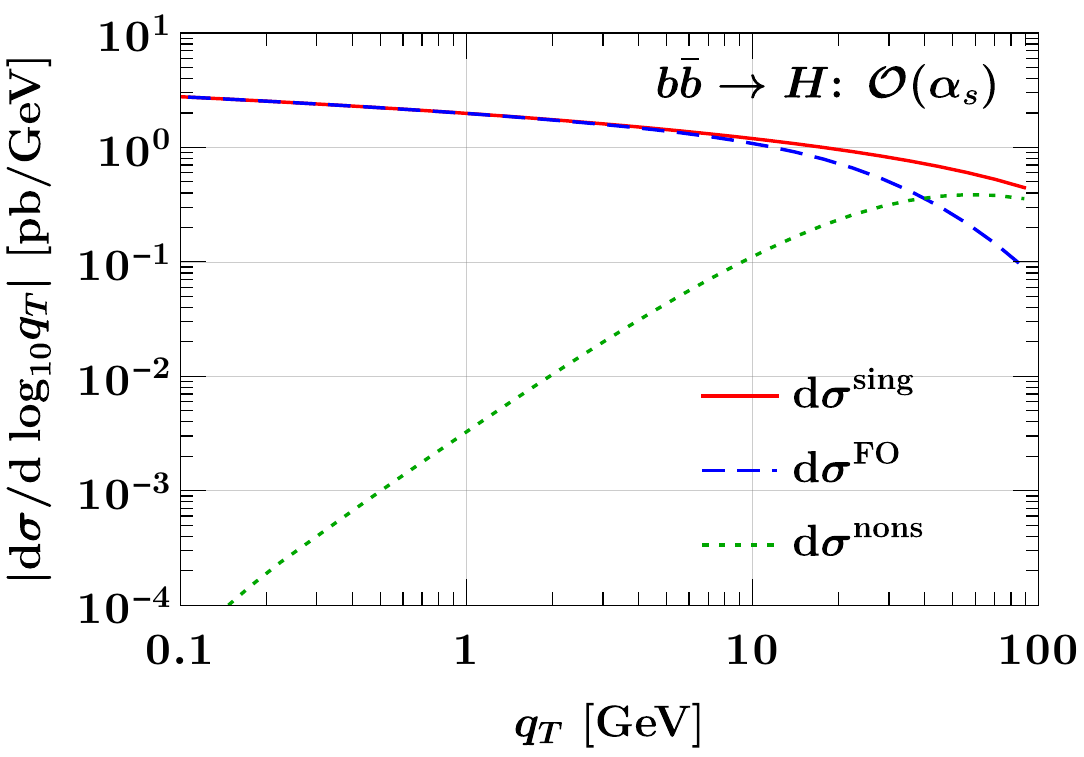}%
\hfill
\includegraphics[width=\WidthTwoSubfigs]{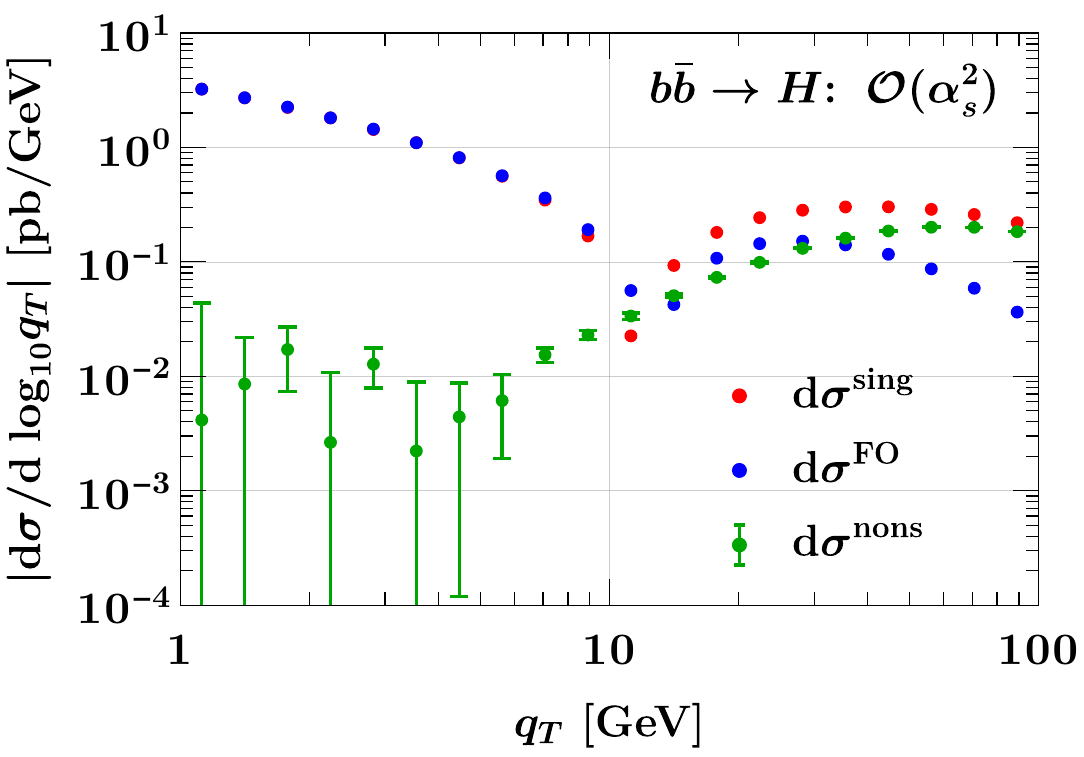}%
\caption{Singular (solid red), full (dashed blue), and nonsingular (dotted green) contributions for $b\bar{b}\to H$ at fixed $\ord{\alpha_s}$ (left) and $\ord{\alpha_s^2}$ (right). The nonsingular exhibit
the expected quadratic power suppression for $q_T\to 0$.}
\label{fig:nonsingcancloglog}
\end{figure}

\begin{figure}
\centering
\includegraphics[width=\WidthTwoSubfigs]{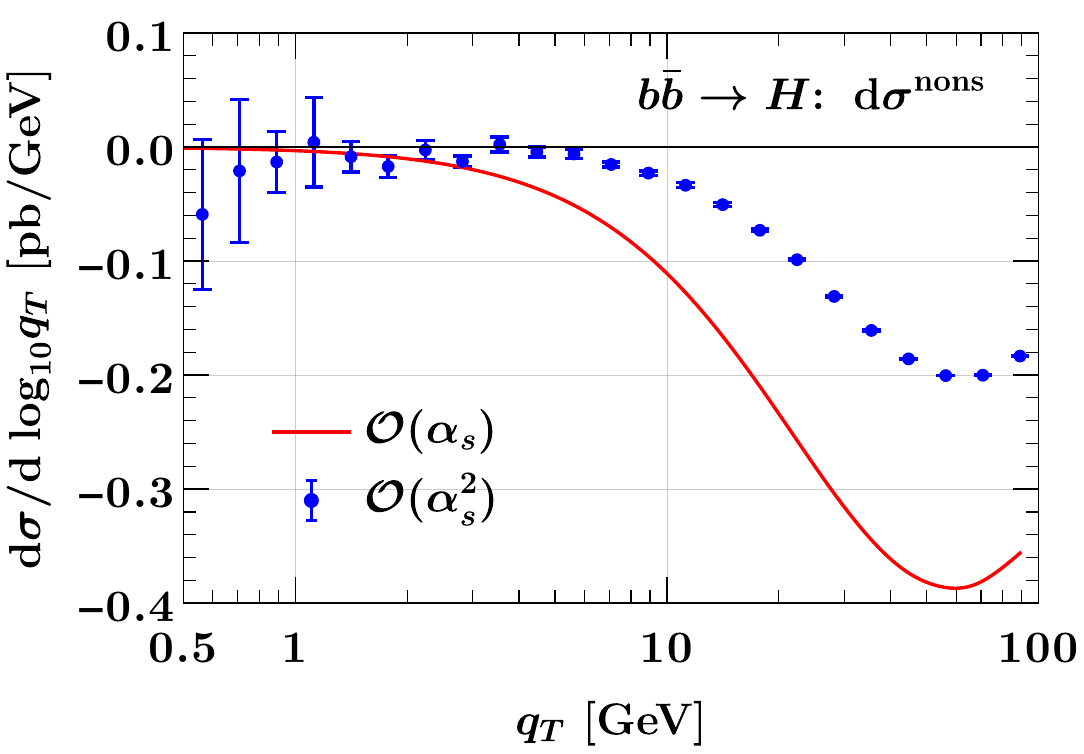}%
\caption{Nonsingular contributions at $\mathcal{O}(\alpha_s)$ (red) and $\mathcal{O}(\alpha_s^2)$ for $b\bar{b}\to H$ on a linear scale, corresponding respectively to the green curves and points in \fig{nonsingcancloglog}.}
\label{fig:nonsingcancloglin}
\end{figure}

A powerful cross check of the fixed-order calculation is provided by the
cancellation of all singular terms in the $q_T\to 0$ limit in \eq{nons}. This is
shown for both the $\ord{\as}$
and $\ord{\as^2}$ corrections in \fig{nonsingcancloglog}. In both cases, the
full (blue) and singular (red) results become essentially equal for small $q_T$,
and the nonsingular (green) given by their difference exhibits the expected power
suppression. Note that these plots show $|\df\sigma/\df\log_{10} q_T|$ on a
log-log scale, for which an $\ord{q_T^2/m_H^2}$ power suppression corresponds to a line
with an asymptotic slope of $-2$ for $q_T\to 0$. This is clearly seen
at $\ord{\as}$. At $\ord{\as^2}$ this is less apparent due to the limited Monte-Carlo
integration precision at very small $q_T$ and because the nonsingular contribution
contains powers of logarithms $\ln^n(q_T^2/m_H^2)$ up to $n \leq 3$, which weaken
the power supression and effectively delay the strictly quadratic scaling to
smaller $q_T$. We nevertheless observe a clear power suppression
from around 30 GeV down to a few GeV until the numerical precision becomes insufficient to actually
resolve the small but nonzero value of the nonsingular. Note that once
this happens, the result for the nonsingular should fluctuate around and be
consistent with zero within the statistical uncertainties. This is confirmed in
\fig{nonsingcancloglin}, which shows the nonsingular from \fig{nonsingcancloglog}
but on a linear $y$ axis and including the sign. Analogous results for $c\bar c\to H$
and $s\bar s\to H$ are provided in \app{singnons}.

%-------------------------------------------------------------------------------
%-------------------------------------------------------------------------------

%-------------------------------------------------------------------------------
%-------------------------------------------------------------------------------

%~~~~~~~~~~~~~~~~~~~~~~~~~~~~~~~~~~~~~~~~~~~~~~~~~~~~~~~~~~~~~~~~~~~~~~~~~~~~~~~
\subsubsection{\texorpdfstring{NNLO$_1$}{NNLO1}}
%~~~~~~~~~~~~~~~~~~~~~~~~~~~~~~~~~~~~~~~~~~~~~~~~~~~~~~~~~~~~~~~~~~~~~~~~~~~~~~~

A calculation for $b\bar{b}\to H+j$ has been achieved at NNLO$_1$
in \refcite{Mondini:2021nck} using $N$-jettiness subtractions~\cite{Gaunt:2015pea, Boughezal:2015dva}.
This calculation uses a cut on the jet-$p_T \geq 30\GeV$, which gives an unbiased result for the
$q_T$ spectrum only for $q_T \geq 60\GeV$. The jet-$p_T$ cut limits the size of
residual power corrections in the $N$-jettiness slicing parameter, which scale with
the inverse of the smallest kinematic scale in the process. It would require
substantial high-performance computing resources to perform the full NNLO$_1$
calculation without a jet cut down to much smaller $q_T$ (this is also what experience has shown
in case of Drell-Yan~\cite{Neumann:2022lft}).
On the other hand, the spectrum at small $q_T$ is entirely dominated by the
resummed singular contributions, while the $\ord{\alpha_s^3}$ nonsingular corrections
only give a very small correction: this does not justify the computational
cost and associated carbon footprint. In addition, we also need the NNLO$_1$ calculations
for charm and strange production, which are not presently available. Therefore,
we find it more prudent to construct an approximate NNLO$_1$ calculation, described
in \sec{aNNLO1}, that is suitable for our purposes and which is designed to give
good agreement with the known result for $b\bar b\to H+j$ at $q_T \geq 60 \GeV$.

%===============================================================================
\subsection{Estimation of matching uncertainties}
\label{sec:matchprocedure}
%===============================================================================

\begin{figure}
\includegraphics[width=0.47\textwidth]{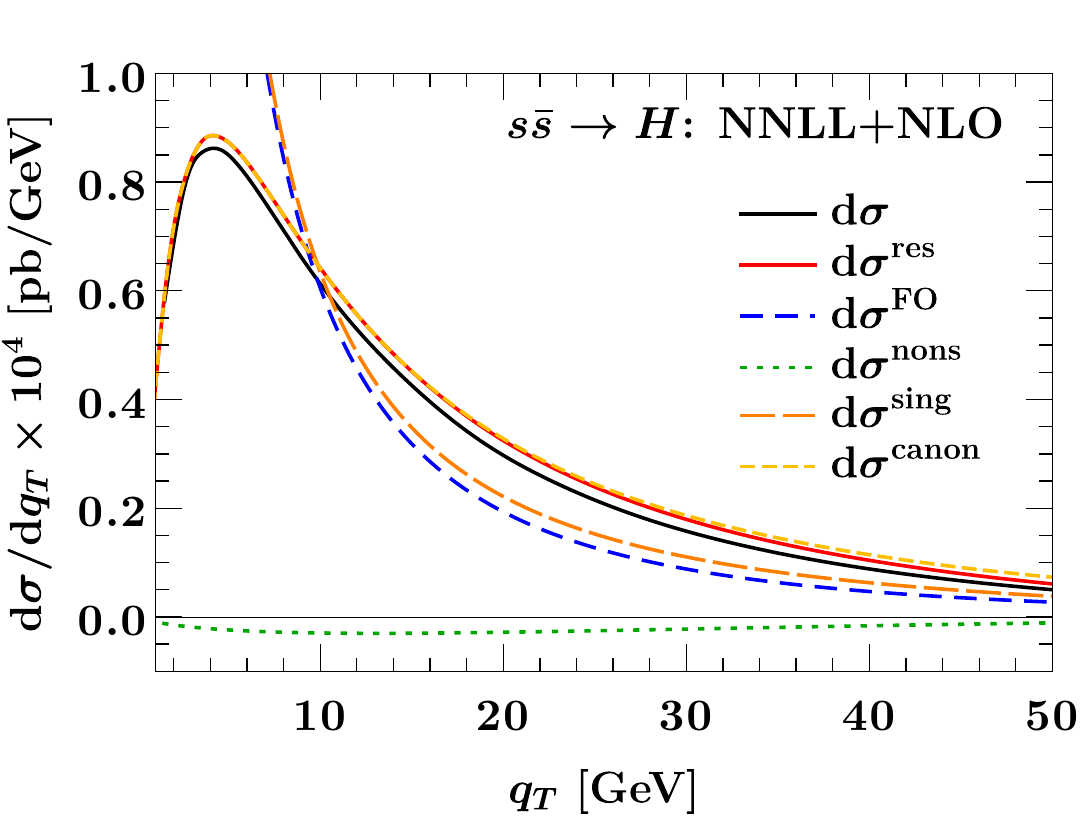}%
\hfill
\includegraphics[width=0.51\textwidth]{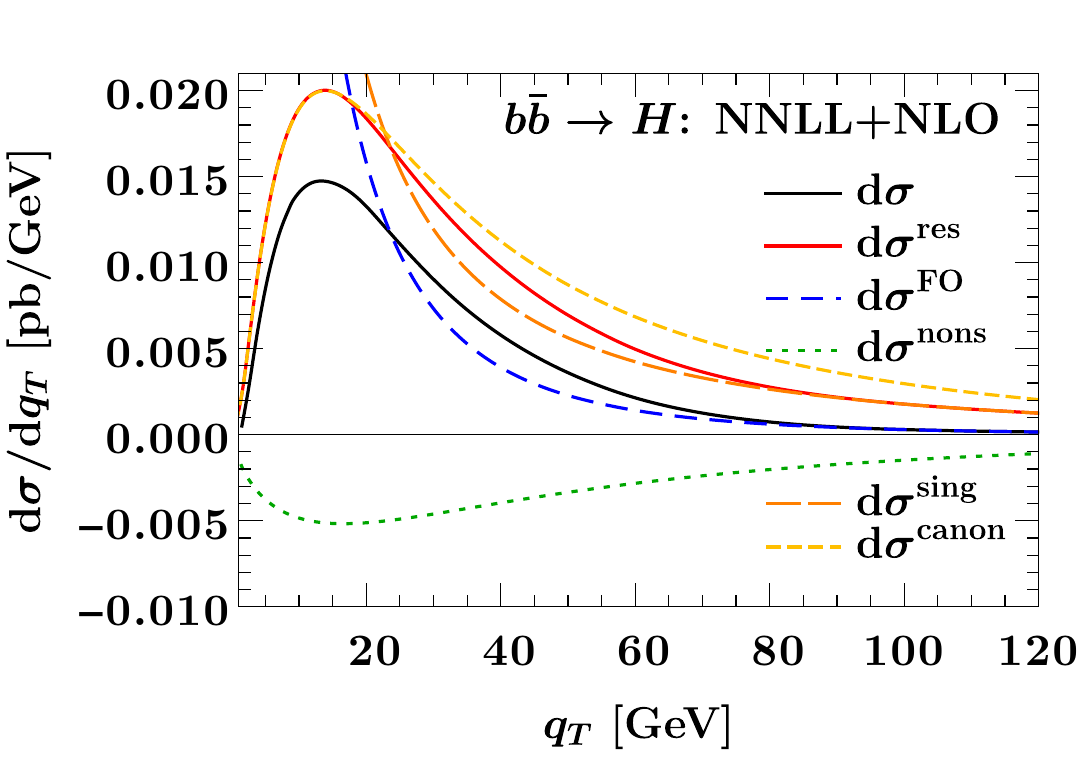}%
\caption{Different contributions entering the matching procedure for $s\bar{s}\to H$ (left) and
$b\bar{b}\to H$ (right) at NNLL$+$NLO. The final matched result $\df\sigma$ (solid black)
is the sum of the nominal resummed $\df\sigma^\res$ (solid red) and
the nonsingular $\df\sigma^\nons$ (dashed green). The fixed-order result $\df\sigma^\FO$ (dashed blue)
is the sum of the fixed-order singular $\df\sigma^\sing$ (long-dashed orange)
and $\df\sigma^\nons$ (dashed green).
Here, the canonically resummed result is denoted as $\df\sigma^\canon$ (short-dashed yellow).
The nominal resummed $\df\sigma^\res$ transitions from $\df\sigma^\canon$ at small $q_T$ to $\df\sigma^\sing$
at large $q_T$.}
\label{fig:matching}
\end{figure}

Na\"{i}vely, one might expect the $q\bar{q}\to H$ process to share many features
in its numerical behaviour with the Drell-Yan process. Indeed, both are
quark-initiated at Born level and produce a single heavy colour-singlet state
in the $s$ channel. Nevertheless, inspecting \fig{matching}, we see that this is not
quite the case. The figure shows the various contributions
entering in the matching procedure for both $s\bar{s}\to H$ and
$b\bar{b}\to H$. It shows that the numerical importance of the different contributions
strongly depends on the incoming flavour. The $s\bar s$ channel indeed behaves
very similar to Drell-Yan (see e.g.~\refcite{Ebert:2020dfc}):
it exhibits a very small nonsingular contribution $\df\sigma^\nons$ (dotted green),
such that the final matched result (solid black) is almost the same as the
nominal resummed result $\df\sigmares$ (solid red). Furthermore, the transition of
$\df\sigma^\res$, using profile scales, from the canonically resummed result $\df\sigma^\canon$ (short-dashed yellow) at small $q_T$ towards the fixed-order singular $\df\sigma^\sing$ (long-dashed orange) at large $q_T$ is very gentle. The $b\bar b$ channel instead features a
much larger nonsingular contribution, and the transition that $\df\sigma^\res$
has to undergo from canonical resummation to fixed-order singular is very pronounced.
The result of this is a much increased sensitivity to the precise choice of the
transition points $x_i$ compared to the Drell-Yan case.

%-------------------------------------------------------------------------------
%-------------------------------------------------------------------------------

This difference between the channels can be understood from the very different
size of the quark PDFs involved. At lowest order, the nonsingular receives
contributions from two different flavour channels, namely $q\bar q\to H g$ and
$gq \to Hq$ (which includes $g\bar q\to H\bar q$ for the sake of this discussion).
In Drell-Yan, these two channels have opposite sign and similar size (see e.g.~\refcite{Ebert:2018gsn}),
and thus partially cancel each other, leading to the relatively small nonsingular corrections
typical for that process. The same also happens for $s\bar{s}\to H$. For
$b\bar b\to H$, however, the very small $b$-quark PDF suppresses the $b\bar b$-induced
contributions. This has two effects leading to the observed behaviour:
first, the nonsingular is dominated by the gluon-induced channels leading to
smaller cancellations. This is compounded by the fact that the leading (NLL) contributions
in the resummed are also $b\bar b$ induced and numerically suppressed. Both of
these effects numerically enhance the nonsingular. The second effect furthermore causes a
larger difference between (canonically) resummed and fixed-order singular.
From this discussion one would expect the
$c\bar c\to H$ process (not shown in \fig{matching}) to exhibit behaviour intermediate between
$s\bar s\to H$ and $b\bar b\to H$, which is indeed the case.

\begin{figure}
\includegraphics[width=\WidthTwoSubfigs]{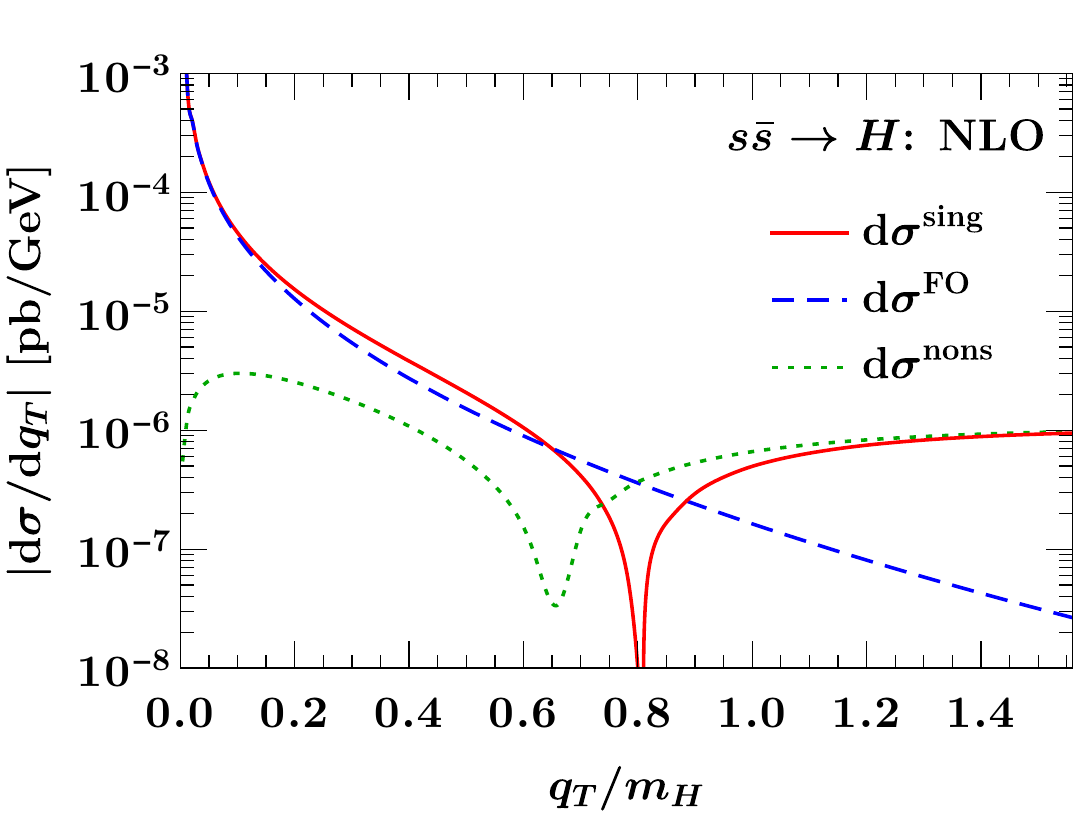}%
\hfill%
\includegraphics[width=\WidthTwoSubfigs]{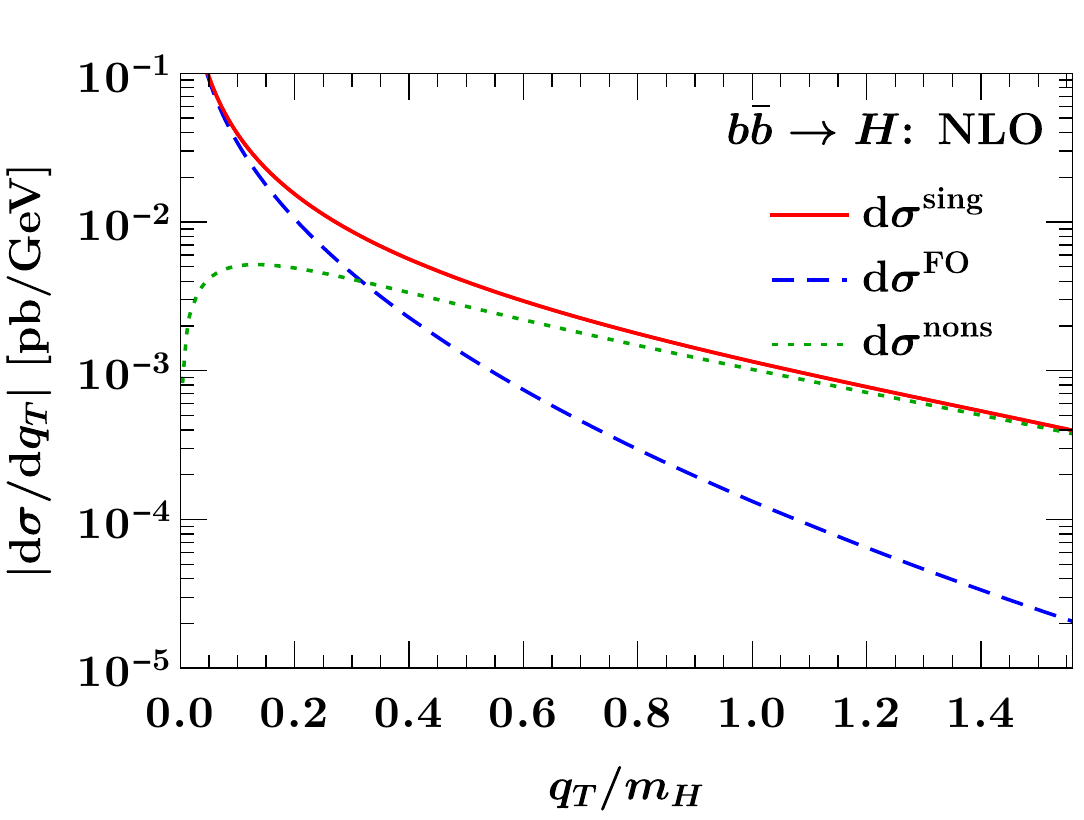}%
\caption{Singular (solid red), full (dashed blue), and nonsingular (dotted green) contributions at fixed $\ord{\alpha_s}$ as a function of $q_T/m_H$ for $s\bar{s}\to H$ (left) and $b\bar{b}\to H$ (right).}
\label{fig:singnons}
\end{figure}

The stronger sensitivity to the (ultimately arbitrary) choice of transition points in
$b\bar{b}\to H$ requires us to take greater care in choosing the transition points
and in estimating the associated matching uncertainty.
Usually, the start and endpoints of the transition, $x_1$ and $x_3$ (see \sec{profile})
are chosen based on examining the relative sizes of the singular and nonsingular pieces as a function of $q_T$, as shown in \fig{singnons} for $s\bar s\to H$ (left) and $b\bar b\to H$ (right).
The rather different behaviour of the channels is seen again here.
The $s\bar s\to H$ channel again looks very similar to Drell-Yan, with the nonsingular
becoming important only at relatively large $q_T/m_H \gtrsim 0.8$, such that a typical choice for the transition points would be $x_1 = 0.3$, $x_3 = 0.9$, $x_2 = (x_1 + x_3)/2 = 0.6$~\cite{Ebert:2020dfc, Billis:2023xxx}.
In contrast, for $b\bar b\to H$ the nonsingular becomes important much earlier.
Based on this plot, one might take sensible central values of $x_1=0.2$, $x_3=0.7$, and
$x_2=(x_1 + x_3)/2 = 0.45$.

%-------------------------------------------------------------------------------
%-------------------------------------------------------------------------------

\begin{figure}
\includegraphics[width=0.49\textwidth]{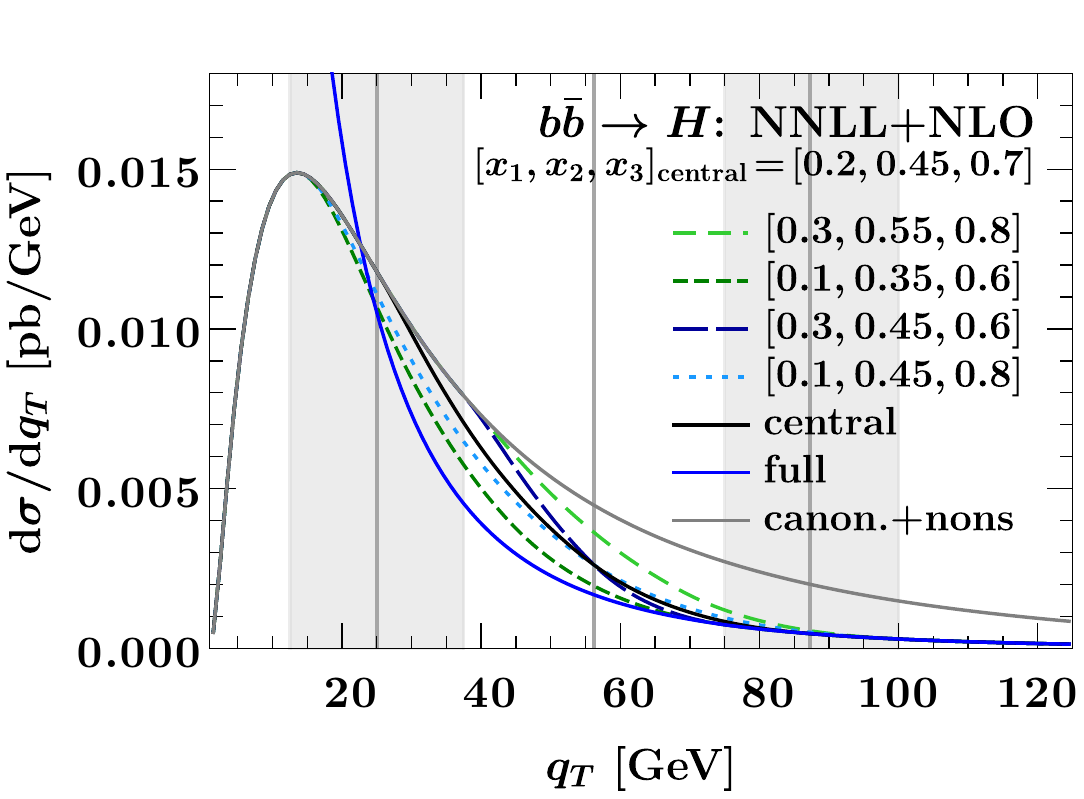}%
\hfill
\includegraphics[width=0.475\textwidth]{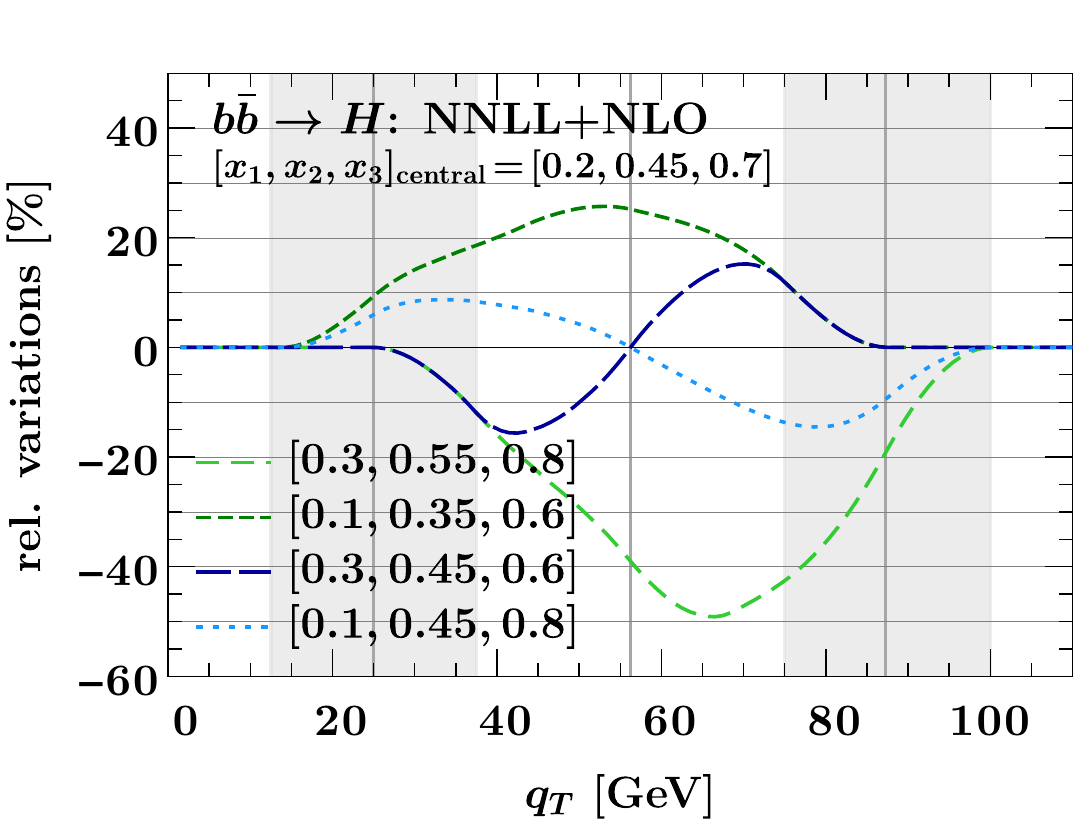}%
\caption{Standard variation of the transition points $x_1$ and $x_3$ for $b\bar
b\to H$ at NNLL$+$NLO. The absolute variation in the spectrum is shown on the
left, and the relative variation with respect to the central result is shown on
the right. The vertical lines indicate the central values $[x_1, x_2, x_3] =
[0.2, 0.45,0.7]$, and the grey bands the variations of $x_1$ and $x_3$ by $\pm
0.1$, where always $x_2 = (x_1 + x_3)/2$.}
\label{fig:standardtransition}
\end{figure}

The matching uncertainty $\Delta_\match$ is related to the ambiguity in these choices. The standard
method to estimate it is to vary $x_1$ and $x_3$, typically by $\pm 0.1$, with $x_2$
given by $(x_1 + x_3)/2$ for any given variation. The resulting variations for $b\bar b\to H$
are shown in \fig{standardtransition}.
We first note that this standard method leads by construction to a one-sided uncertainty
above the central $x_3$ and below the central $x_1$, because varying $x_3$ up or $x_1$ down
can only change the cross section in one direction. In practical applications, e.g.\
when propagating the $x_{1,3}$ variations in a fit, this is a rather undesirable feature.
Furthermore, varying $x_1$ and $x_3$ up (long-dashed green) produces an unreasonably
large uncertainty. The reason for this large variation, as evident from the left plot,
is precisely due to the rather large difference between the canonically resummed
and fixed-order results already discussed, between which the transition must interpolate.

\begin{figure}
\includegraphics[width=0.49\textwidth]{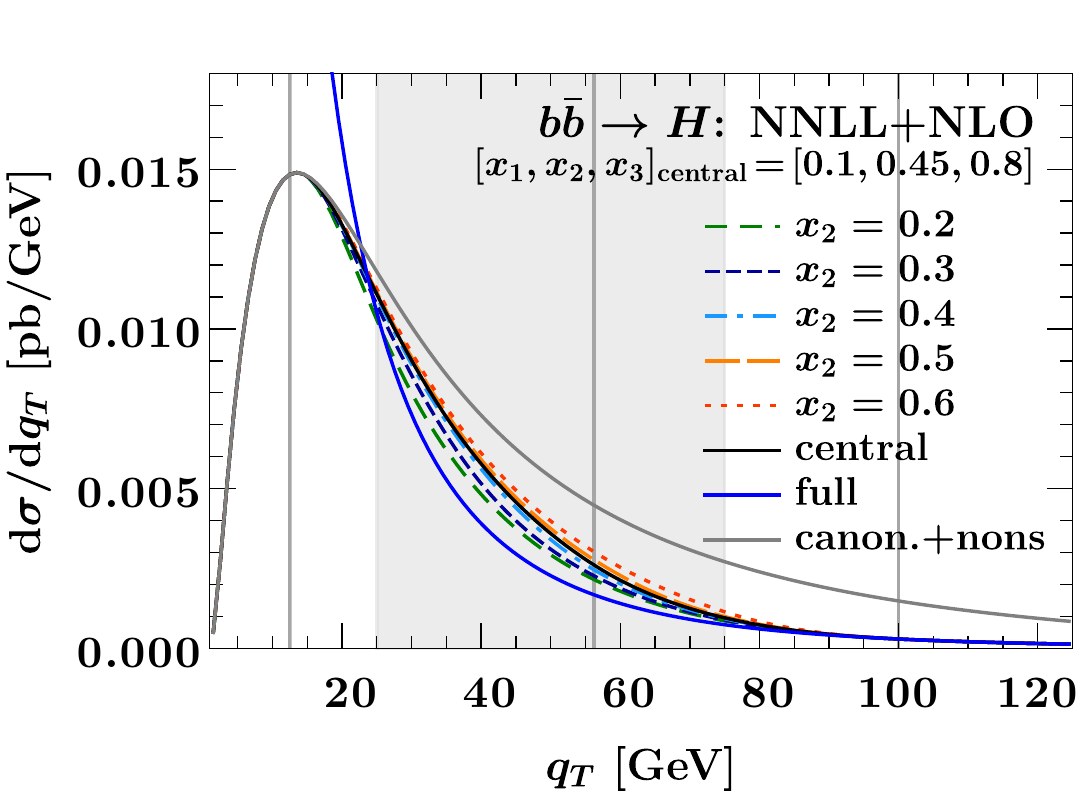}%
\hfill%
\includegraphics[width=0.475\textwidth]{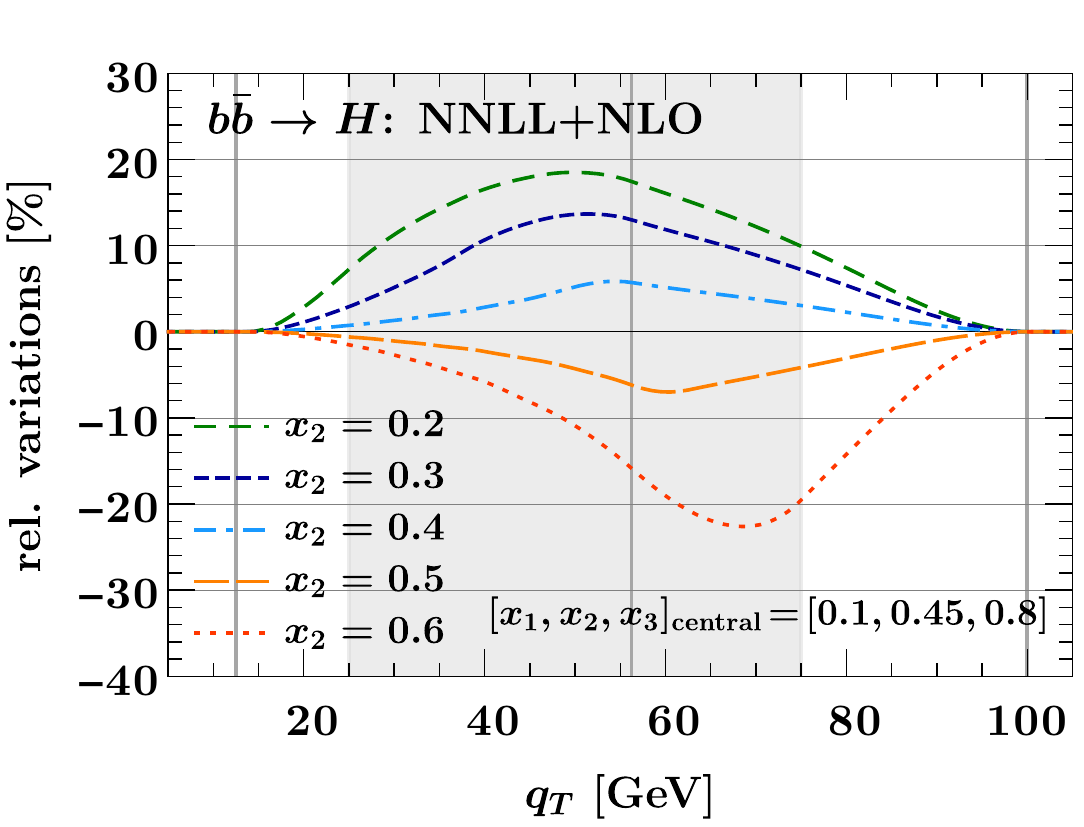}%
\caption{New variation of the transition point $x_2$ for $b\bar b\to H$ at
NNLL$+$NLO. The absolute variation in the spectrum is shown on the left, and the
relative variation with respect to the central result is shown on the right. The
vertical lines indicate the central values $[x_1, x_2, x_3] = [0.1, 0.45,0.8]$,
where $x_1$ and $x_3$ are fixed to their respective minimum and maximum values
used in \fig{standardtransition}. The grey band shows the variation in $x_2$
from $0.2$ to $0.6$. }
\label{fig:newtransition}
\end{figure}

We therefore adopt a somewhat different approach to estimate the matching
uncertainty. We first fix $x_1$ to its lowest and $x_3$ to its highest
reasonable value, e.g. to their respective minimum and maximum values one would
consider in the previous approach (which are $x_1=0.1$ and $x_3=0.8$ in our
example). We then vary the point $x_2$ to estimate the matching uncertainty. Conceptually, this effectively varies whether the transition happens
earlier or later within the maximal window in which the transition should occur.
Here, we take $x_2 = (x_1 + x_3)/2 = 0.45$ as our central value and vary it within
the range $[0.2, 0.6]$. Note that the size of this range is twice that of the
$x_1$ and $x_3$ variations, so the total amount of variation is preserved. The
resulting variations are shown in \fig{newtransition}. 

We begin by observing that this method avoids
the undesired one-sided uncertainties, although the uncertainty is still
somewhat asymmetric at any given $q_T$. This is practically unavoidable, since it is inherent to the nature of the
matching uncertainty. We can, however, choose the
$x_2$ range of variation such that the maximum up and down variations in the
cross section are of similar size, which is why we vary it further down than up.
Furthermore, the $x_2$ variation yields a much more reasonable size for the
matching uncertainty. Finally, this method has the added benefit that
the matching uncertainty is now parameterized by a single variable. This makes
it much easier to propagate in practice, as it avoids having to take envelopes
of different parameter variations.

For our final numerical results, the matching uncertainty $\Delta_\match$ is still
obtained as the maximum of the
absolute impact of varying $x_2$ down to $0.2$ and up to $0.6$. However, this is
now just for ease of presentation and not a requirement.
Since $s\bar s\to H$ and $c\bar c\to H$ are less sensitive to the precise
transition, we will use the same central values and $x_2$ variations for simplicity.

%-------------------------------------------------------------------------------
%-------------------------------------------------------------------------------

%-------------------------------------------------------------------------------
%-------------------------------------------------------------------------------

%===============================================================================
\subsection{Decorrelation of singular and nonsingular contributions}
\label{sec:decorrelation}
%===============================================================================

As discussed in \sec{fullFO}, we wish to construct an approximate result for
$\df\sigma^\FO$ at NNLO$_1$, which we can consistently match to $\df\sigma^\res$
at \nnnllp. As discussed in
\sec{profile}, this requires that the NNLO$_1$ cross section
contains the correct singular terms $\df\sigmasing$, which are part of
the \nnnllp\ result. We must therefore approximate the remaining nonsingular
part of the full NNLO$_1$.
However, as also discussed in \sec{profile}, at large $q_T$ there is a strong cancellation
between singular and nonsingular, which means the two pieces are strongly
correlated and only the full fixed-order result is meaningful there. Hence, at
large $q_T$ we should do the opposite and approximate the full result, considering
the nonsingular as a derived quantity given by the difference of full and singular.
To satisfy these competing requirements,
we introduce a general method to decorrelate the singular and nonsingular
contributions, which we will then use in the next subsection to construct the
actual approximation.

The basic idea behind the decorrelation of the singular and nonsingular contributions at
large $q_T$ involves shifting a correlated piece between the two~\cite{Dehnadi:2022prz},
%%%
\begin{align} \label{eq:decorrelation}
\df\sigma^\FO(q_T) &= \df\sigmasing(q_T) + \df \sigmanons(q_T)
\nn \\
&=\underbrace{ \df\sigmasing(q_T) + \df\sigmacorr(q_T)}_{\df\sigmasingtilde}
+ \underbrace{\df\sigmanons(q_T) - \df\sigmacorr(q_T)}_{\df\sigmanonstilde}
\equiv \df\sigmasingtilde(q_T) + \df\sigmanonstilde(q_T)
\,,\end{align}
%%%
where here and below we use the notation $\df\sigma(q_T) \equiv \df\sigma/\df q_T$
to make the $q_T$ dependence explicit.
We call $\df\sigmasingtilde(q_T)$ and $\df\sigmanonstilde(q_T)$ the
decorrelated singular and nonsingular contributions. The correlated piece
$\df\sigmacorr(q_T)$ is as of yet unspecified.

To achieve the desired decorrelation, we require the decorrelated nonsingular
to become equal to the full fixed-order result toward large $q_T$, and as
a consequence the decorrelated singular to vanish,
%%%
\begin{align}
\df\sigmanonstilde(q_T \to m_H) \to \df\sigma^\FO(q_T)
\,, \qquad
\df\sigmasingtilde(q_T \to m_H) \to 0
\,.\end{align}
%%%
This guarantees that no cancellations occur between them.
At the same time, the decorrelated nonsingular must remain power suppressed for
$q_T \ll m_H$, such that the decorrelated singular still contains all singular terms,
%%%
\begin{align}
\frac{\df\sigmanonstilde(q_T)}{\df\sigmasing(q_T)} \sim \ORd{\frac{q_T^2}{m_H^2}}
\,,\qquad
\df\sigmasingtilde(q_T) = \df\sigmasing(q_T)\biggl[1 +  \ORd{\frac{q_T^2}{m_H^2}}\biggr]
\,.\end{align}
%%%
These two conditions are equivalent to the following two conditions on $\df\sigmacorr(q_T)$,
%%%
\begin{align}
\df\sigmacorr(q_T \to m_H) \to -\df\sigma^\sing(q_T)
\,,\qquad
\frac{\df\sigmacorr(q_T)}{\df\sigmasing(q_T)} \sim \ORd{\frac{q_T^2}{m_H^2}}
\,.\end{align}
%%%

The easiest way to satisfy these conditions might be simply to take $\df\sigmacorr(q_T)$
to be a constant, $\df\sigmacorr(q_T) = -\df\sigma^\sing(m_H)$.
This is equivalent to what was used in \refcite{Dehnadi:2022prz}, where the analogous
decorrelation was used in a similar context. In that particular case, the phase space was strictly
bounded to the equivalent of $q_T \leq m_H$. In contrast, this is no longer possible in our case:
the phase  space does not have such a strict boundary, and the decorrelation
condition in \eq{decorrelation} must hold not only at the single point $q_T = m_H$
but for any $q_T \gtrsim m_H$. In other words, we require not only that
$\df\sigmasingtilde(q_T)$ crosses through $0$ at $q_T = m_H$, but also that it remains
zero for any larger $q_T$. Furthermore, a constant value for $\df\sigmacorr(q_T)$ only
corresponds to a linear power suppression of $\ord{q_T/m_H}$. To obtain the correct
quadratic power suppression of $\ord{q_T^2/m_H^2}$, the correct extension of
\refcite{Dehnadi:2022prz} to our case is to take $\df\sigmacorr/\df q_T^2$ to be
a constant.

To achieve this, let us denote $s(q_T) \equiv \df\sigma/\df q_T^2$ and choose
$\df\sigmacorr(q_T)$ more generally such that
%%%
\begin{alignat}{9}
s^\mathrm{corr}(q_T) &= - s^\sing(q_T) \quad &&\text{for} \quad q_T \sim m_H
\,, \nn \\
s^\mathrm{corr}(q_T) &=  -s^\sing(\kappa m_H) = \text{const.}
\quad &&\text{for} \quad q_T \ll m_H
\,.\end{alignat}
%%%
That is, $s^\mathrm{corr}(q_T)$ is given by $-s^\sing(q_T)$ at large $q_T$
and freezes to a constant $-s^\sing(\kappa m_H)$ at small $q_T$,
where $\kappa \sim 1$ is a constant of our choice.
To make this a smooth transition, we can reuse our profile functions and take
%%%
\begin{align} \label{eq:scorr}
s^\mathrm{corr}(q_T)= -s^\sing[\qttilde(q_T)]
\,,\end{align}
%%%
where $\qttilde(q_T)$ is a function of $q_T$ that transitions from $\kappa m_H$ to $q_T$,
%%%
\begin{align}
\qttilde (q_T)= \kappa\, m_H\, \grun(q_T/m_H) + q_T [1-\grun (q_T/m_H)]
\,,\end{align}
%%%
and $\grun (q_T)$ is defined as in \eq{def_g_run}. For simplicity, we will use the
same transition points $[0.1,0.45,0.8]$ which we use for turning off the resummation
(see \sec{profile}).
Using \eq{scorr}, we arrive at our final choice for $\df\sigmacorr(q_T)$,
%%%
\begin{align} \label{eq:sigmacorr}
\df\sigmacorr(q_T)= - 2q_T\, s^\sing[\qttilde(q_T)]
= - \frac{q_T}{\tilde q_T(q_T)}\,\df\sigmasing[\tilde q_T(q_T)]
\,.\end{align}
%%%

%-------------------------------------------------------------------------------
\begin{figure}
\includegraphics[width=\WidthTwoSubfigs]{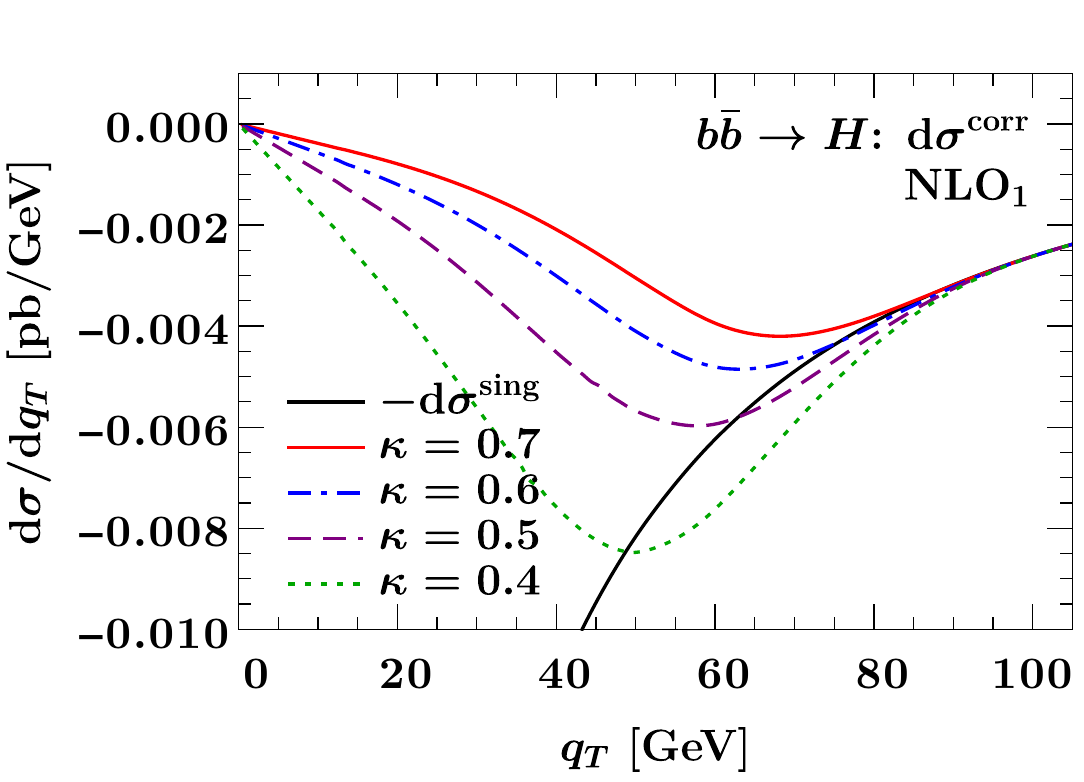}%
\hfill%
\includegraphics[width=\WidthTwoSubfigs]{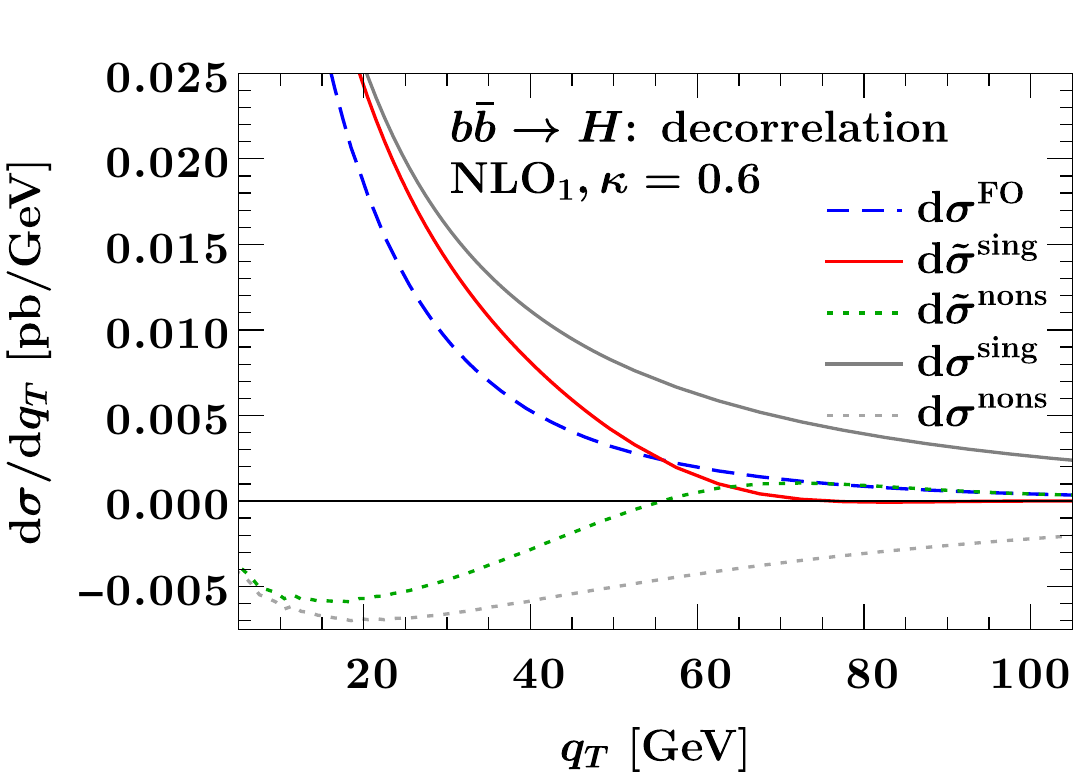}%
\caption{Decorrelation of singular and nonsingular at \nloone.
Left: Correlated contribution for different choices of $\kappa$ (coloured lines)
and its large-$q_T$ asymptotic value $-\df\sigmasing$ (black).
Right: Decorrelated singular (solid red), full, (dashed blue), and decorrelated
nonsingular (dotted green) for $\kappa = 0.6$ at fixed \nloone. The original,
correlated singular and nonsingular are shown by the dotted and solid grey lines.}
\label{fig:decorrelationNLO1}
\end{figure}
%-------------------------------------------------------------------------------

In \fig{decorrelationNLO1}, we study the decorrelation procedure at \nloone,
where the fixed-order result is fully known. The left panel of the figure shows
the correlated piece $\df\sigmacorr(q_T)$ for different choices of $\kappa$
alongside $-\df\sigmasing(q_T)$. For $q_T \geq x_3 m_H = 100 \GeV$,
$\df\sigmacorr(q_T)$ exactly equals $-\df\sigmasing(q_T)$, while going to lower
$q_T$ it starts to deviate and eventually turn around and vanish linearly
for $q_T\to 0$ as required by \eq{sigmacorr}. The correlated
contribution itself depends strongly on the choice of $\kappa$, which determines
where it effectively freezes out and turns around toward $0$.
Note also that by construction this dependence cancels exactly, such that the
full result at this order is independent of $\kappa$. The actual choice of
$\kappa$ could in principle influence our NNLO$_1$ approximation, but essentially
does not do so, as we shall see in the following subsection.

The right panel of \fig{decorrelationNLO1} shows both the original, correlated
singular (solid grey) and nonsingular (dotted grey) as well as the decorrelated
singular (solid red) and nonsingular (dotted green) for $\kappa = 0.6$. Since they each sum to the
fixed-order result (dashed blue), the correlated terms clearly exhibit a large
cancellation for large $q_T\sim m_H$. In contrast, the decorrelated singular
goes to zero for $q_T \sim m_H$, while the decorrelated nonsingular (dotted
green) becomes equal to the full fixed order. This confirms that the
decorrelation works as expected, and that $\df\sigmanonstilde$ and
$\df\sigmasingtilde$ no longer exhibit strong cancellations. We will therefore
use $\kappa = 0.6$ for $b\bar b\to H$. Since 
the strong cancellations between singular and nonsingular occur successively
later for $c\bar c\to H$ and $s\bar s\to H$, as we saw in \sec{matchprocedure}, we will use higher values
$\kappa = 0.7$ for $c\bar c\to H$ and $\kappa = 0.8$ for $s\bar s\to H$.

%===============================================================================
\subsection{\texorpdfstring{aNNLO$_1$}{aNNLO1}}
\label{sec:aNNLO1}
%===============================================================================

Using the decorrelation method explained in the previous section, we are now in
a position to construct an approximate \nnloone\ result as
%%%
\begin{align} \label{eq:aNNLO1approx}
\df\sigma^\FO(q_T)
= \df\sigmasingtilde(q_T) + \df\sigmanonstilde(q_T)
= \df\sigmasing(q_T) + \df\sigmacorr(q_T; \kappa)
+ \df\sigmanonstilde(q_T; \kappa)
\,,\end{align}
%%%
where we made the dependence on $\kappa$ in the last two terms explicit.
We now need to approximate the unknown $\ord{\as^3}$ contribution of
$\df\sigmanonstilde(q_T)$. To do so, we decompose $\df \tilde{\sigma}^\nons$
at the fixed scale $\mu_R = \mu_F = m_H$ in terms of perturbative coefficients $\tilde{c}_i(q_T)$,
%%%
\begin{align}
\df \sigmanonstilde(q_T)
= |y_b(m_H)|^2 \Bigl[{\alpha_s(m_H)}\, \tilde{c}_1(q_T) +\alpha_s^2(m_H)\, \tilde{c}_2(q_T) + \alpha_s^3(m_H)\, \tilde{c}_3(q_T)  \Bigr]
\,,\end{align}
%%%
where $\tilde{c}_1(q_T)$ and $\tilde{c}_2(q_T)$ are known, and our goal is to approximate $\tilde{c}_3(q_T)$.
To get the correct power of logarithms for $\tilde c_3(q_T)$, we perform a Pad\'e-like approximation
%%%
\begin{align}\label{eq:defc3}
\tilde{c}_3^{\rm approx}(q_T) = K\, \frac{[\tilde{c}_2(q_T)]^2}{\tilde{c}_1(q_T)}
\,,\end{align}
%%%
where we use the constant factor $K$ to rescale this result such that its overall size agrees with \refcite{Mondini:2021nck}. When using this approximation in \eq{aNNLO1approx}, we refer
to the result as aNNLO$_1$.

\begin{figure}
\includegraphics[width=\WidthTwoSubfigs]{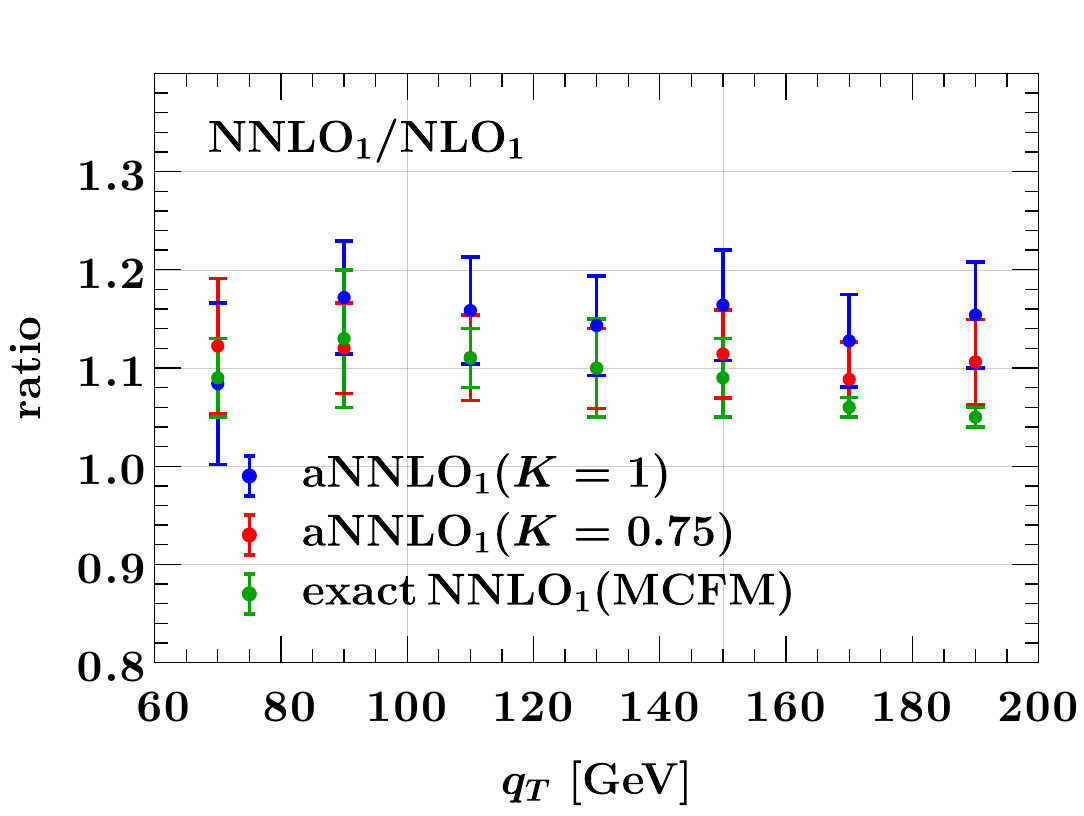}%
\hfill%
\includegraphics[width=\WidthTwoSubfigs]{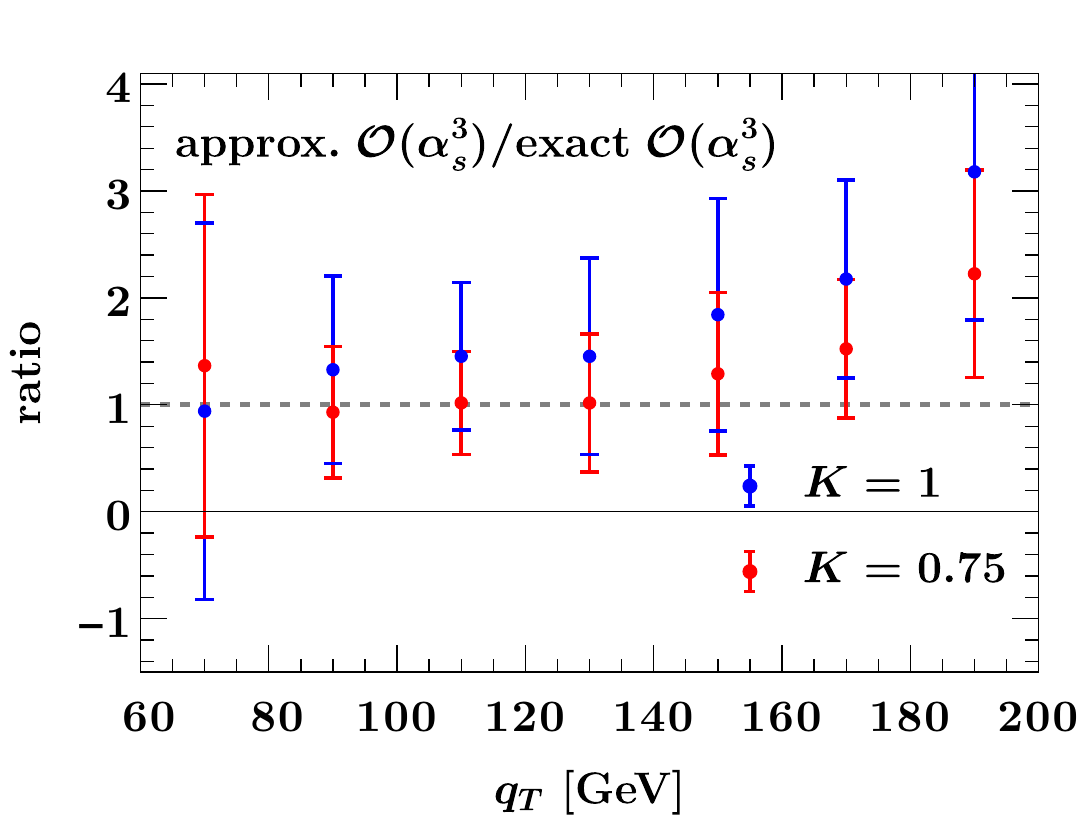}%
\caption{Comparison of the exact NNLO$_1$ results from \refcite{Mondini:2021nck}
at large $q_T$ to our approximation. Left: Ratio of NNLO$_1/$NLO$_1$ for the exact result (green) and the approximate result with $K=1$ (blue) and $K=0.7$ (red).
Right: Ratio of the approximate and exact $\ord{\alpha_s^3}$ contribution for $K=1$ (blue) and $K=0.7$ (red).
The uncertainties in all cases correspond to the scale variations.}
\label{fig:Kfac}
\end{figure}

To determine an appropriate value for $K$, we consider the ratio of our approximate $\ord{\alpha_s^3}$ coefficient to the exact result shown on the right in \fig{Kfac}.
The exact $\ord{\alpha_s^3}$ coefficient is obtained by subtracting NNLO$_1-$NLO$_1$, where we can read off the ratio NNLO$_1/$NLO$_1$ from the results shown in \refcite{Mondini:2021nck}.
As already mentioned, \refcite{Mondini:2021nck} uses a cut on the leading jet $p_T \geq 30 \GeV$, so we can only use their results for $q_T \geq 60 \GeV$.
For this purpose we use the same inputs as \refcite{Mondini:2021nck}, i.e., $\mu_F=m_H/2$ and the \texttt{CT14nnlo} PDF set.
To extract a value for $K$, we perform a simple $\chi^2$ fit
to this ratio as a function of $K$, requiring that the ratio is unity.
Since the kinematic region we are interested in is $q_T \lesssim m_H$, we use
the first four points for $q_T \in [60, 140] \GeV$. Note that in this region the ratio is
well approximated by a constant, which shows that the approximation in \eq{defc3}
captures the $q_T$ dependence well. We find $K = 0.75$, which we will use as our
default value.
We refrain from including an uncertainty on $K$, since it would be negligible
compared to the nominal perturbative uncertainties.

The left panel of \fig{Kfac} shows the ratio NNLO$_1/$NLO$_1$ for the exact NNLO$_1$ result from \refcite{Mondini:2021nck} (green) as well as for our approximate aNNLO$_1$ for $K=1$ (blue) and our default $K=0.75$ (red).
The uncertainties correspond to varying $\mu_R$ and $\mu_F$ by a factor of 2.
For our default $K$, we find very good agreement in the region of interest $q_T \lesssim m_H$
between our approximation and the exact results.

We then use the same
value of $K$ also at lower values of $q_T$ as well as for the $s\bar s\to H$
and $c\bar c\to H$ channels and our default PDF set. That is, we effectively use
our approximation to extrapolate the exact results from \refcite{Mondini:2021nck}
to lower $q_T$ and the other channels and PDF.

%-------------------------------------------------------------------------------
%-------------------------------------------------------------------------------

\begin{figure}
\includegraphics[width=\WidthTwoSubfigs]{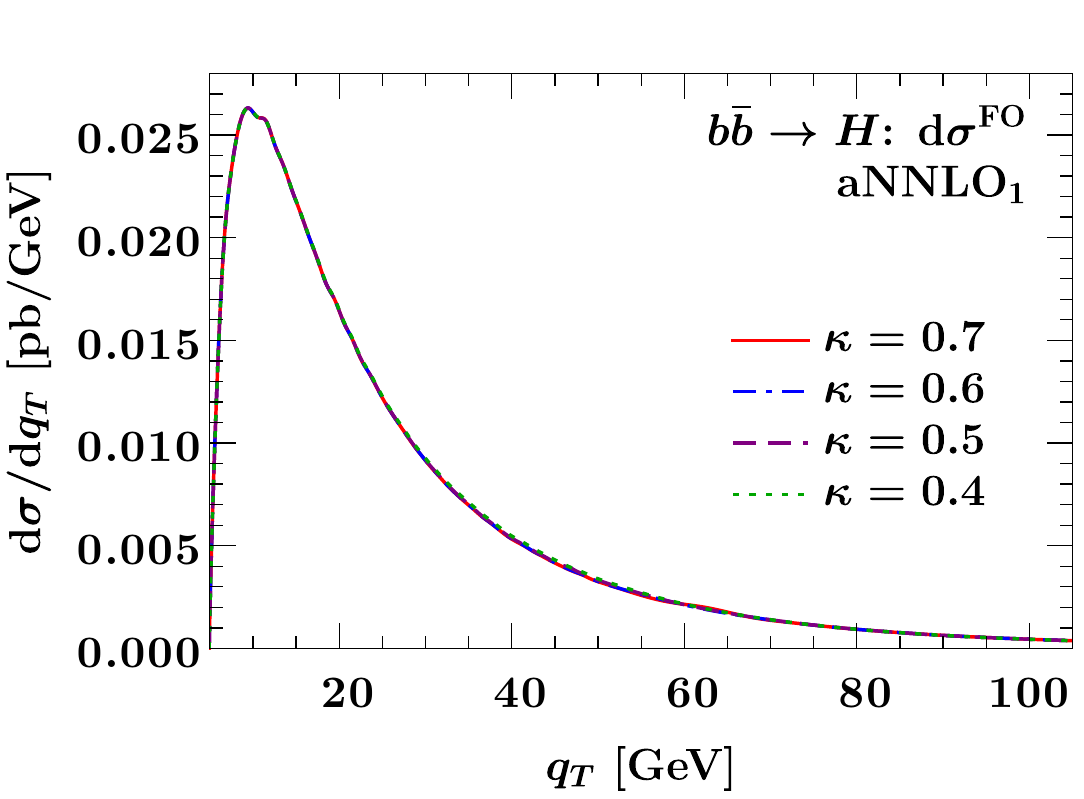}%
\hfill%
\includegraphics[width=\WidthTwoSubfigs]{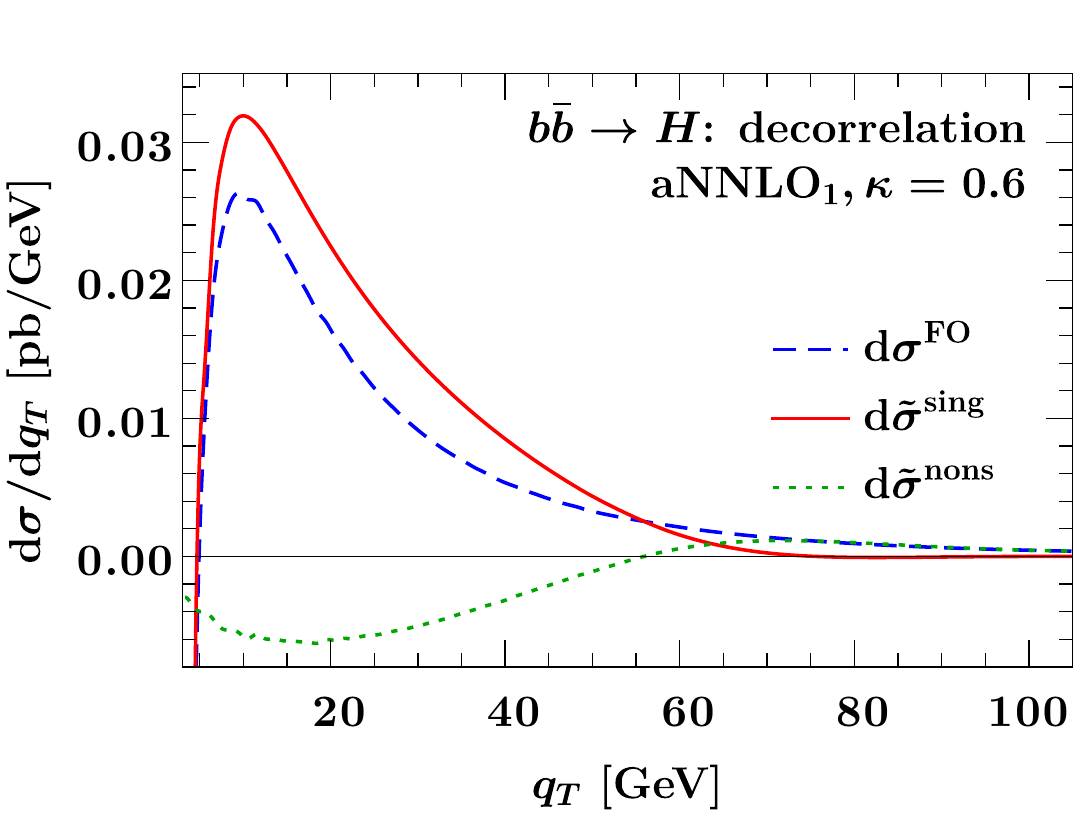}%
\caption{Decorrelation at aNNLO$_1$. Left: The resulting full fixed-order order cross section
for different values of $\kappa$. Right: Decorrelated singular (solid red), full (dashed blue),
and decorrelated nonsingular (dotted green) for $\kappa=0.6$.}
\label{fig:aN3LO}
\end{figure}

The coefficients $\tilde{c}_i(q_T)$ depend on the choice of $\kappa$. For the
exact coefficients, the $\kappa$ dependence exactly cancels between the last
two terms in \eq{aNNLO1approx}. However, when using the approximate
$\tilde c_3^{\rm approx}(q_T)$, the $\kappa$ dependence will no longer cancel exactly. The residual $\kappa$ dependence of the aNNLO$_1$ result is shown in
the left panel of \fig{aN3LO}. Happily, we find that the approximate result is practically
independent of the value of $\kappa$.
In the right panel of \fig{aN3LO}, we illustrate the decomposition of the
approximated full result for our default choice $\kappa = 0.6$ into the decorrelated
singular and nonsingular pieces.

%-------------------------------------------------------------------------------
%-------------------------------------------------------------------------------

To obtain the correct scale dependence for the approximated result,
we re-expand $\alpha_s(m_H)$ and $y_b(m_H)$ in terms of $\alpha_s(\mu_R)$, which yields
%%%
\begin{align}\label{eq:anlo_muR}
\frac{\df \tilde{\sigma}^\nons}{\df q_T}
&= |y_b(\mu_R)|^2\biggl\{
  \alpha_s(\mu_R)\, \tilde{c}_1(q_T)
+ \alpha_s^2(\mu_R) \biggl[\tilde{c}_2(q_T) + \frac{\tilde{c}_1}{2 \pi}(\beta_0-\gamma_0) \ln\frac{\mu_R}{m_H} \biggr]
\nn \\ & \quad
+ \alpha_s^3(\mu_R) \biggl[\tilde{c}_3(q_T) + \Bigl[ \frac{\tilde{c}_1(q_T) }{8 \pi^2} (\beta_1- \gamma_1)
+ \frac{\tilde{c}_2(q_T)}{2 \pi} (2\beta_0-\gamma_0 ) \Bigr] \ln\frac{\mu_R}{m_H}
\nn \\ & \qquad\qquad\qquad
+ \frac{\tilde{c}_1(q_T)}{8 \pi^2} (2 \beta_0^2 - 3\beta_0 \gamma_0 +\gamma_0^2) \ln^2\frac{\mu_R}{m_H}\biggr]\biggr\}
\,,
\end{align}
%%%
where $\beta_n$ and $\gamma_n$ are the relevant coefficients of the QCD beta
function and the Yukawa anomalous dimension,
%%%
\begin{alignat}{9}
\beta_0&=\frac{11}{3} C_A - \frac{4}{3} T_F n_f
\,,\qquad &
\beta_1&= \frac{34}{3} C_A - \left(\frac{20}{3}C_A+ 4 C_F\right) T_F n_f
\,, \nn \\
\gamma_0 &= - 6 C_F
\,,\qquad &
\gamma_1 &= - 2 C_F \left(\frac{3}{2} C_F + \frac{97}{6} C_A - \frac{10}{3} T_F n_f\right)
\,.\end{alignat}
%%%
The $\mu_R$ dependence in
the approximated result is therefore exact, and we are able to vary $\mu_R$
without further approximation.

For the $\mu_F$ dependence, for simplicity we perform the approximation for
$\tilde c_3(q_T)$ in \eq{defc3} in terms of $\tilde{c}_1(q_T)$ and
$\tilde{c}_2(q_T)$ at any given $\mu_F$, using the same rescaling factor $K$ as
for the central $\mu_F$ choice. This means we will only have an approximate
$\mu_F$ dependence at $\ord{\alpha_s^3}$ that only approximately cancels up to
higher $\ord{\alpha_s^4}$ terms. This will lead to slightly larger $\mu_F$
variations compared to the exact $\ord{\alpha_s^3}$ $\mu_F$ dependence, which we
can simply consider as an additional uncertainty due to the approximation.

%%%%%%%%%%%%%%%%%%%%%%%%%%%%%%%%%%%%%%%%%%%%%%%%%%%%%%%%%%%%%%%%%%%%%%%%%%%%%%%%
\section{Results}
\label{sec:results}
%%%%%%%%%%%%%%%%%%%%%%%%%%%%%%%%%%%%%%%%%%%%%%%%%%%%%%%%%%%%%%%%%%%%%%%%%%%%%%%%

%-------------------------------------------------------------------------------
\begin{figure}
\includegraphics[width=\WidthTwoSubfigs]{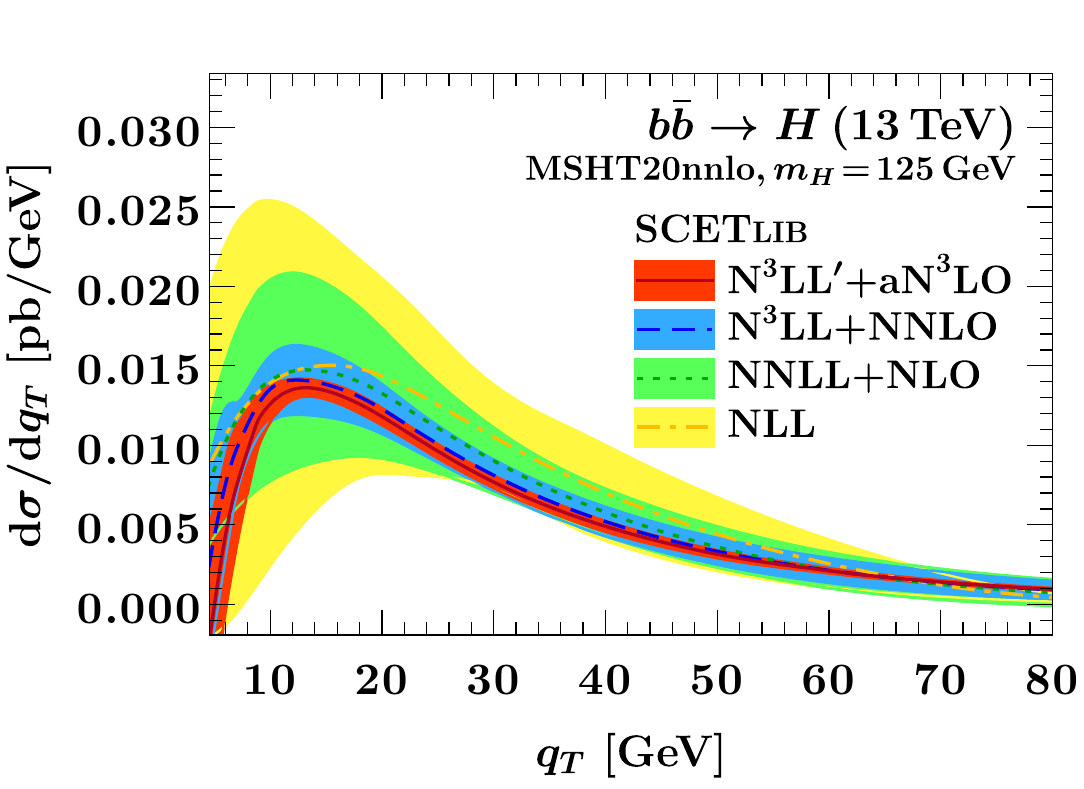}%
\hfill%
\includegraphics[width=\WidthTwoSubfigs]{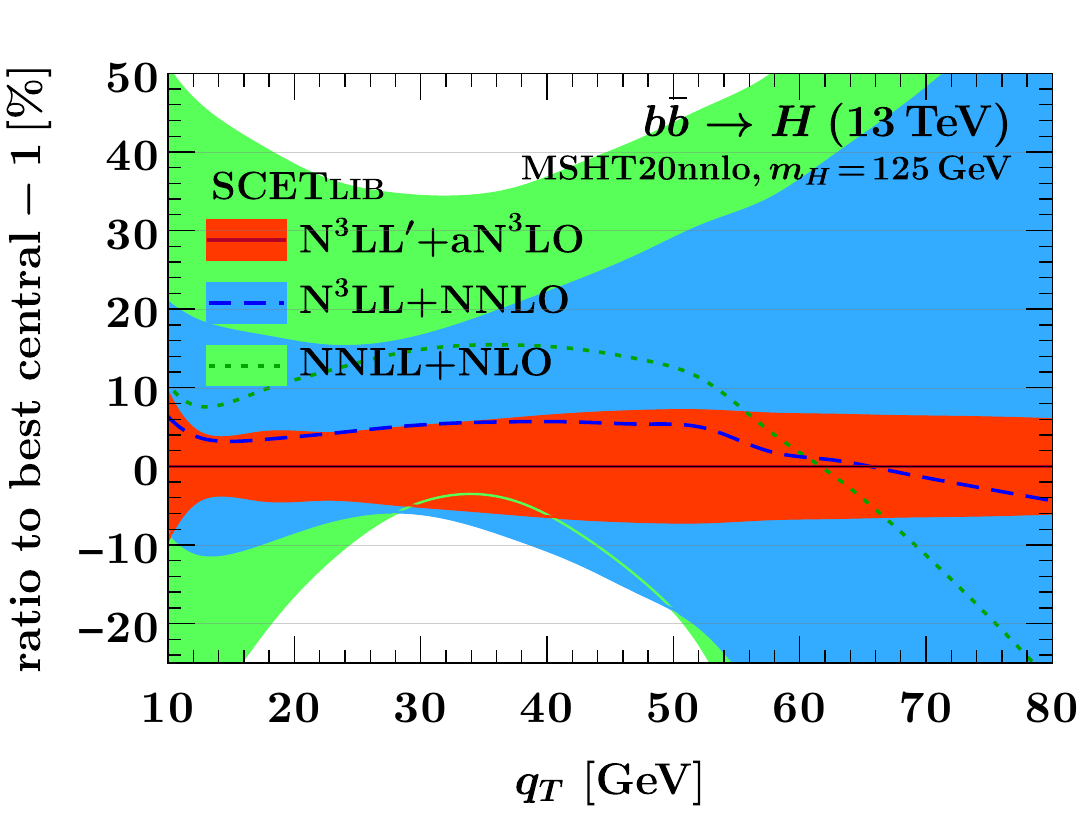}%
\\
\includegraphics[width=\WidthTwoSubfigs]{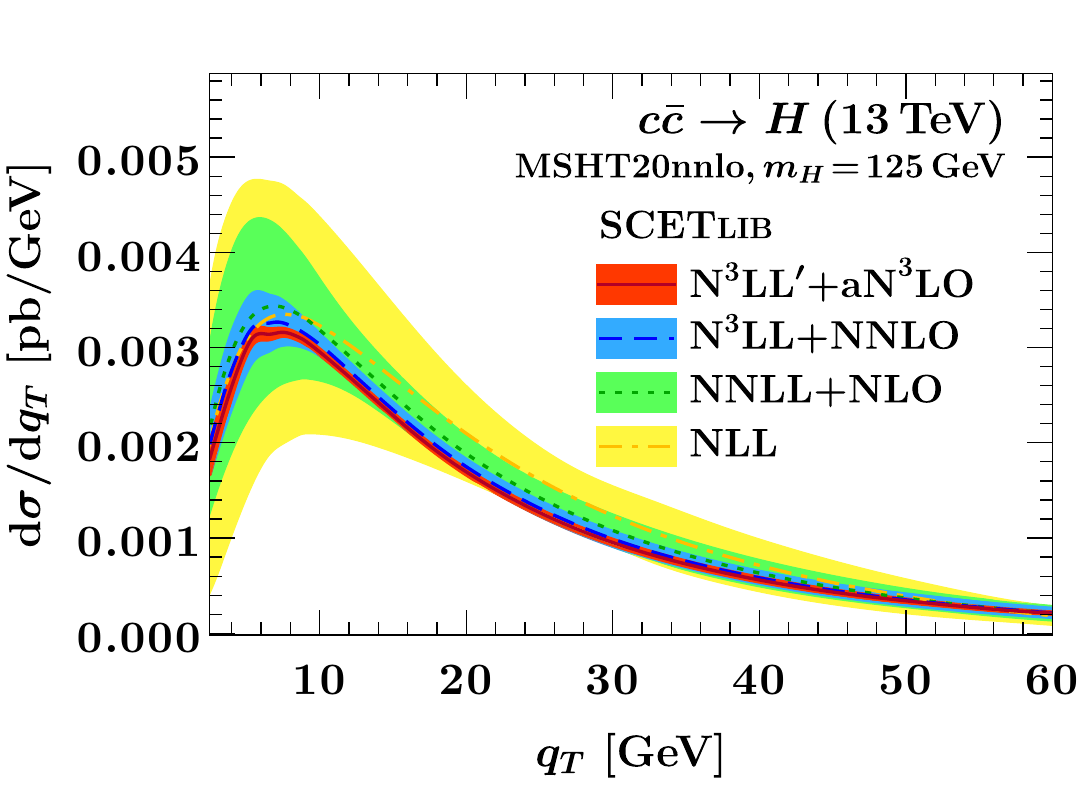}%
\hfill%
\includegraphics[width=\WidthTwoSubfigs]{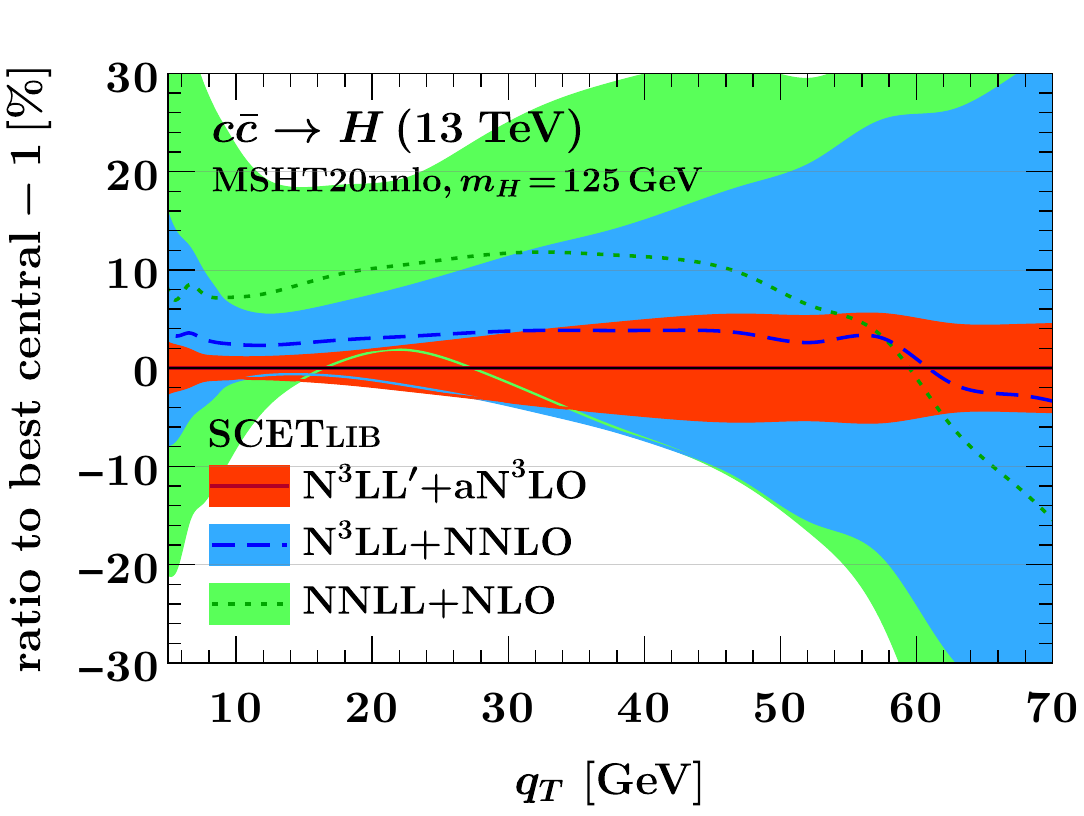}%
\\
\includegraphics[width=\WidthTwoSubfigs]{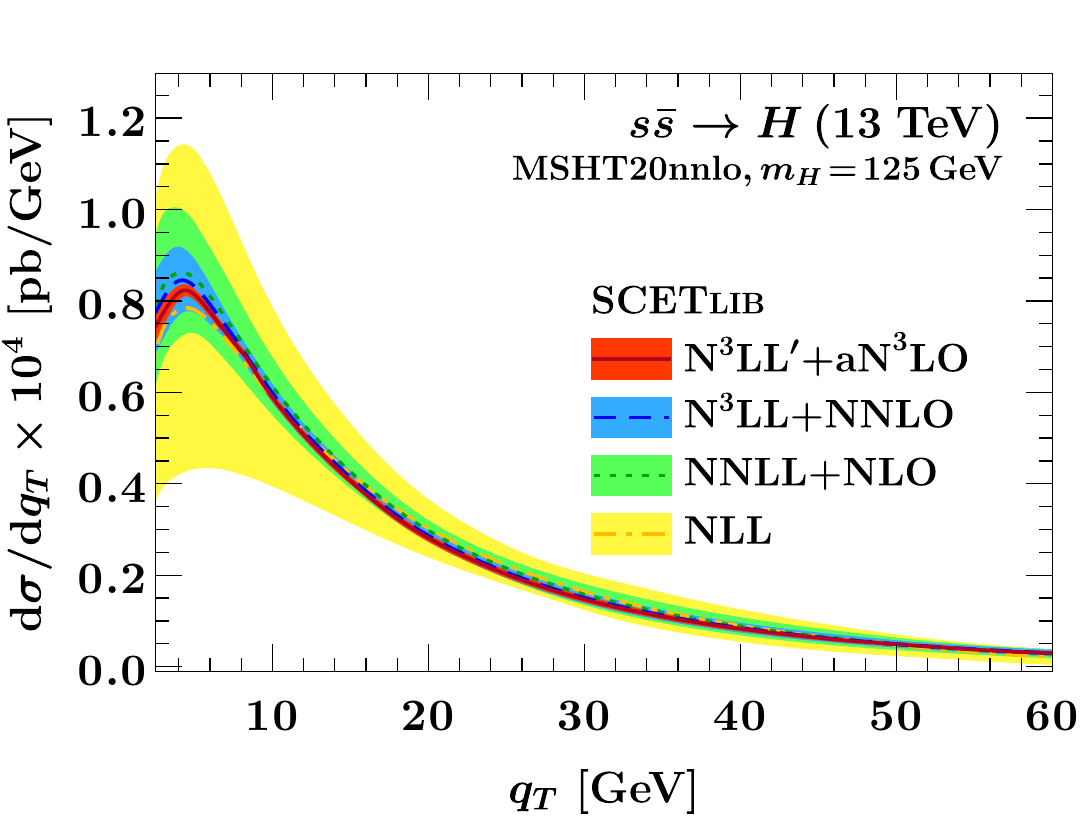}%
\hfill%
\includegraphics[width=\WidthTwoSubfigs]{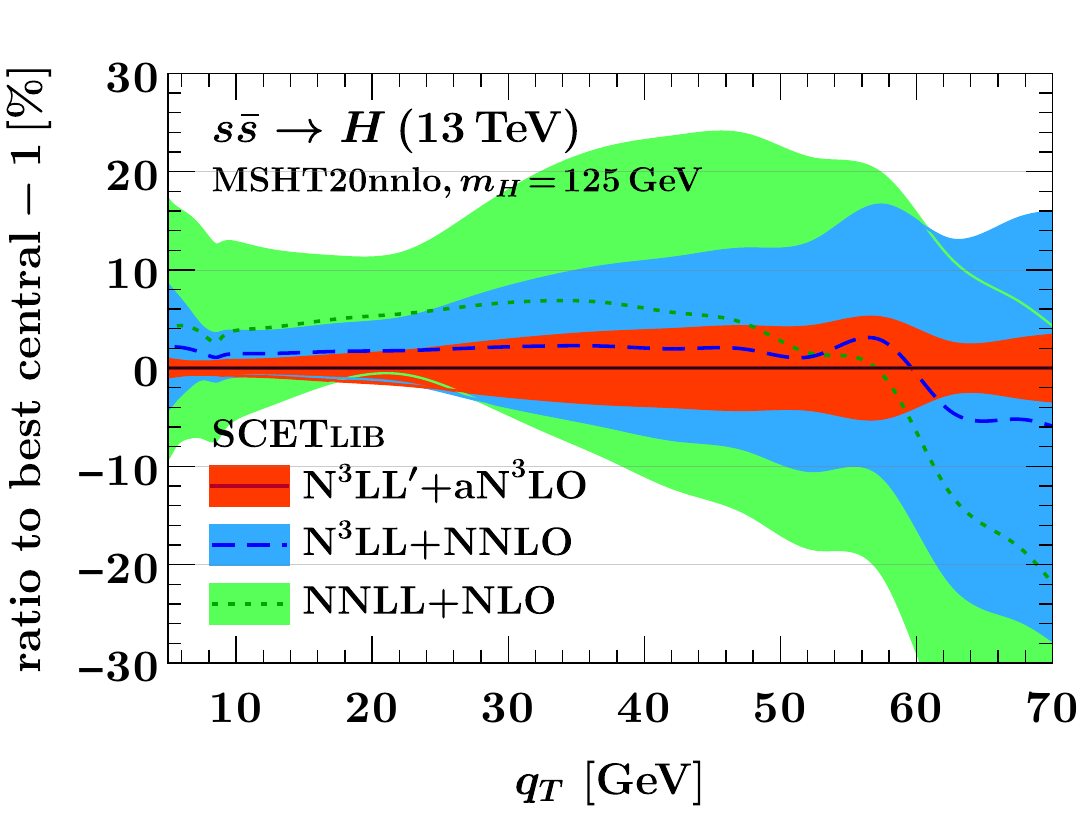}%
\caption{Resummed $q_T$ spectrum up to N$^3$LL$'+$aN$^3$LO for $b\bar{b}\to H$
(top row),  $c\bar{c}\to H$ (middle row), and $s\bar{s}\to H$ (bottom row).
The results for the spectrum are shown on the left. The results normalized
relative to the best central value at N$^3$LL$'+$aN$^3$LO are shown on the right.}
\label{fig:convergence}
\end{figure}
%-------------------------------------------------------------------------------

%-------------------------------------------------------------------------------
\begin{figure}
\centering%
\includegraphics[width=\WidthTwoSubfigs]{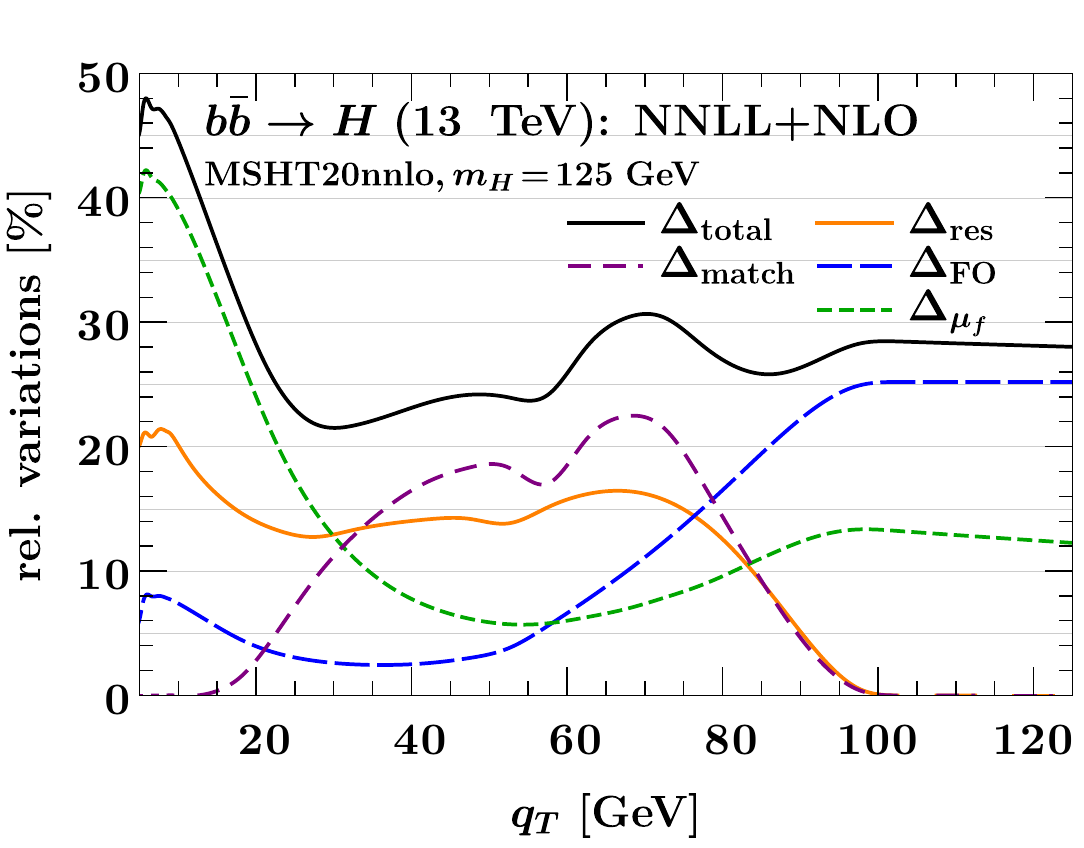}%
\\
\includegraphics[width=\WidthTwoSubfigs]{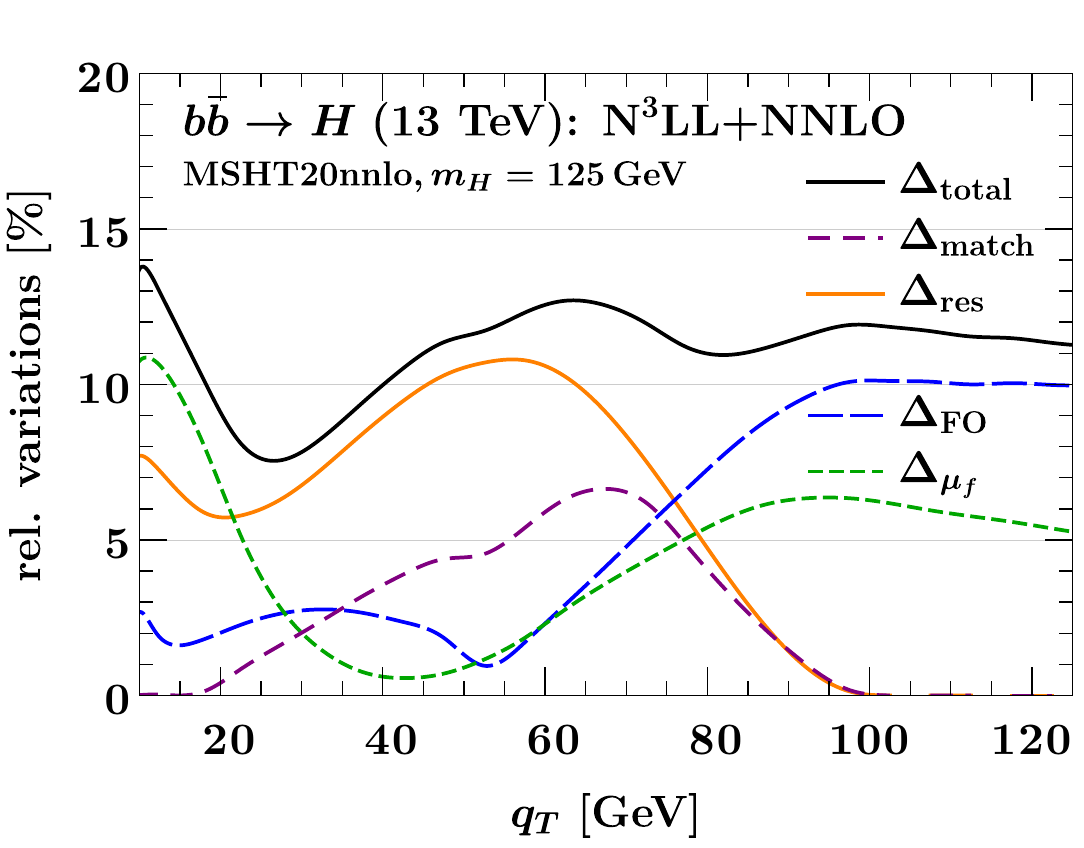}%
\hfill%
\includegraphics[width=\WidthTwoSubfigs]{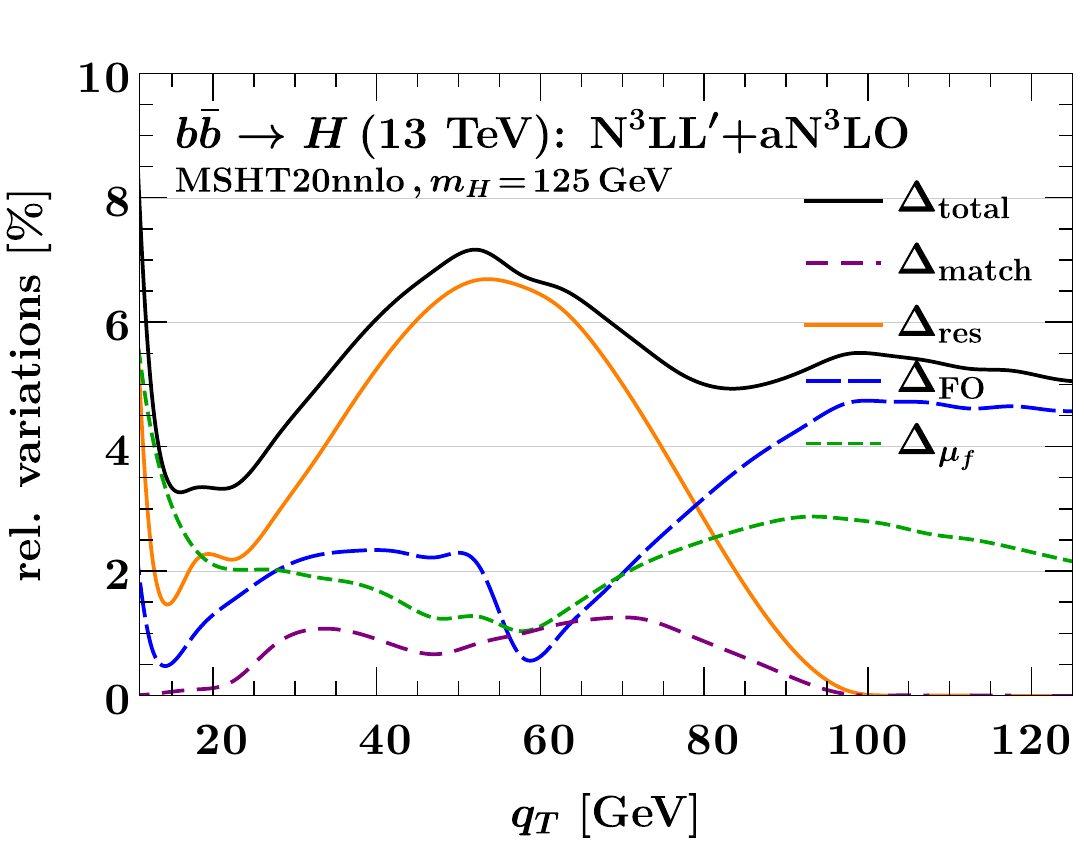}%
\caption{Breakdown of the perturbative uncertainties into its different
contributions for $b\bar{b} \to H$ at NNLL$+$NLO (top), N$^3$LL$+$NNLO (bottom
left), and N$^3$LL$'+$aN$^3$LO (bottom right). Note the different scales on the
vertical axes for each plot.}
\label{fig:bbHimpact}
\end{figure}
%-------------------------------------------------------------------------------

In this section, we present our numerical result for the $q\bar q \to H$ $q_T$
spectrum. We use $\Ecm = 13\TeV$, $m_H=125\GeV$, and the \texttt{MSHT20nnlo} PDF
set~\cite{Bailey:2020ooq} with $\alpha_s(m_Z) = 0.118$.
We assess the impact of changing the PDF set to \texttt{MSHT20an3lo}~\cite{McGowan:2022nag}
in \app{plots}. For the Yukawa coupling we evolve
$\overline{m}_q(\mu_q)$ to $\mu_\FO$ where $\mu_{b,c}=\overline{m}_{b,c}$ for
the bottom and charm quarks and $\mu_s=2 \GeV$ for the strange quark. The input
$\overline{\mbox{MS}}$ quark masses are $\overline{m}_b(\overline{m}_b)
=4.18\GeV$, $\overline{m}_c(\overline m_c) =1.27\GeV$, and
$\overline{m}_s(2\GeV)=93.4\MeV$~\cite{Workman:2022ynf}, and we use $v =
246.22\GeV$ for the Higgs vev to convert the masses into Yukawa couplings. Our
scale choices are described in \sec{profile}. All our numerical results for the
resummed and fixed-order singular contributions are obtained with
\scetlib~\cite{scetlib}. The full fixed-order results are obtained as discussed
in \secs{fullFO}{aNNLO1}. For the aNNLO$_1$ result we use the parameters $\kappa
= 0.6$ and $K = 0.7$.

In \fig{convergence}, we show the resummed $q_T$ spectrum for $b\bar{b}\to H$ (top), $c\bar{c}\to H$ (middle) and $s\bar{s}\to H$ (bottom) at different resummation orders up to the highest N$^3$LL$'+$aN$^3$LO. The bands show the perturbative uncertainty estimate as discussed in \sec{pertunc}. We observe excellent perturbative convergence for all channels, with reduced uncertainties at each higher order. The perturbative uncertainties increase in general from $s\bar{s}\to H$, to $c\bar c\to H$, to $b\bar b\to H$.
Comparing the ratio plots for $s\bar{s} \to H$ and $b\bar{b} \to H$ it is evident that the relative uncertainties for $b\bar{b} \to H$ at a given order are of similar size as those for $s\bar{s} \to H$
at one lower order.
As already mentioned in \sec{matchprocedure}, the main difference between the channels is the relative size of the PDF luminosities. Since for $b\bar b \to H$, the $b\bar b$ Born channel is numerically suppressed by the small $b$-quark PDFs, the gluon-induced PDF channels which start at one higher order play a much more prominent role. This explains the observed pattern of uncertainties for the different cases.

Our default choice for the PDF scale $\mu_f$ corresponds to taking $\mu_F = m_H$
in the fixed-order limit.
Fixed-order predictions for $b\bar b\to H$ traditionally use a lower
scale of $\mu_F = m_H/2$ or $\mu_F = m_H/4$, so one might wonder
whether the uncertainties for $b\bar b\to H$ might be reduced by choosing a
lower central value for $\mu_F$. In \app{plots} we provide analogous resummed
results for these lower $\mu_F$ choices, which show that in fact the opposite is
the case: by lowering $\mu_F$, the perturbative convergence for the resummed $q_T$
spectrum gets noticeably worse, justifying our default choice for $\mu_F$.

A detailed breakdown of the uncertainty estimate for $b\bar{b}\to H$ is shown in
\fig{bbHimpact}. The $\Delta_{\mu_f}$ uncertainty (short-dashed green) dominates
up to $q_T \lesssim 20 \GeV$ before tending to a constant for $q_T \gtrsim 80
\GeV$. At NNLL$+$NLO the matching uncertainty $\Delta_\match$ (long-dashed
purple) is largest for $30 \GeV \lesssim q_T \lesssim 80 \GeV$ and vanishes
outside of the transition region as it should. At higher orders, the resummation
uncertainty $\Delta_\res$ (solid orange) dominates in this region before going
to zero as the resummation is turned off toward large $q_T$. As one might
expect, at the same time the fixed-order uncertainty $\Delta_\FO$ (dashed blue)
increases and becomes the dominant uncertainty in the fixed-order region.

%-------------------------------------------------------------------------------
\begin{figure}
\includegraphics[width=\WidthTwoSubfigs]{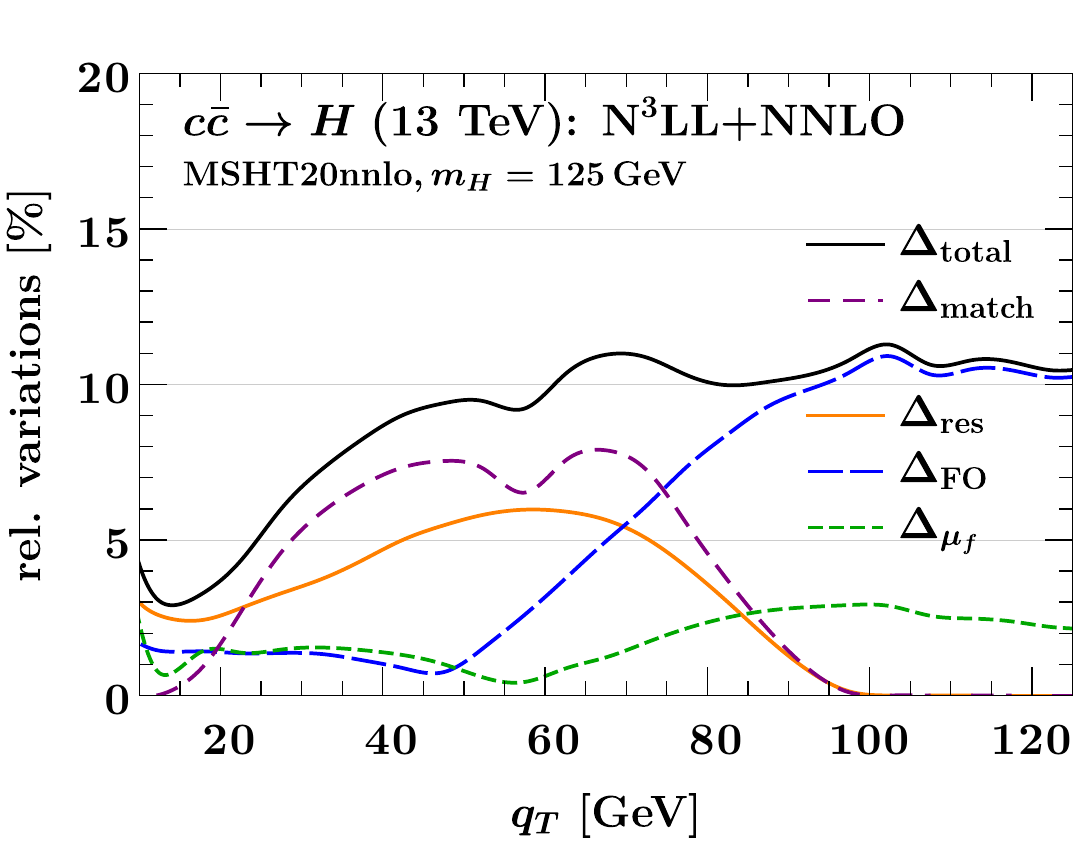}%
\hfill%
\includegraphics[width=\WidthTwoSubfigs]{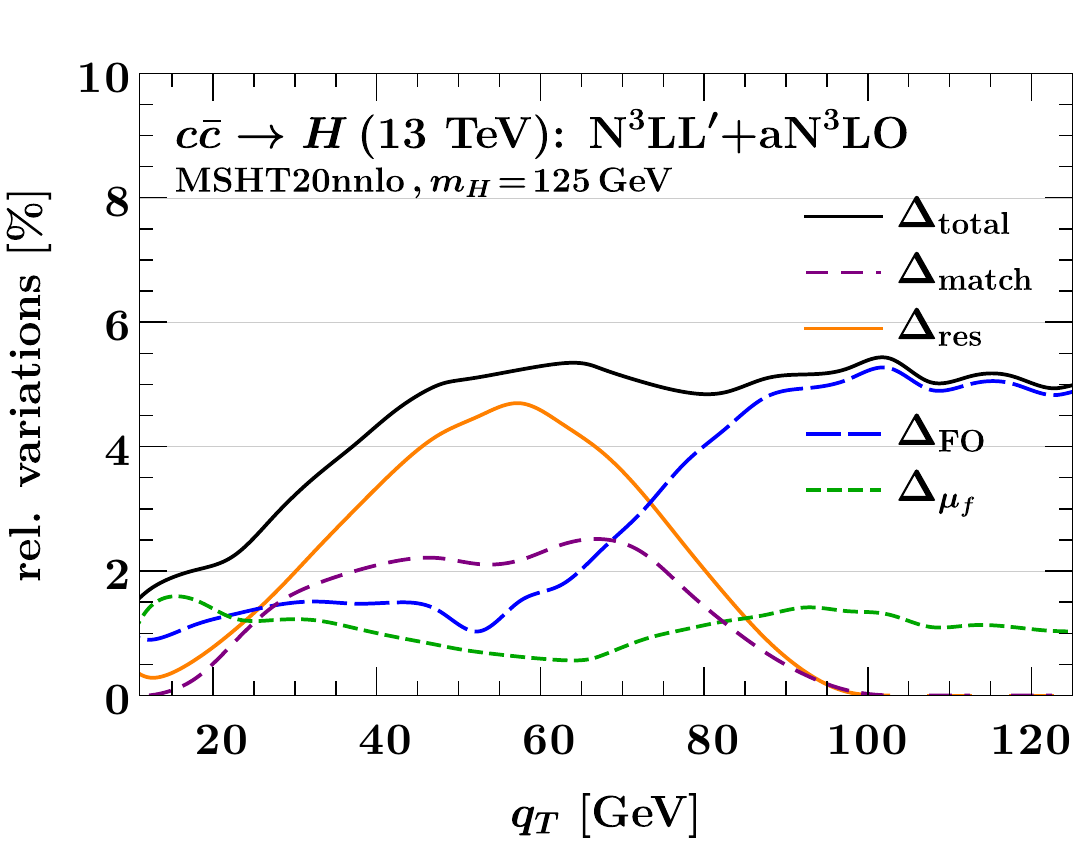}%
\\
\includegraphics[width=\WidthTwoSubfigs]{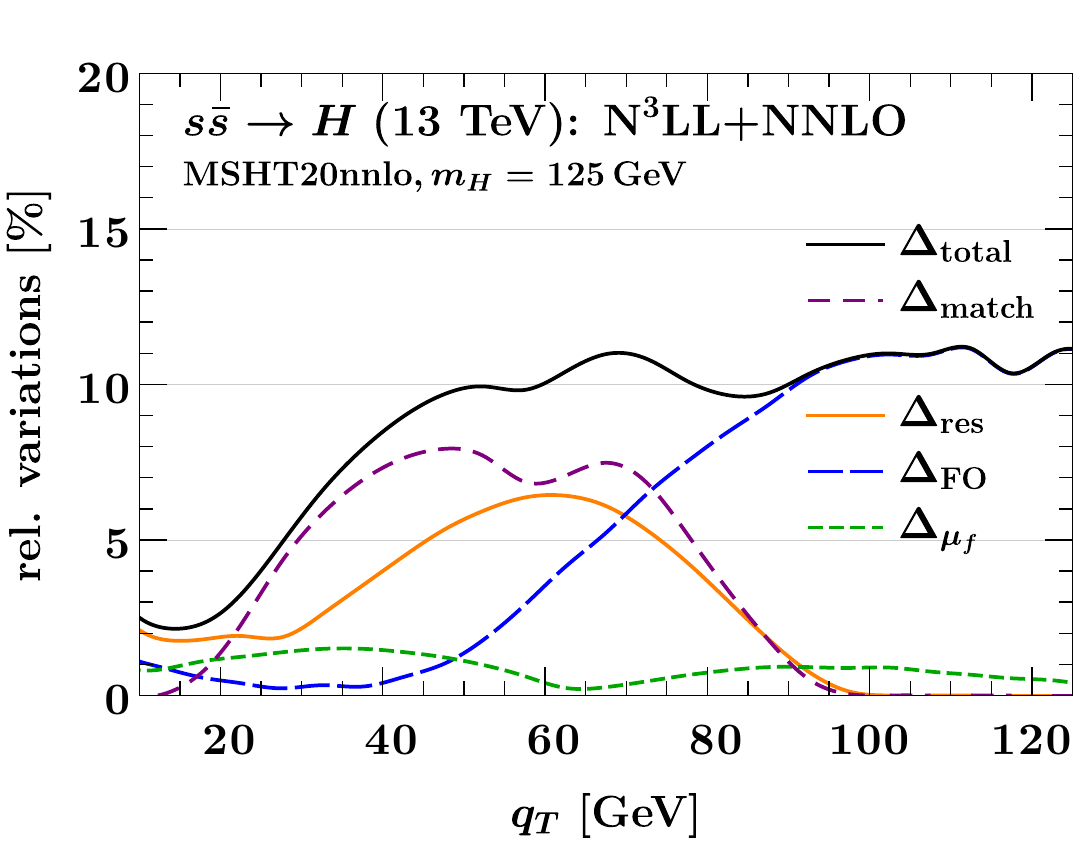}%
\hfill%
\includegraphics[width=\WidthTwoSubfigs]{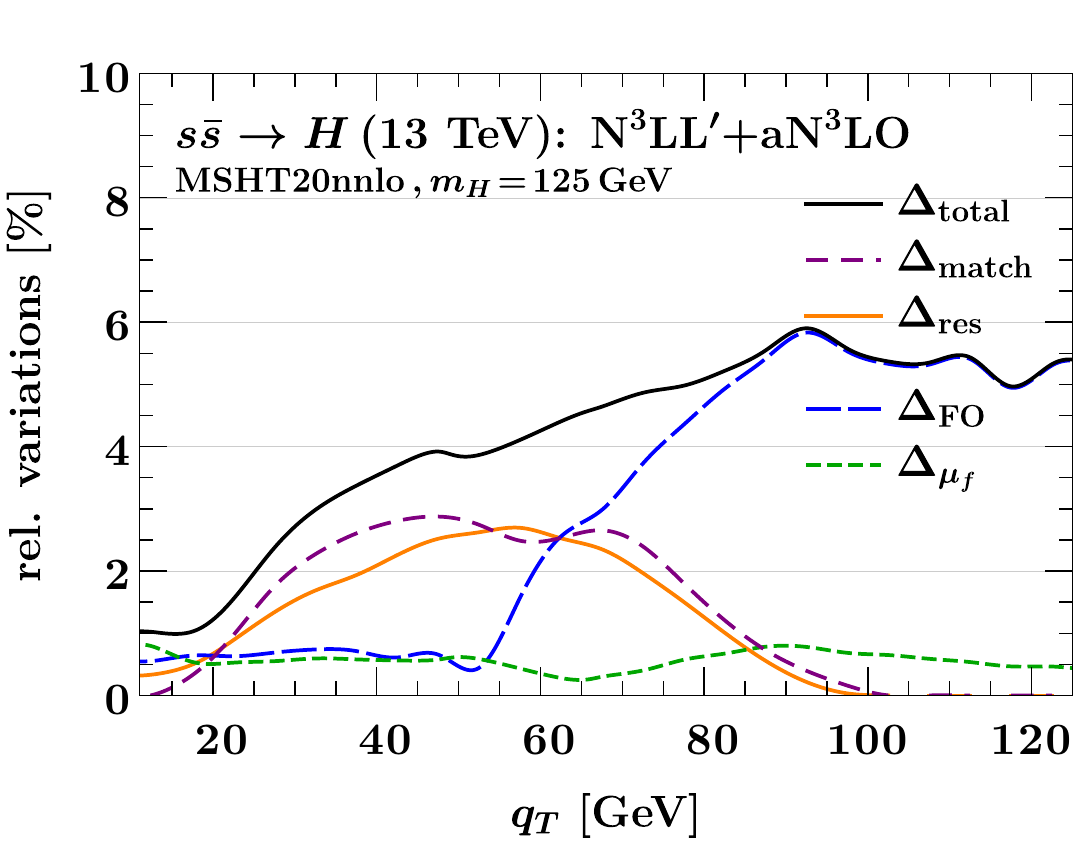}%
\caption{Breakdown of the perturbative uncertainties into its different
contributions for $c\bar{c} \to H$ (left) and $s\bar{s} \to H$ (right) at
N$^3$LL$+$NNLO (top) and N$^3$LL$'+$aN$^3$LO (bottom).}
\label{fig:ccHssHimpact}
\end{figure}
%-------------------------------------------------------------------------------

In \fig{ccHssHimpact}, we show the impact of the individual uncertainties for
$c\bar{c}\to H$ and $s\bar{s}\to H$. In general, they display a behaviour very
similar to $b\bar{b}\to H$. The main difference between the processes is the
size of the resummation and the matching uncertainty. The matching uncertainty
is slightly smaller for $b\bar{b}\to H$; this is to be expected, since we chose
our transition points in \sec{matchprocedure} for this specific case. On the
other hand, $\Delta_\res$ is smaller for $c\bar{c}\to H$ and $s\bar{s}\to H$.
The total uncertainty for $s\bar{s}\to H$ is therefore dominated by
$\Delta_\match$ for $30 \GeV \lesssim q_T \lesssim 70 \GeV$. For $c\bar{c}\to
H$, $\Delta_\match$ has the largest impact at N$^3$LL$+$NNLO whereas at
N$^3$LL$'+$aN$^3$LO $\Delta_\res$ contributes the most. The fixed-order
uncertainty starts to dominate the total uncertainty slightly earlier than for
$b\bar{b}\to H$ as the other contributions are in general smaller.

\begin{figure}
\includegraphics[width=\WidthTwoSubfigs]{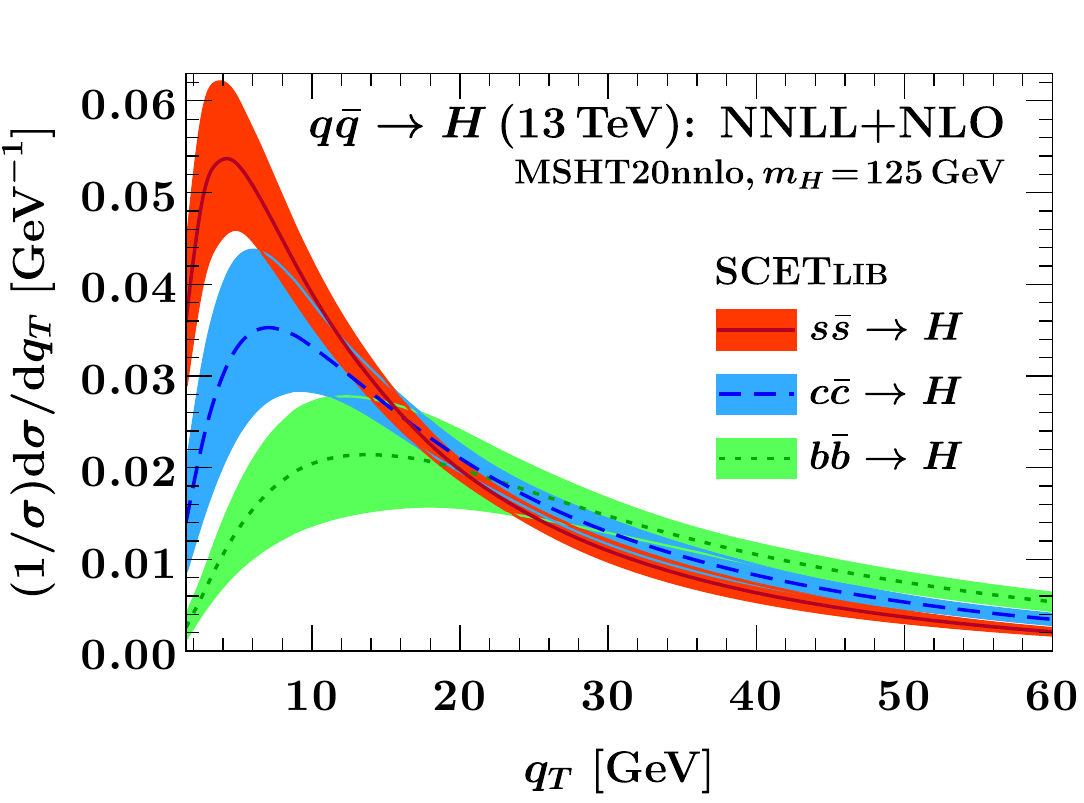}%
\hfill%
\includegraphics[width=\WidthTwoSubfigs]{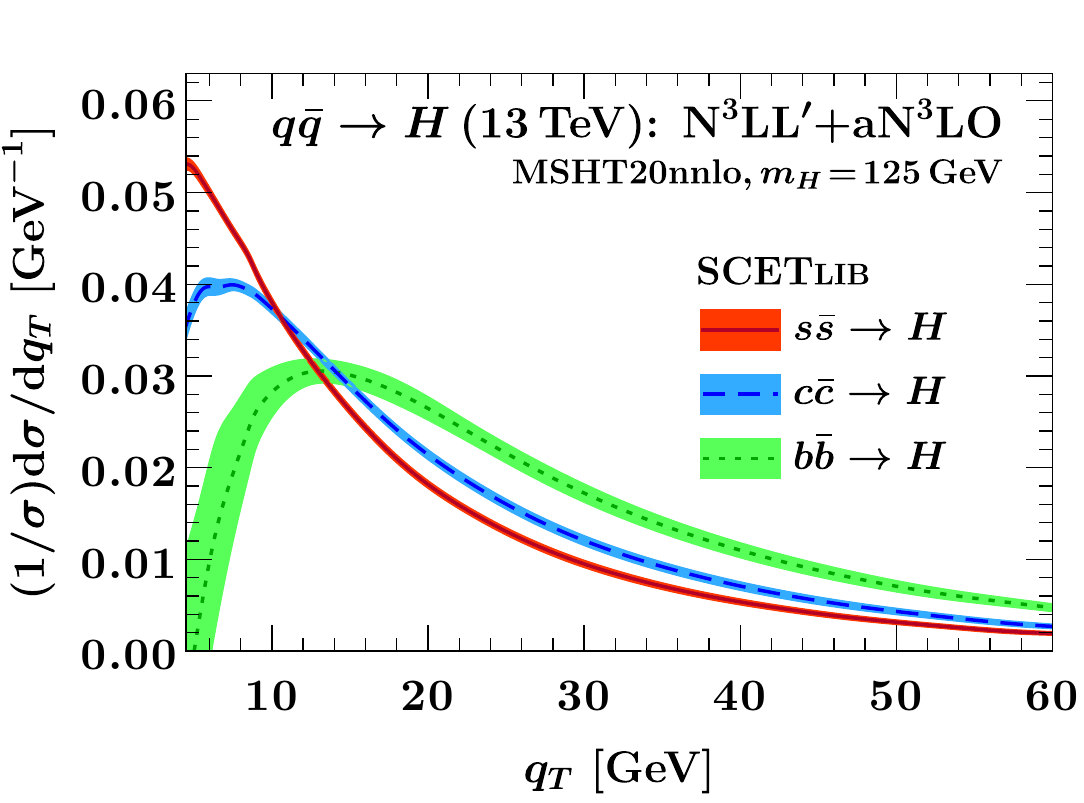}%
\caption{Comparison of the normalized $q_T$ spectrum for $s\bar{s}\to H$,
$c\bar{c}\to H$, and $b\bar{b}\to H$ at NNLL$+$NLO (left) and  N$^3$LL$'+$aN$^3$LO (right).}
\label{fig:separation}
\end{figure}

In \fig{separation}, we compare the normalized $q_T$ spectra for $s\bar{s}\to H$ (red), $c\bar{c}\to H$ (blue), and $b\bar{b}\to H$ (green). The left panel shows the spectra at NNLL$+$NLO, where one can already see that the spectra of the channels exhibit different shapes. At this order however the uncertainties largely overlap in the peak region. The right panel shows the spectra at N$^3$LL$'+$aN$^3$LO. Here, the uncertainties are significantly smaller and we can clearly distinguish the channels by their different shapes in $q_T$. Our predictions could therefore be used to improve the determination of quark Yukawa coupling from the shape of the measured Higgs $q_T$ spectrum -- such an analysis has already been performed in \refscite{ATLAS:2022qef,CMS:2023gjz}, using measurements in the $H\to ZZ^* \to 4 \ell$ and $H\to\gamma \gamma$ decay channels.

%-------------------------------------------------------------------------------
%-------------------------------------------------------------------------------

%%%%%%%%%%%%%%%%%%%%%%%%%%%%%%%%%%%%%%%%%%%%%%%%%%%%%%%%%%%%%%%%%%%%%%%%%%%%%%%%
\section{Conclusions}
\label{sec:conclusions}
%%%%%%%%%%%%%%%%%%%%%%%%%%%%%%%%%%%%%%%%%%%%%%%%%%%%%%%%%%%%%%%%%%%%%%%%%%%%%%%%

In this work, we have studied the transverse momentum spectrum of the Higgs
boson produced in heavy quark annihilation, $q\bar q\to H$ with $q = s, c, b$.
This is an interesting process, as
it has the potential to constrain the Yukawa couplings of charm, bottom, and
possibly strange quarks. We have used soft-collinear effective theory to resum
large logarithms of $q_T/m_H$ up to N$^3$LL$'$ order and matched these results to
fixed-order calculations. For $b\bar{b}\to H$ and to a
lesser extent $c\bar c\to H$, the large size of the
nonsingular terms requires extra care in the matching procedure and
the estimation of matching uncertainties. Accordingly, we introduced some
refinements to the standard method when using profile-scale variations, which
could also be useful in other contexts. It consists of fixing the
extreme profile function transition points and varying instead the central point
over a wider range. This leads to an uncertainty estimate without one-sided
uncertainties, and which in our case avoids being overly conservative.

We have constructed an approximation of the $q_T$ spectrum at fixed $\ord{\as^3}$,
which we have used to extrapolate from existing NNLO$_1$ results for $b\bar b\to H+j$
for $q_T\geq 60\GeV$ to smaller $q_T$ and other flavour channels. This is based on
introducing a decorrelation procedure to ensure the correct cancellation
between singular and nonsingular terms at scales $q_T\sim m_H$, and then
approximating the $\mathcal{O}(\as^3)$ nonsingular piece. This allows us to
achieve a final accuracy of N$^3$LL$'$+aN$^3$LO for the $q_T$ spectrum.
Our results display good convergence properties from order to order, and
constitute the highest available accuracy for these processes.
As we have seen in \fig{separation}, at the highest available order the
uncertainties are significantly reduced, such that the different flavour channels
are clearly distinguishable by their different shapes in $q_T$. Our predictions
could therefore be used to improve the determination of Higgs Yukawa couplings
from the Higgs $q_T$ spectrum as carried out in \refscite{ATLAS:2022qef,CMS:2023gjz}.

Our treatment of the $q\bar{q}\to H$ process in this work has neglected finite
quark-mass effects, which are relevant for $q_T \sim m_q$ and are thus an
important consideration especially for $b\bar b\to H$.
The inclusion of these terms in the resummation formalism
has been derived for the Drell-Yan process in \refcite{Pietrulewicz:2017gxc},
and the extension to our case would be relatively straightforward.
It would also be interesting to investigate in more detail the impact of the
resummation of time-like logarithms in the $q\bar qH$ hard function on the
resummed $q_T$ spectrum, as it has been shown to have a nontrivial impact on the
inclusive $b\bar b \to H$ cross section~\cite{Ebert:2017uel}. Finally, we have
only considered the $q_T$ spectrum for inclusive Higgs production here.
Experimentally required cuts on the Higgs decay products induce fiducial power
corrections~\cite{Ebert:2019zkb, Ebert:2020dfc}, which were found to be
important in case of $gg\to H$ production~\cite{Billis:2021ecs}.
It would thus be interesting to investigate their importance also in case of
$q\bar q \to H$. We leave these topics to future work.

%%%%%%%%%%%%%%%%%%%%%%%%%%%%%%%%%%%%%%%%%%%%%%%%%%%%%%%%%%%%%%%%%%%%%%%%%%%%%%%%
\acknowledgments
We thank Jonas Lindert for providing us with the \texttt{OpenLoops} libraries necessary
for this work and Johannes Michel for useful discussions. We are grateful to Lawrence Berkeley National Laboratory and the MIT CTP for their hospitality during the completion of this work. This project has received funding from the European Research Council (ERC)
under the European Union's Horizon 2020 research and innovation programme
(Grant agreement 101002090 COLORFREE). MAL acknowledges support from the Deutsche
Forschungsgemeinschaft (DFG) under Germany's Excellence Strategy -- EXC 2121 ``Quantum Universe''-- 390833306, and from the UKRI guarantee scheme for the Marie Sk\l{}odowska-Curie postdoctoral fellowship, grant ref. EP/X021416/1.
%%%%%%%%%%%%%%%%%%%%%%%%%%%%%%%%%%%%%%%%%%%%%%%%%%%%%%%%%%%%%%%%%%%%%%%%%%%%%%%%

%%%%%%%%%%%%%%%%%%%%%%%%%%%%%%%%%%%%%%%%%%%%%%%%%%%%%%%%%%%%%%%%%%%%%%%%%%%%%%%%
\appendix
%%%%%%%%%%%%%%%%%%%%%%%%%%%%%%%%%%%%%%%%%%%%%%%%%%%%%%%%%%%%%%%%%%%%%%%%%%%%%%%%

%%%%%%%%%%%%%%%%%%%%%%%%%%%%%%%%%%%%%%%%%%%%%%%%%%%%%%%%%%%%%%%%%%%%%%%%%%%%%%%%
\section{Analytic \texorpdfstring{LO$_1$}{LO1} calculation}
\label{app:lo1}
%%%%%%%%%%%%%%%%%%%%%%%%%%%%%%%%%%%%%%%%%%%%%%%%%%%%%%%%%%%%%%%%%%%%%%%%%%%%%%%%

We consider the production of an on-shell Higgs boson, measuring its rapidity $Y$ and the magnitude of its transverse momentum $q_T^2 = |\vec{q}_T|^2$. The underlying partonic process is
\begin{align}
a(p_a) + b(p_b) \to H(q) + X(k_1, ....)
\end{align}
where $a$ and $b$ are incoming partons and $X$ denotes additional QCD radiation. Following \refcite{Ebert:2018gsn}, the cross section can be written as
\begin{align} \label{eq:sigma1}
 \frac{\df\sigma}{ \df Y \df q_T^2} &
 = \int_{0}^{1}\!\! \df \zeta_a \df \zeta_b\, \frac{f_a(\zeta_a)\, f_b(\zeta_b)}{2 \zeta_a \zeta_b \Ecm^2}
   \int\!\biggl(\prod_i \frac{\df^d k_i}{(2\pi)^d} (2\pi) \delta_+(k_i^2) \biggr)
   \int\!\!\frac{\df^d q}{(2\pi)^d} \, |\mathcal{M}(p_a, p_b; \{k_i\}, q)|^2
   \nn\\* &\quad\times
   (2\pi) \delta(q^2-m_H^2) \, (2\pi)^d \delta^{(d)}(p_a + p_b - k - q)
   \, \delta\biggl(Y - \frac{1}{2}\ln\frac{q^-}{q^+}\biggr)
   \, \delta\bigl(q_T^2 - |\vec{k}_T|^2\bigr)
\,.\end{align}
Here, $k=\sum_i k_i$ denotes the total outgoing hadronic momentum, and in particular, $\vec{k}_T= \sum_i \vec k_{i,T}$ is the vectorial sum of the transverse momenta of all emissions. Moreover, the incoming momenta are given by
\begin{align} \label{eq:p_ab}
 p_a^\mu = \zeta_a \Ecm \frac{n^\mu}{2}
\,,\qquad
 p_b^\mu = \zeta_b \Ecm\frac{\bar{n}^\mu}{2}
\,.\end{align}
The $\delta$-functions in \eq{sigma1} set the Higgs boson on-shell and measure its rapidity, fixing the incoming momentum fractions to be
\begin{align} \label{eq:zeta_ab}
 \zeta_a(k) &= \frac{1}{\Ecm} \Bigl(k^- +  e^{+Y} \sqrt{m_H^2 + k_T^2} \Bigr)
\,,\quad
 \zeta_b(k) = \frac{1}{\Ecm} \Bigl(k^+ +  e^{-Y} \sqrt{m_H^2 + k_T^2} \Bigr)
\,,\end{align}
and allowing us to simplify \eq{sigma1} to
\begin{align} \label{eq:sigma}
 \frac{\df\sigma}{\df Y \df q_T^2} &
 = \int\!\biggl(\prod_i \frac{\df^d k_i}{(2\pi)^d} (2\pi) \delta_+(k_i^2) \biggr)
   \f{\pi}{ \zeta_a \zeta_b  \Ecm^4} f_a(\zeta_a)\, f_b(\zeta_b)
   A(Y; \{k_i\}) \, \delta\bigl(q_T^2 - |\vec{k}_T|^2\bigr)
\,. \end{align}
where $A(Y; \{k_i\} )$ denotes the squared matrix-element
\begin{align} \label{eq:Msquared}
 A(Y; \{k_i\}) \equiv |\mathcal{M}(p_a, p_b, \{k_i\}, q=p_a+p_b - k)|^2
\,.\end{align}

%-------------------------------------------------------------------------------
\begin{figure}[t]
\begin{subfigure}{0.33\textwidth}
\includegraphics[width=\textwidth]{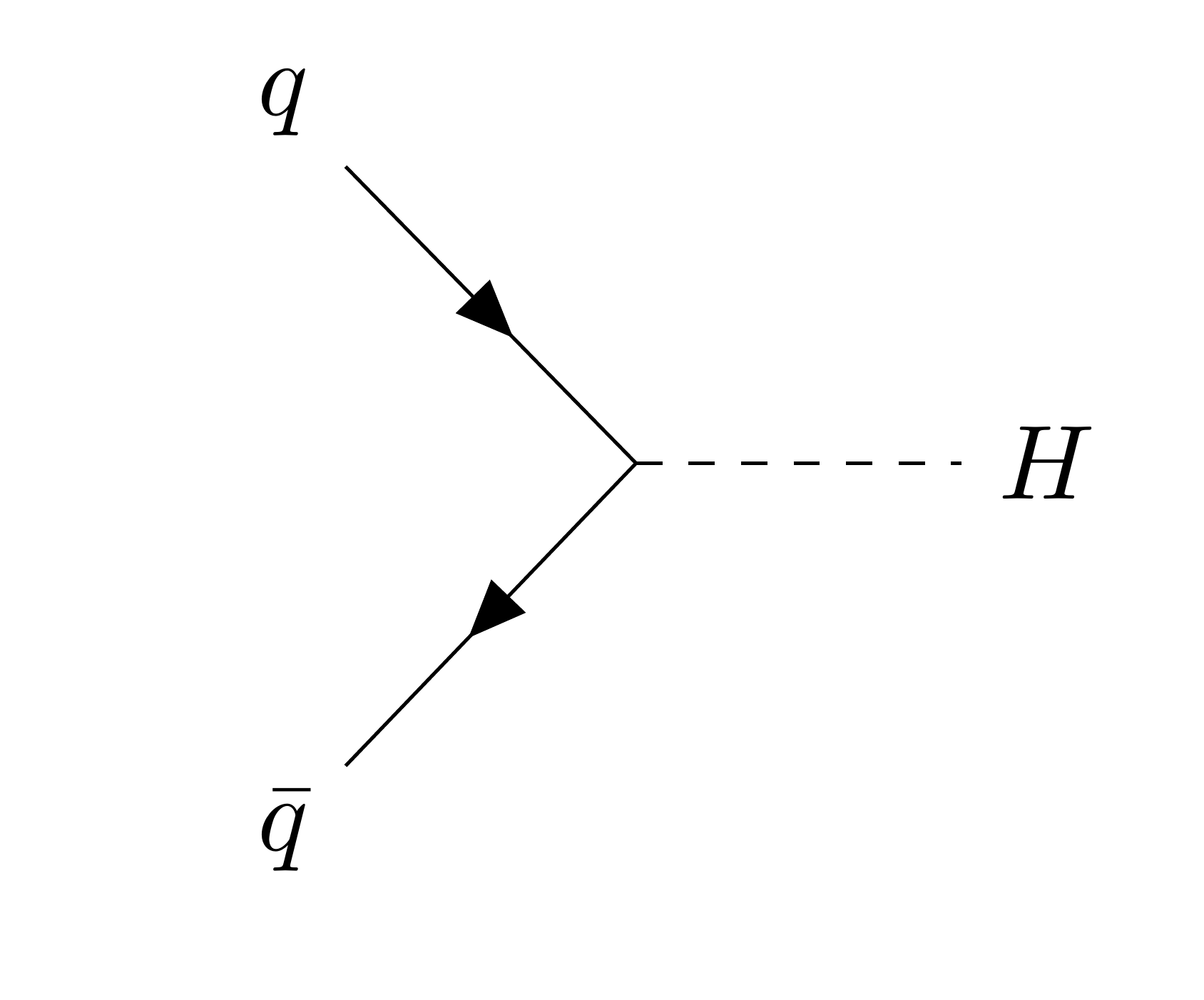}
\caption{ $q \bar{q} \to H$}  \label{fig:BornProcess}
\end{subfigure}%
\begin{subfigure}{0.33\textwidth}
\includegraphics[width=\textwidth]{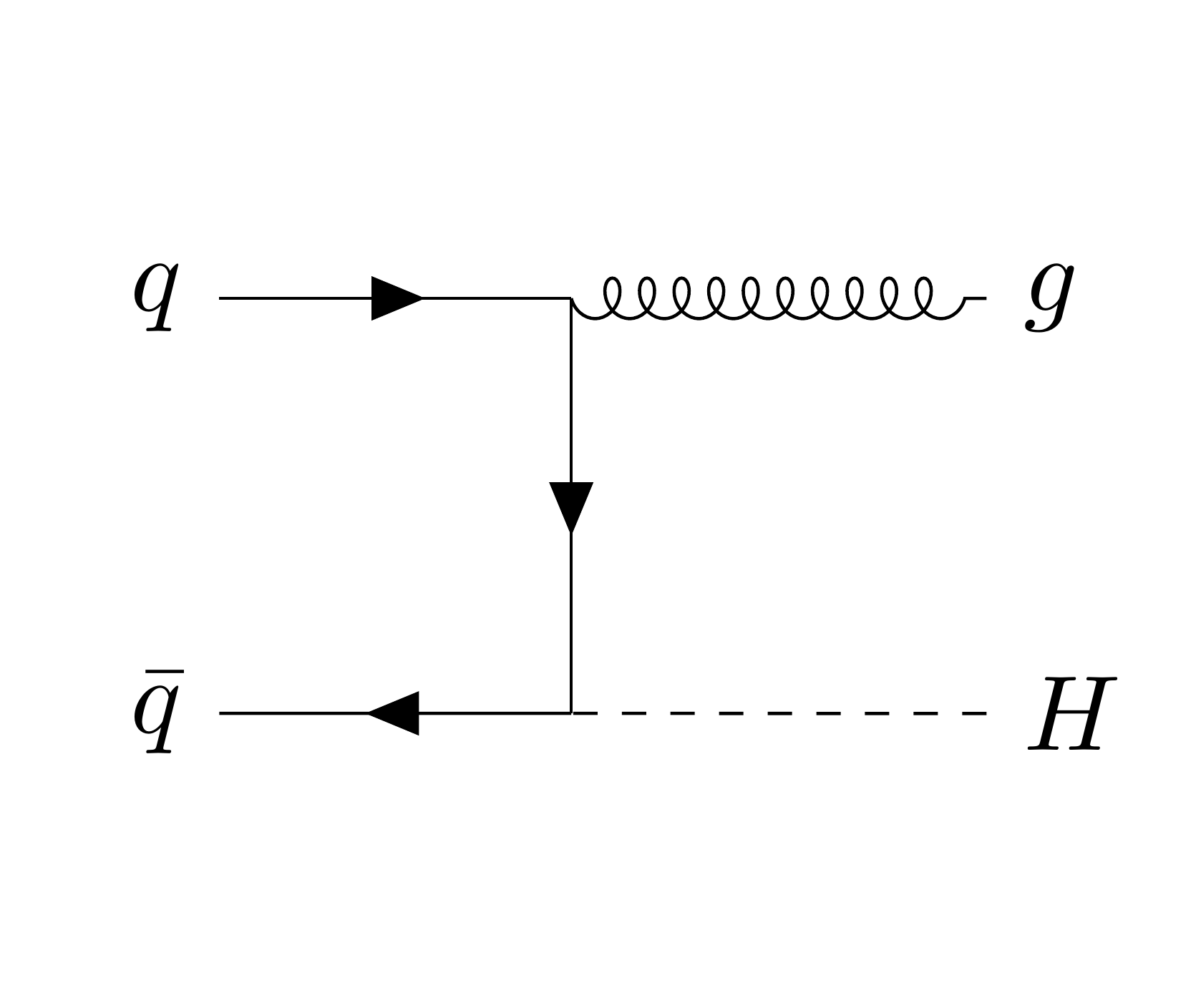}
\caption{$q\bar{q} \to Hg$} \label{fig:qq to gH}
\end{subfigure}%
\begin{subfigure}{0.33\textwidth}
\includegraphics[width=\textwidth]{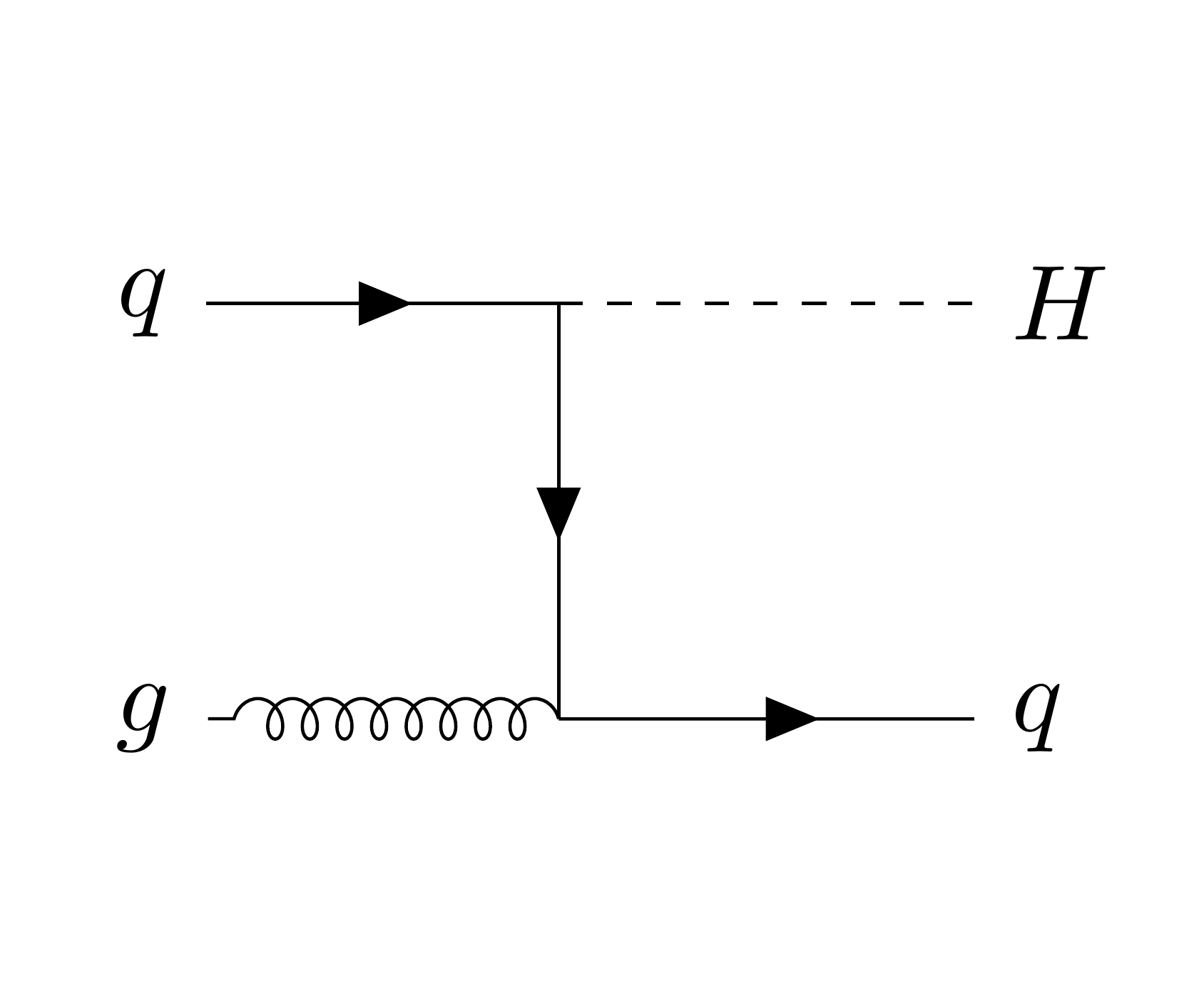}
\caption{$q g \to H q$}  \label{fig:qg to qH}
\end{subfigure}%
\caption{(a): Born process,  (b) and (c): Feynman diagrams contributing to LO$_1$}
\end{figure}
%-------------------------------------------------------------------------------

For reference, we start with the LO$_0$ cross section, i.e. the cross-section for the Born process $q \bar{q}\to H$ without any QCD radiation, which can be seen in \fig{BornProcess}. Since there is no extra emission, the Higgs has no transverse momentum and the cross section is proportional to $\delta(q_T^2)$. Following from \eq{sigma} we obtain
\begin{align} \label{eq:sigmaLO}
 \frac{\df\sigma^{(0)}}{\df Y \df q_T^2} &
 =  \f{\pi}{ x_a x_b  \Ecm^4} f_a(x_a)\, f_b(x_b)  A^{(0)}(Y)\, \delta\bigl(q_T^2\bigr)
\,,\end{align}
where
\begin{align} \label{eq:xab}
 x_a = \frac{m_H \, e^Y}{\Ecm} \,,\quad x_b = \frac{m_H \, e^{-Y}}{\Ecm}
\,,\end{align}
and the squared matrix element is given by
\begin{align}
A^{(0)}(Y) =\f{m_q^2 \, x_a\,  x_b \Ecm^2}{ 2 v^2 N_c}
\,,\end{align}
For convenience we also define the partonic Born cross section $\hat \sigma^{(0)}$ through
\begin{align} \label{eq:sigmaLOpartonic}
 \frac{\df\sigma^{(0)}}{\df Y \df q_T^2} &
 = \int \df x_a \, \df x_b \, \hat{\sigma}^{(0)} \delta(x_a x_b \Ecm^2 -m_H^2) \, \delta\bigg[Y- \f{1}{2} \ln\bigg(\f{x_a}{x_b}\bigg) \bigg] \delta(q_T^2)
\,,\end{align}
yielding
\begin{align}
\hat{\sigma}^{(0)} = \f{\pi m_q^2}{2 v^2 N_c}.
\end{align}

At LO$_1$, we have one QCD emission so the boson will have a finite transverse momentum. From \eq{sigma} we obtain
\begin{align} \label{eq:sigmaLO_1}
 \frac{\df\sigma^{(1)}}{\df Y \df q_T^2} &
 =   \int\! \frac{\df^d k}{(2\pi)^d}\, (2\pi) \delta_+(k^2)\,
      \f{\pi}{ \zeta_a \zeta_b  \Ecm^4} f_a(\zeta_a)\, f_b(\zeta_b)\,
   A^{(1)}(Y; k) \, \delta\bigl(q_T^2 - |\vec{k}_T|^2\bigr)
\nn\\&
 = \frac{q_T^{-2\eps}}{(4\pi)^{2-\eps} \Gamma(1-\eps)} \int_0^\infty \frac{\df k^-}{k^-}
    \f{\pi}{ \zeta_a \zeta_b  \Ecm^4} f_a(\zeta_a)\, f_b(\zeta_b)\,
   A^{(1)}(Y; k)\bigg|_{k^+ = q_T^2/k^-}
\,.\end{align}

The type of diagrams contributing to the squared matrix element can be seen in \figs{qq to gH}{qg to qH}. We decompose $A^{(1)}(Y; k) \equiv \sum_{a,b = q, \bar{q}, g} A^{(1)}_{ab}$ into its contributing channels

\begin{align}
A_{q\bar{q}}(k^-, q_T^2, Y)= A_{\bar{q}q}(k^-, q_T^2, Y) &= \as C_F \f{ 4  \pi   m_b^2  }{N_c v^2}  \bigg(  \f{s_{ab}^2+m_H^4}{s_{ak} s_{bk}}  \bigg)\, ,  \nn\\
A_{g q}(k^-, q_T^2, Y)= A_{g \bar{q}}(k^-, q_T^2, Y) &= \as C_F \f{ 4  \pi   m_q^2  }{(N_c^2-1) v^2}  \bigg(  \f{s_{bk}^2+m_H^4}{-s_{ab} s_{ak}}  \bigg)\,, \nn\\
A_{q g}(k^-, q_T^2, Y)= A_{\bar{q} g}(k^-, q_T^2, Y) &=\as C_F \f{ 4  \pi   m_q^2  }{(N_c^2-1) v^2}  \bigg(  \f{s_{ak}^2+m_H^4}{-s_{ab} s_{bk}}  \bigg)\,   ,
\end{align}
where the $s_{ab}, s_{ak}$ and $s_{bk}$ are kinematic invariants that can be written in terms of $k^-, q_T^2$ and $Y$ as

\begin{align}
	s_{ab} \equiv \,\,\, 2 p_a \cdot p_b&= m_H^2+ 2 q_T^2 + \left(k^+ e^{Y}+ k^- e^{-Y}\right)\sqrt{m_H^2 + q_T^2}\, , \nn\\
	s_{ak} \equiv - 2 p_a \cdot k &= -q_T^2 - k^+e^{+Y}\sqrt{m_H^2 + q_T^2}\, , \nn\\
	s_{bk} \equiv - 2 p_b \cdot k &= -q_T^2 - k^-e^{-Y}\sqrt{m_H^2 + q_T^2}\, ,
\end{align}
with $k^+= q_T^2/k^-$.  The limits of the  $k^-$ integral are found by constraining the PDF argument to be between zero and one, yielding
\begin{align}
    k_\mathrm{min}^-&= \frac{q_T^2}{\Ecm}-e^{-Y} \sqrt{q_T^2+m_H^2}\, ,\nn \\
    k_\mathrm{max}^-&=\Ecm-e^{+Y}\sqrt{ q_T^2+m_H^2}\, .
\end{align}

%%%%%%%%%%%%%%%%%%%%%%%%%%%%%%%%%%%%%%%%%%%%%%%%%%%%%%%%%%%%%%%%%%%%%%%%%%%%%%%%
\section{Nonsingular validation for \texorpdfstring{$c\bar c\to H$}{ccH}
and \texorpdfstring{$s\bar s\to H$}{ssH}}
\label{app:singnons}
%%%%%%%%%%%%%%%%%%%%%%%%%%%%%%%%%%%%%%%%%%%%%%%%%%%%%%%%%%%%%%%%%%%%%%%%%%%%%%%%

%-------------------------------------------------------------------------------
\begin{figure}
\includegraphics[width=\WidthTwoSubfigs]{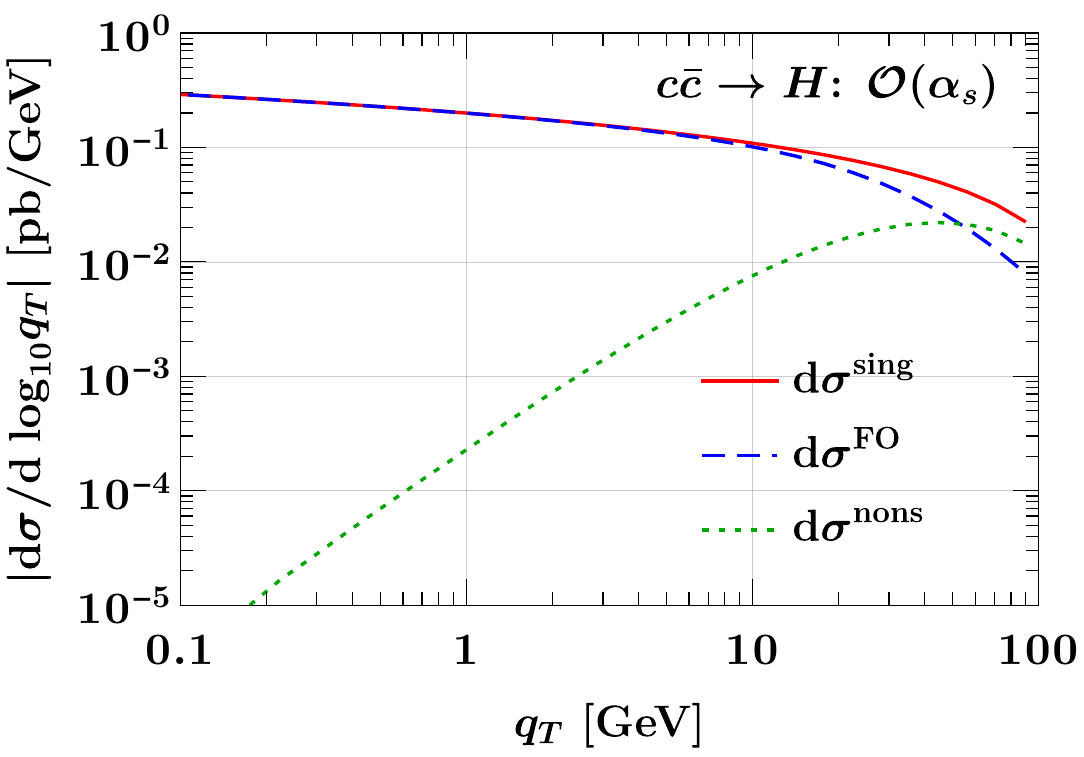}%
\hfill%
\includegraphics[width=\WidthTwoSubfigs]{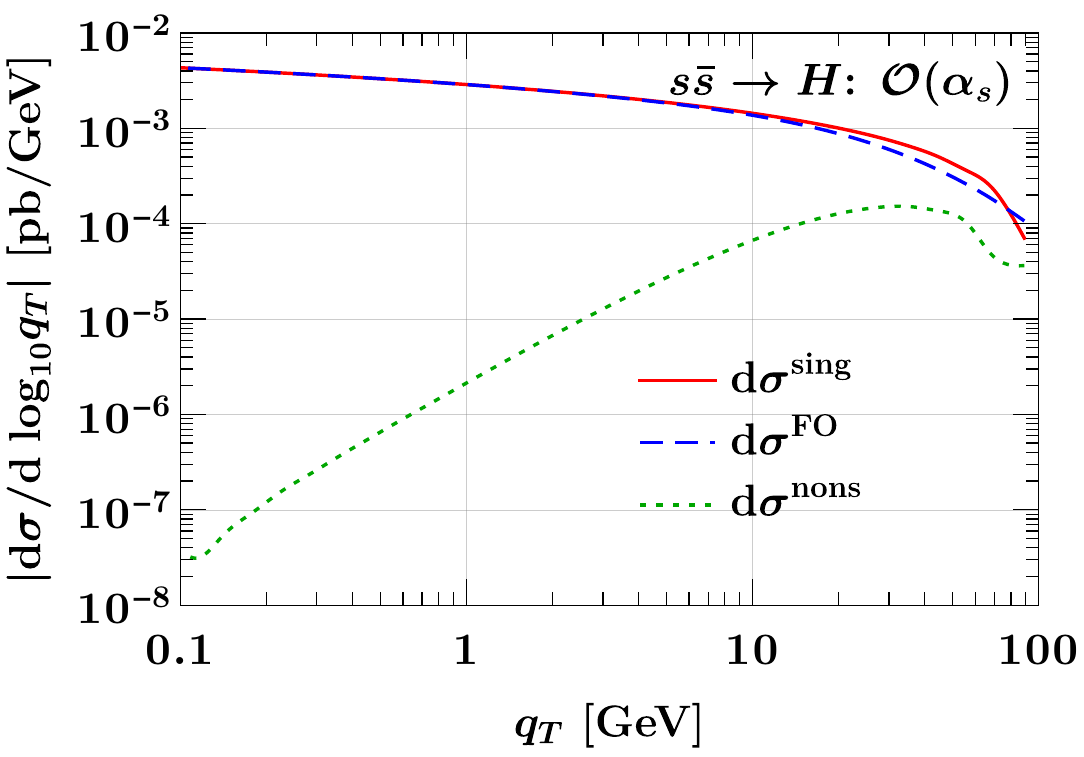}%
\\
\includegraphics[width=\WidthTwoSubfigs]{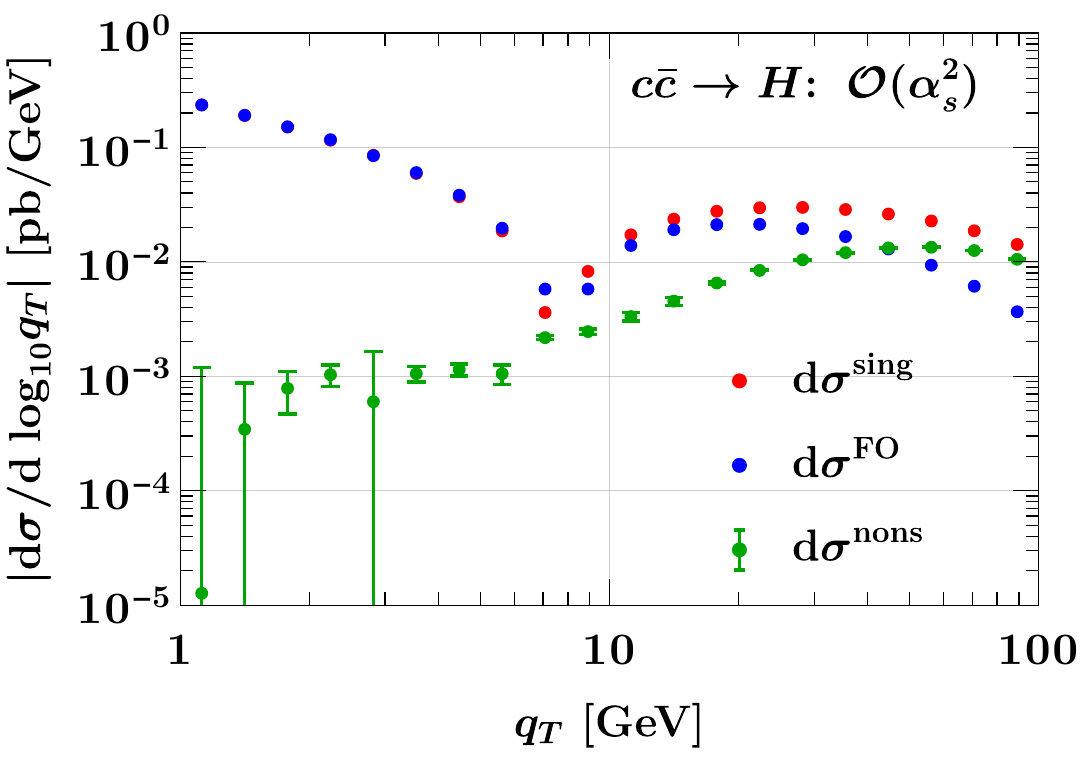}%
\hfill%
\includegraphics[width=\WidthTwoSubfigs]{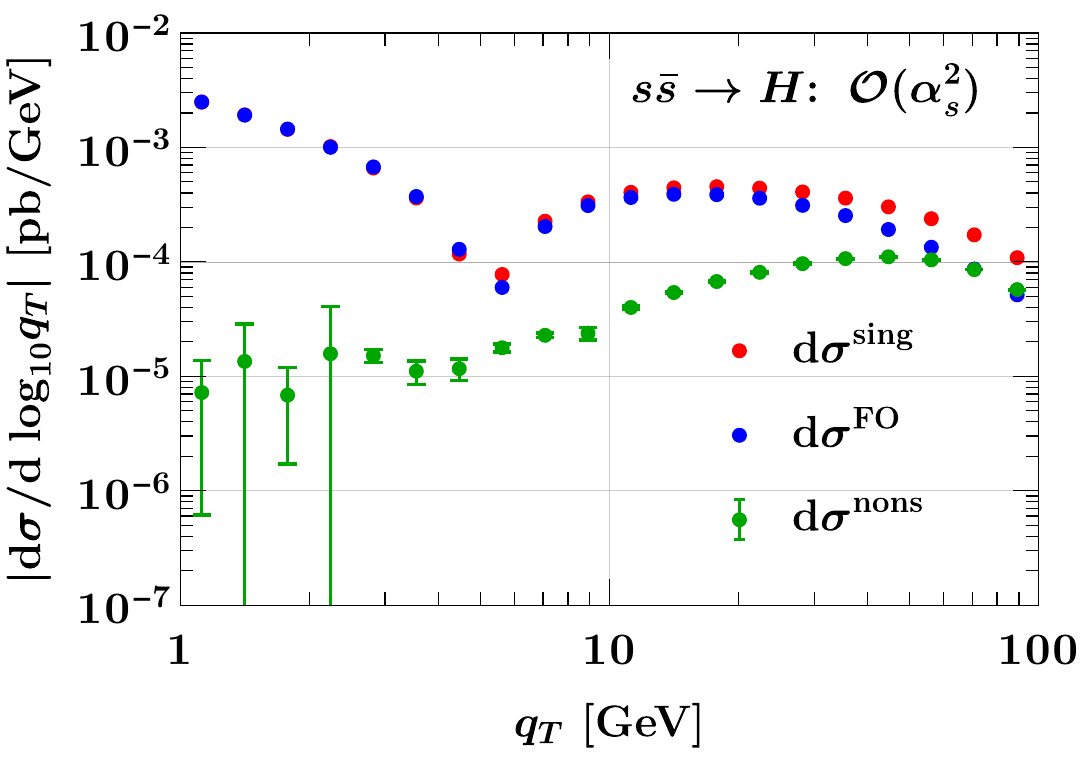}%
\caption{Singular (solid red), full (dashed blue), and nonsingular (dotted green)
contributions for $c\bar{c}\to H$ (left) and  $s\bar{s}\to H$ (right) at fixed
$\ord{\alpha_s}$ (top) and $\ord{\alpha_s^2}$ (bottom).}
\label{fig:nonsingcancloglogccHssH}
\end{figure}
%-------------------------------------------------------------------------------

%-------------------------------------------------------------------------------
\begin{figure}
\includegraphics[width=\WidthTwoSubfigs]{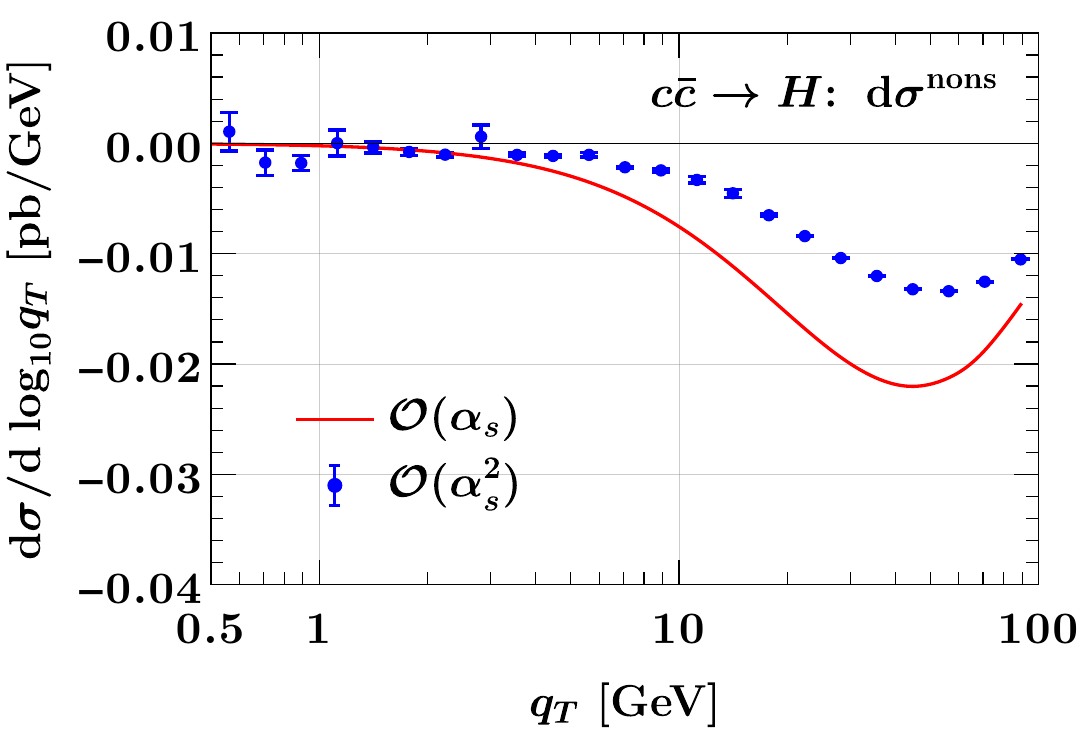}%
\hfill%
\includegraphics[width=\WidthTwoSubfigs]{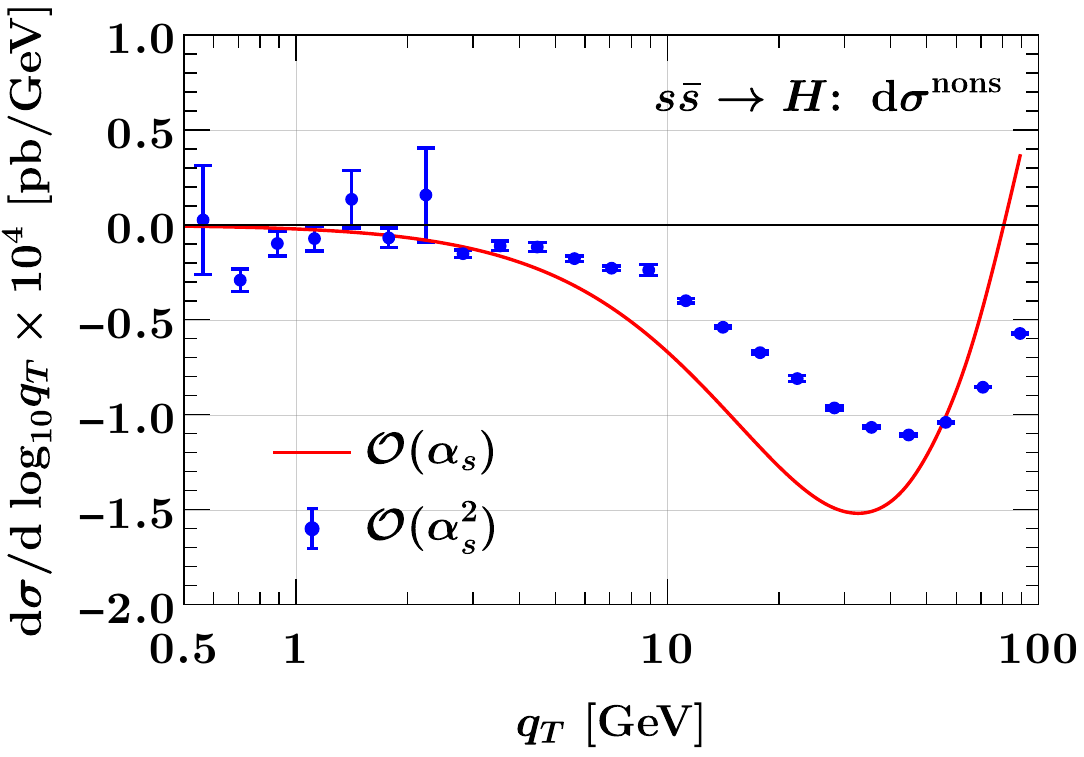}%
\caption{Nonsingular contributions at $\mathcal{O}(\alpha_s)$ (red) and $\mathcal{O}(\alpha_s^2)$ (blue) for $c\bar{c}\to H$ (left) and $s\bar{s}\to H$ (right), corresponding respectively to the green
curves and points in \fig{nonsingcancloglogccHssH}.}
\label{fig:nonsingcancloglinccHssH}
\end{figure}
%-------------------------------------------------------------------------------

In \figs{nonsingcancloglog}{nonsingcancloglin} in \sec{fullFO}, we showed the $\ord{\alpha_s}$ and $\ord{\as^2}$ nonsingular
corrections for $b\bar{b}\to H$. For completeness, here we provide the analogous
plots for $c\bar{c}\to H$ and $s\bar{s}\to H$ on a logarithmic scale in
\fig{nonsingcancloglogccHssH} and on a linear scale in \fig{nonsingcancloglinccHssH}.
In both cases we observe the expected power suppression of the nonsingular similar
to $b\bar b\to H$, which provides an important validation of our implementation of
the LO$_1$ and NLO$_1$ fixed-order results.

%%%%%%%%%%%%%%%%%%%%%%%%%%%%%%%%%%%%%%%%%%%%%%%%%%%%%%%%%%%%%%%%%%%%%%%%%%%%%%%%
\section{Impact of factorization scale and PDF choices}
\label{app:plots}
%%%%%%%%%%%%%%%%%%%%%%%%%%%%%%%%%%%%%%%%%%%%%%%%%%%%%%%%%%%%%%%%%%%%%%%%%%%%%%%%

\begin{figure}
\includegraphics[width=\WidthTwoSubfigs]{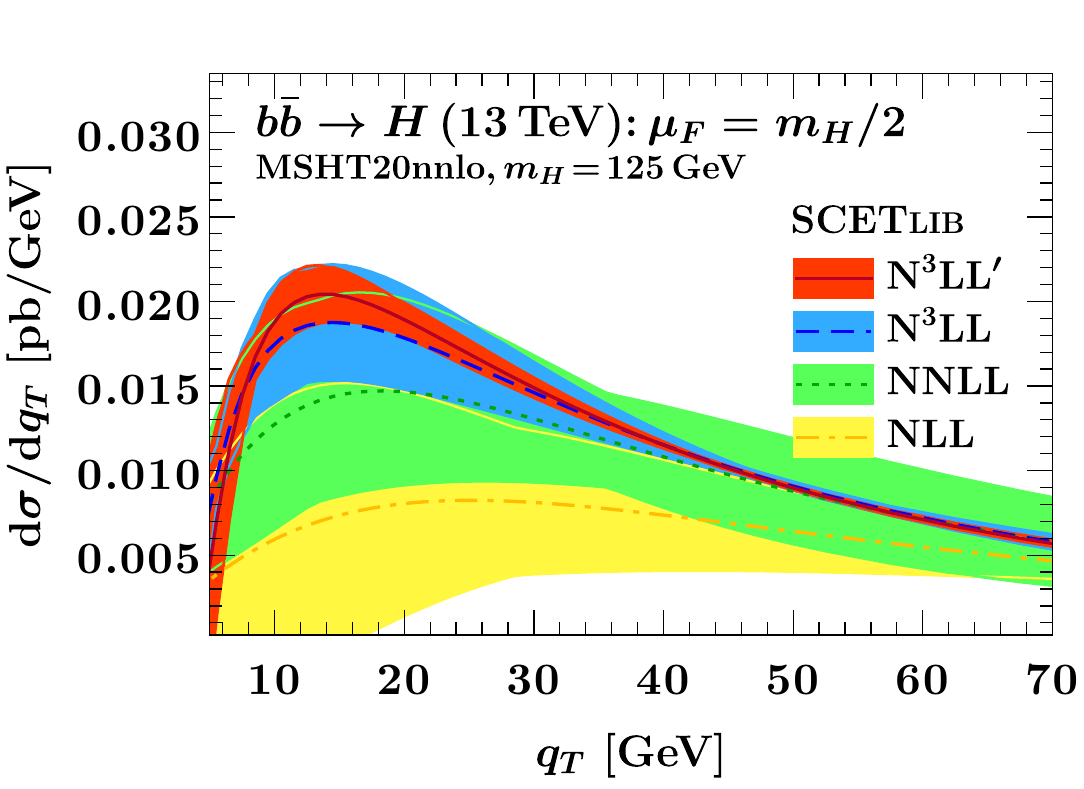}%
\hfill%
\includegraphics[width=\WidthTwoSubfigs]{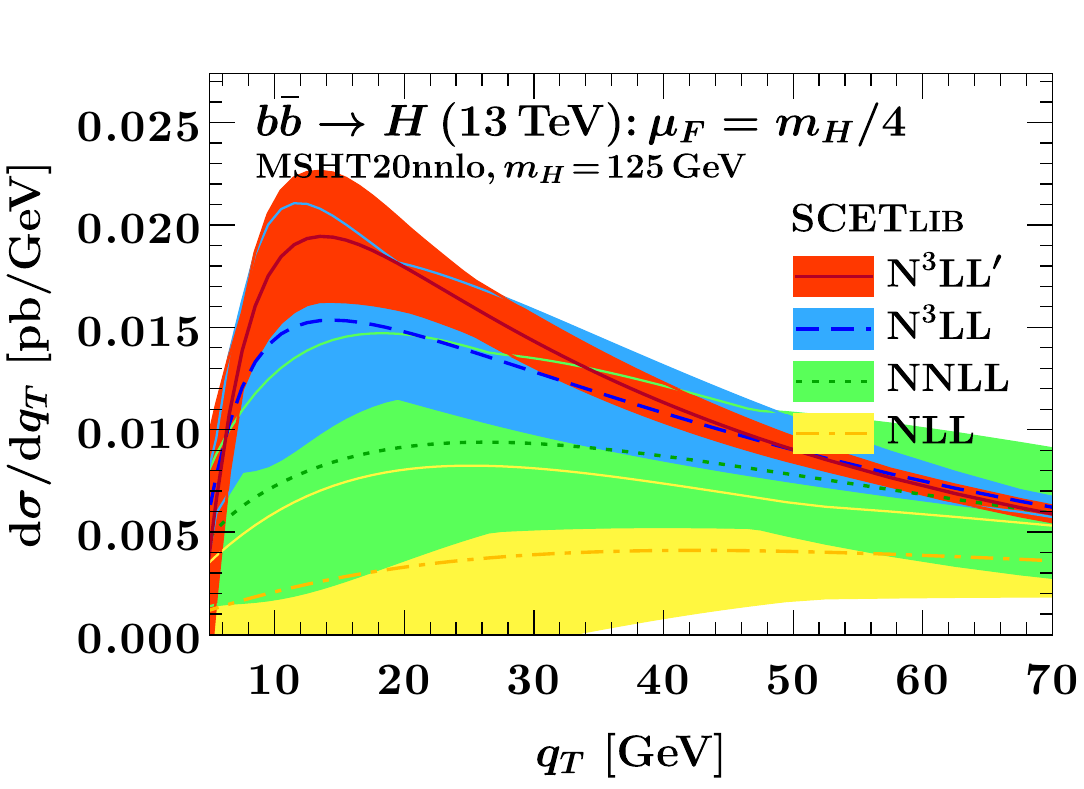}%
\caption{The purely resummed result for $b \bar{b}\to H$ at different orders for
different values of $\mu_F$.}
\label{fig:convergence_mFvals}
\end{figure}

In the context of $b\bar b\to H$, fixed-order predictions often use a low
factorization scale $\mu_F = m_H/2$ or even $\mu_F = m_H/4$. For completeness,
we therefore also give results for $b\bar{b}\to H$ at these lower values for
the central factorization scale. We implement this by taking $w_F =
-1$ or $w_F = -2$ as central choice in \eq{profile_vars}.
\Fig{convergence_mFvals} shows the convergence of the resummed contribution to
the $q_T$ spectrum at NLL (yellow), NNLL (green), N$^3$LL (blue), and N$^3$LL$'$
(red) for $\mu_F=m_H/2$ (left) and $\mu_F=m_H/4$ (right). While the convergence
pattern of subsequent orders is acceptable in all cases, both the corrections
and perturbative uncertainties are somewhat larger for $\mu_F = m_H/2$ than for
$\mu_F = m_H$, and substantially larger for $\mu_F = m_H/4$. In our
context these lower choices are therefore clearly less preferable.

\begin{figure}
\includegraphics[width=\WidthTwoSubfigs]{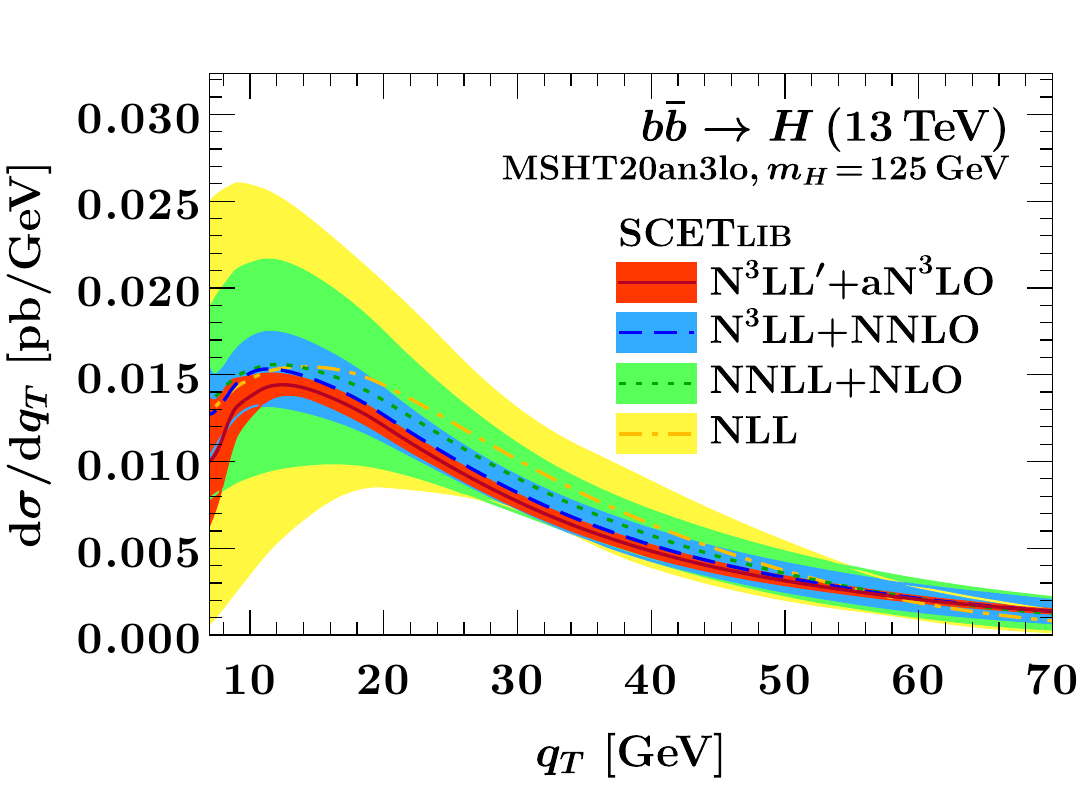}%
\hfill%
\includegraphics[width=\WidthTwoSubfigs]{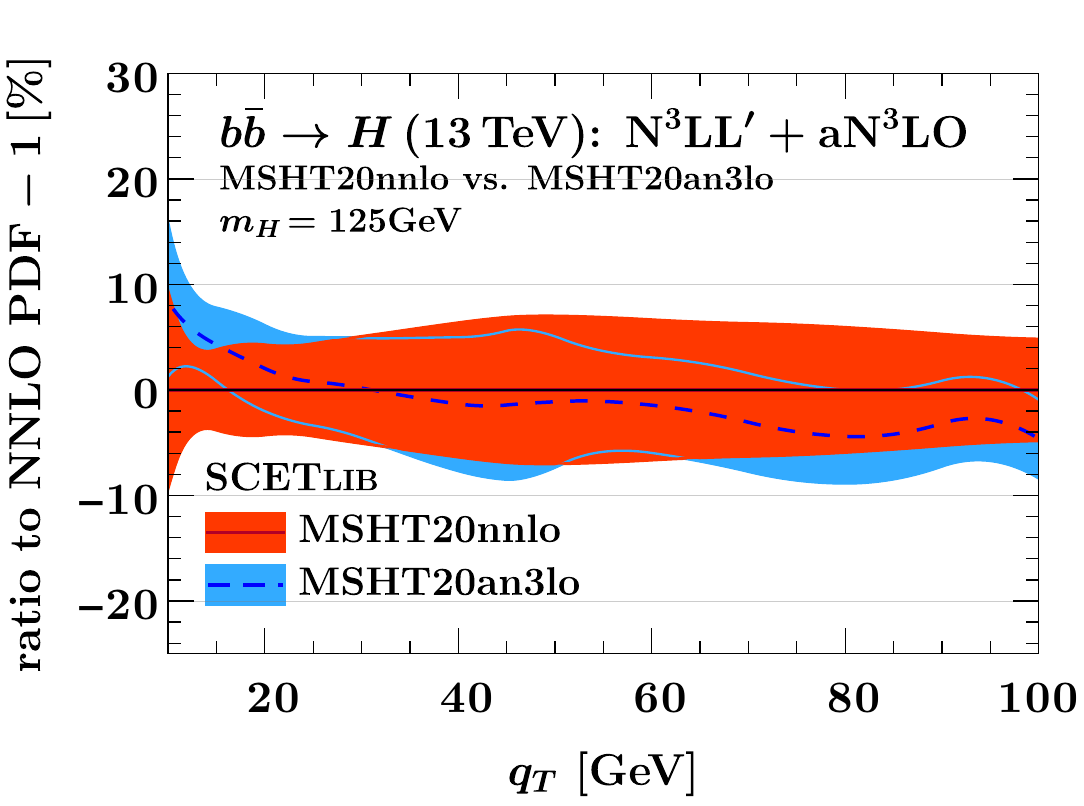}%
\caption{Comparison of the \texttt{MSHT20nnlo} and the \texttt{MSHT20an3lo} PDF sets for $b \bar{b}\to H$.}
\label{fig:an3loPDF}
\end{figure}

In \fig{an3loPDF}, we assess the impact of changing the PDF set from
\texttt{MSHT20nnlo} to \texttt{MSHT20an3lo}. The left panel shows the $q_T$
spectrum at different resummation orders for the \texttt{MSHT20an3lo} PDF set.
As expected, the perturbative convergence is equally good as for our default choice in
\fig{convergence}. The right panel shows the highest order (N$^3$LL$'$+N$^3$LO)
for both PDF sets normalized to our default \texttt{MSHT20nnlo} set. Although
the difference for the $b$-quark PDF between the two sets is quite significant,
this is not reflected in our final results for the $q_T$ spectrum. The largest
difference appears below $q_T \lesssim 20\GeV$, where the \texttt{MSHT20an3lo}
PDF leads to a 5-10\% increase in the spectrum.

%-------------------------------------------------------------------------------
%-------------------------------------------------------------------------------

%-------------------------------------------------------------------------------
%-------------------------------------------------------------------------------

%%%%%%%%%%%%%%%%%%%%%%%%%%%%%%%%%%%%%%%%%%%%%%%%%%%%%%%%%%%%%%%%%%%%%%%%%%%%%%%%
\addcontentsline{toc}{section}{References}
\bibliographystyle{jhep}
\bibliography{refs}

\end{document}